\documentclass[prd,aps,onecolumn,preprintnumbers, showpacs,
nofootinbib,superscriptaddress,notitlepage]{revtex4-1}
\usepackage{amssymb, amsthm}
\usepackage{amsmath}    
\usepackage{graphicx}   
\usepackage{footnote}   
\usepackage[colorinlistoftodos]{todonotes}
\usepackage{multirow}
\usepackage{slashed}    
\usepackage{hyperref}   
\usepackage{physics}
\usepackage[utf8]{inputenc}
\usepackage{color}
\usepackage[normalem]{ulem}

\DeclareMathOperator*{\argmin}{arg\,min}

\begin{document}

\preprint{MSUHEP-19-021}

\title{Machine-Learning Prediction for Quasi-PDF Matrix Elements}

\author{Rui Zhang}
\affiliation{Department of Physics and Astronomy, Michigan State University, East Lansing, MI 48824}

\author{Zhouyou Fan}
\affiliation{Department of Physics and Astronomy, Michigan State University, East Lansing, MI 48824}

\author{Ruizi Li}
\affiliation{Department of Physics and Astronomy, Michigan State University, East Lansing, MI 48824}

\author{Huey-Wen Lin}
\affiliation{Department of Physics and Astronomy, Michigan State University, East Lansing, MI 48824}
\affiliation{Department of Computational Mathematics, Science \& Engineering, Michigan State University, East Lansing, MI 48824}

\author{Boram Yoon}
\affiliation{Computer, Computational, and Statistical Sciences CCS-7, Los Alamos National Laboratory, Los Alamos, NM 87545, USA}

\begin{abstract}

There have been rapid developments in the direct calculation in lattice QCD (LQCD) of the Bjorken-$x$ dependence of hadron structure through large-momentum effective theory (LaMET). 
LaMET overcomes the previous limitation of LQCD to moments (that is, integrals over Bjorken-$x$) of hadron structure, allowing LQCD to directly provide the kinematic regions where the experimental values are least known. 
LaMET requires large-momentum hadron states to minimize its systematics and allow us to reach small-$x$ reliably. This means that very fine lattice spacing to minimize lattice artifacts at order $(P_z a)^n$ will become crucial for next-generation LaMET-like structure calculations.
Furthermore, such calculations require operators with long Wilson-link displacements, especially in finer lattice units, increasing the communication costs relative to that of the propagator inversion. In this work, 
we explore whether machine-learning (ML) algorithms can make  predictions of correlators
to reduce the computational cost of these LQCD calculations. 
We consider two algorithms, gradient-boosting decision tree and linear models, applied to LaMET data,
the matrix elements needed to determine the kaon and $\eta_s$ unpolarized parton distribution functions (PDFs), meson distribution amplitude (DA), and the nucleon gluon PDF. 
We find that both algorithms can reliably predict the target observables with different prediction accuracy and systematic errors.  The predictions from smaller displacement $z$ to larger ones work better than those for momentum $p$ due to the higher correlation among the data.
\end{abstract}

\maketitle

\section{Introduction}

In the early days, probing hadron structure with lattice QCD (LQCD) was limited to only the first few moments, due to complications arising from the breaking of rotational symmetry by the discretized Euclidean spacetime.
The nonzero lattice spacing breaks the symmetry group of Euclidean spacetime from $O(4)$ to the discrete hypercubic subgroup $H(4)$. Due to the reduced symmetry, the required operators are more complicated and often either suffer from divergences or mix with other operators under renormalization. This is treatable but complicated. 
As a result, even with increasing computational resources becoming available to the lattice-QCD community, LQCD hadronic structure calculations were limited to the lowest few moments (see Ref.~\cite{Aoki:2019cca,Lin:2017snn} and references within for more details).
Although modeling the $x$-dependence to reproduce the calculated lattice moments to gain information on the $x$-dependence~\cite{Detmold:2001dv} was attempted, this will only give the combinations of the difference between quark and antiquark contributions rather than individual (anti)quark contributions. 
Experiments such as E665 at FNAL can probe nucleon sea flavor asymmetry, meaning that lattice QCD would be excluded if it could only apply traditional moment calculations. Similarly, STAR at RHIC is probing the polarized (anti)quark structure of nucleon. The future electron-ion collider (EIC) will further study sea structure. 
Facing these challenges, LQCD required a new computationally friendly approach to extend its applicability to calculations of PDFs and catch up with ongoing experimental efforts.

Large-momentum effective theory (LaMET)~\cite{Ji:2013dva} is one of the most widely adopted new methods for calculating the full $x$ dependence of hadron structure. In the LaMET framework,
we take an operator containing an integral of gluonic field strength along a line and boost the nucleon momentum toward the speed of light, tilting the spacelike line segment toward the light-cone direction. The time-independent, nonlocal (in space) correlators at finite $P_z$ can be directly evaluated on the lattice. For example, the quark unpolarized distribution of a hadron can be calculated via
\begin{equation}
\label{eq:qlat}
q_\text{lat}(x,\mu,P_z)= \int \frac{dz}{4\pi} e^{izk} \times 
  \left\langle \vec{P} \right| \bar{\psi}(z) \Gamma
      \left( \prod_n U_z(n\hat{z})\right) \psi(0) \left| \vec{P} \right\rangle,
\end{equation}
where $U_z$ is a discrete gauge link in the $z$ direction, 
$\Gamma=\gamma_t$, 
$x=k/P_z$, $\mu$ is the renormalization scale and $\vec{P}$ is the momentum of the hadron, taken such that $P_z\rightarrow\infty$. 
The $q_\text{lat}(x,\mu,P_z)$, often called the ``quasi-PDF''~\cite{Detmold:2019ghl}, is related to the light-cone PDF through a factorization theorem, where the former can be factorized into a perturbative matching coefficient and the latter, up to power corrections suppressed by the nucleon momentum. This factorization theorem is founded in LaMET~\cite{Ji:2013fga,Ji:2013dva,Hatta:2013gta,Ji:2014gla,Ji:2014lra}, where the matching coefficient can be calculated exactly in perturbation theory.
Lattice-QCD results using LaMET already include the isovector quark PDF of the nucleon~\cite{Lin:2014zya,Alexandrou:2015rja,Chen:2016utp,Alexandrou:2016jqi,Lin:2017ani}, the pion genearlized parton distribution(GPD)~\cite{Chen:2019lcm}, the meson DAs~\cite{Zhang:2017bzy,Chen:2017gck} and the nonperturbative renormalization in the regularization-independent momentum subtraction (RI/MOM) scheme~\cite{Chen:2017mzz,Alexandrou:2017huk}. Certain technical issues regarding the nonperturbative renormalization were raised and addressed in Refs.~\cite{Constantinou:2017sej,Alexandrou:2017huk,Green:2017xeu,Chen:2017mzz,Chen:2017mie,Lin:2017ani,Chen:2017lnm}. The finite volume effect in nucleon quasi-PDF was studied in~\cite{Lin:2019ocg}.

Even with these promising results published and efforts ongoing, much work remains to be done. For example, most work so far has been limited to a single ensemble; more detailed studies incorporating the systematic errors from lattice artifacts, such as finite volume and lattice spacing, is necessary to reach precision LQCD PDFs. Larger boost momentum in the hadron is important to suppress finite-momentum corrections, as well as getting the antiquark distribution and small-$x$ quark distribution corrections. 
Ensembles with smaller lattice spacing ($a^{-1} > 4$~GeV) will become the crucial factors in the next generation of LaMET calculations. To reach larger boost momentum $P_z$, a smaller lattice spacing is needed to control the $(P_z a)^n$ lattice artifacts, similar to how heavy-quark studies must control the heavy-quark mass artifacts at order $(m_q a)^n$. Likely, more than $O(100,000)$ calculations will be needed to get a good signal-to-noise for the three-point correlators and allow us to disentangle the excited states from the ground state. 
A larger number of degrees of freedom will be necessary to keep the finite-volume systematic within the consensus optimal region, $M_\pi L \approx 4$. More communication costs will be incurred transporting the Wilson link from one side of the lattice to the other, which can easily become a dominating cost for the calculation.
Although optimizing communication efficiency may address the latter problem, we are hoping to find a method that will work for both the large-momentum and Wilson-link displacement issues that are characteristic of LaMET and its similar approaches. 

Recently, authors of Ref.~\cite{Yoon:2018krb} introduced a machine-learning (ML) approach predicting observables by taking advantage of the correlations between lattice QCD observables. 
Two types of data with high-statistics measurements, $O(100,000)$, were used in the studies:
nucleon isovector charges and 
the CP-violating phase induced by the quark chromoelectric dipole moment interactions. 
The authors found a reduction in
the computational cost by 7\%--35\%, depending on observable, showing very promising potential
for lattice-QCD applications. In this work, we are interested in finding out how well the ML approach would work with the difficult computational situation in LaMET-type observables and whether computational cost can be saved for future large lattices (say, $64^3\times 192$ and above). Although this paper focuses on the discussion with quasi-PDF, what we learn here also applies to the pseudo-PDF\footnote{Another paper~\cite{Karpie:2019eiq} applied the neural network(NN) algorithm to the inversion problem to reconstruct PDF from pseudo-PDF matrix elements, though the model was trained and tested on mock datasets instead of real lattice data.}~\cite{Orginos:2017kos,Radyushkin:2017cyf,Joo:2019bzr,Joo:2019jct} correlators since the building blocks of matrix elements are the same.

The structure of this paper is as follows: In Sec.~\ref{sec:ML} we briefly describe the two ML algorithms used in this work.  Section~\ref{sec:lattceME} demonstrates the application of both algorithms to LaMET-type observables, including the correlators from the meson distribution amplitude, kaon and $\eta_s$ parton distribution functions and nucleon gluon parton distribution function. We compare results of the ML predictions. 
We summarize the conclusions and future prospects of this work in Sec.~\ref{sec:summary}.

\section{Machine-Learning Algorithm}\label{sec:ML}

ML works by optimizing a prediction model mapping between input and output data, creating a function approximating the relationship between them, inferred from data. The model is built from a set of data whose label (output) is known, and it is applied to make predictions of the labels for a new set of data whose label is unknown,
assuming that there exists a consistent mapping function between the input and output data. 
In this study, we use regression algorithms, a class of ML approaches, to make quantitative predictions of lattice-QCD measurements. Specifically, the supervised ML regression algorithms we use are the simple linear regression and the gradient boosting tree (GBT) algorithms~\cite{GBT} implemented in the Python scikit-learn package~\cite{scikit-learn}. 
Although all the data we use in this test have labels, we divide them into the labeled and unlabeled sets, by hiding the label for the unlabeled dataset. Then, the labeled dataset is used for training and bias correction procedure, while the unlabeled dataset is used for the test of the trained regression algorithm.


Gradient boosting is one of the techniques creating a strong model from an ensemble of weak prediction models \cite{Friedman00greedyfunction, Friedman:2002:SGB:635939.635941}. For GBT, the shallow decision trees are used as the weak learners (nested mappings as elements of a more complicated function approximation) in series and building the active model:
\begin{equation}
\label{Eqn1}
f_k(x) = \sum_{i=1}^{k} r_i h_i(x), \qquad f(x) = f_{N_\text{est}}(x)
\end{equation}
where $N_\text{est}$ is the number of estimators,
$r_i$ is the learning rate, 
and $h_i(x)$ is the function used in the base decision tree to minimize the loss function $L$:
\begin{equation}
h_i(x) = \argmin_h \sum_j L\left(y_j, f_{i-1}(x_j) + h(x_j)\right)
\end{equation}
where the subscript $j$ iterates over the training-data samples. 
In this work, the loss function is chosen to be the mean squared error, and the depth of the decision tree is fixed at 3.
To optimize the ML predictions, we must choose model parameters in Eq.~\eqref{Eqn1} within the proper range.
Two parameters are tuned explicitly in this process: 
the learning rate $r$,
and the number of estimators $N_\text{est}$.

The prediction accuracy of GBT is compared with those of the linear regression model
\begin{equation}
f^\text{lin}(\vec{x})=\theta_0 + \vec{\theta}\cdot\vec{x}
\end{equation}
for the same set of data. Quantitatively, the quality of the prediction accuracy of the regression models is represented by the fit variance $F_v$ defined as
\begin{align}
F_v &= 1 - \left(\langle(C_\text{ul} - C_\text{pred})^2\rangle - \langle C_\text{ul} - C_\text{pred}\rangle^2\right)/\sigma^2,
\end{align}
where $C_\text{ul}$ and $C_\text{pred}$ are the observed and predicted measurements on unlabeled dataset, respectively, and
$\sigma^2$ is the variance of the observed measurements. The higher value of $F_v$ indicates the better fit quality, and the maximum value of $F_v$ is 1, which shows a perfect prediction, $C_\text{pred} = C_\text{ul}$. In practical calculations, the $F_v$ can be calculated on the bias correction dataset of the labeled dataset, which is described below. However, we use the unlabeled dataset for the calculation of the $F_v$, because $C_\text{ul}$ are available in this test study.

Prediction from a ML algorithm may have bias due to prediction error. We follow the bias correction strategy introduced in Ref.~\cite{Yoon:2018krb} to remove the bias in our  estimate and define the bias corrected prediction as: 
\begin{eqnarray} 
\langle C_{\text{pred},\text{BC}}\rangle=\langle C_{\text{pred}}\rangle_\text{ul}+ \langle C_\text{BC} - C_\text{pred}\rangle_\text{BC},
\end{eqnarray}
where the brackets with subscripts ``ul'' and ``BC'' denote averages over the unlabeled and bias-correction datasets, respectively. After bias correction, the expectation value of the prediction becomes the same as the expectation value of the ground truth, and its statistical error includes the systematic error due to inaccurate predictions. After the bias correction, therefore, our main concern is reducing the statistical error of the final estimate.

We normalize the labeled data so that the standard deviation of each input measurement becomes 1 before we pass it to the ML algorithms. Each subset of the data (training, bias-correction, and unlabeled datasets) described in Ref.~\cite{Yoon:2018krb} are chosen such that the configurations are evenly distributed. The convention of notations throughout this work is:

\begin{table}[h]
\centering
\begin{tabular}{@{} |l|l| @{}}
\hline
\textbf{Subscript} & \textbf{Convention}\\\hline
in & input to the model\\\hline
pred & prediction of the model\\\hline
pred,BC & bias-corrected prediction\\\hline
tr & labeled training data\\\hline
BC & labeled bias correction data\\\hline
lb & all labeled data\\\hline
ul & unlabeled data\\\hline
\end{tabular}
\caption{The convention for the subscripts we use in this work. }
\end{table}

The errors of the predictions are estimated using the bootstrap method. 
We randomly pick the bootstrap samples for labeled and unlabeled datasets, and partition the labeled one into training and BC datasets. We train the model and estimate the bias correction on each bootstrap sample of labeled data. We make prediction on the corresponding sample of unlabeled data and calculate the average of the results for unlabeled data. The error is then estimated over all bootstrap samples.

\section{Application to Lattice Quasi-PDF Matrix Elements}\label{sec:lattceME}

\subsection{Predictions of meson quasi-DA measurements}

Meson distribution amplitudes (DAs) $\phi_M$ are important universal quantities appearing in many factorization theorems, which allow for the description of exclusive processes at large momentum transfers $Q^2 \gg \Lambda_\text{QCD}^2$~\cite{Beneke:1999br,Beneke:2001ev}.
Such quantities can be calculated using large-momentum effective theory (LaMET)~\cite{Ji:2013dva,Ji:2014gla} by calculating the time-independent spatial correlators (the quasi-DA) on the lattice, followed by a matching procedure with corrections suppressed by the hadron momentum.
The light-cone meson DA
\begin{equation}
\phi_M(x,\mu)=\frac{i}{f_M}\int\frac{d\xi}{2\pi}e^{i(x-1)\xi n\cdot P}\langle M(P)| \bar{\psi}_1(0)n\cdot\gamma\gamma_5 U (0, \xi n)\psi_2(\xi n)|0\rangle
\end{equation}
can be extracted from the quasi-DA
\begin{equation}
\tilde{\phi}_M(x,\mu_R,P_z)=\frac{i}{f_M}\int\frac{dz}{2\pi}e^{i(x-1)zP_z}\langle M(P)| \bar{\psi}_1(0)\gamma_z\gamma_5 \prod_{x=0}^{z-1} U_z (x,t)\psi_2(z)|0\rangle
\end{equation}
through the matching~\cite{Ji:2015qla}
\begin{equation}
\tilde{\phi}_M(x,\mu_R,P_z)=\int dy\,Z_\phi(x,y,\mu,\mu_R,P_z),\phi_M(y,\mu)+\mathcal{O}\left(\frac{\Lambda_\text{QCD}}{P_z^2},\frac{m_M^2}{P_z^2}\right).
\end{equation}
according to LaMET.
The quasi-DA can be obtained by computing the following correlators for $K^-$ and $\eta_s$, as presented in the Refs.~\cite{Zhang:2017bzy,Chen:2017gck}:
\begin{equation}
C_\text{2pt}(z,P,t)=\langle 0| \int d^3y\, e^{i\vec{P}\cdot\vec{y}}\bar{\psi}_1(\vec{y},t)\gamma_z\gamma_5 \prod_{x=0}^{z-1} U_z (y+x\hat{z},t)\psi_2(\vec{y}+ z\hat{z},t)\bar{\psi}_2(0,0)\gamma_5\psi_1(0,0) |0 \rangle
\end{equation}
where $\{\psi_1,\psi_2\}$ are $\{u,s\}$ for $K^-$ and $\{s,s\}$ for $\eta_s$, $U(\vec{x},\vec{x}+z)$ is the Wilson line connecting lattice site $\vec{x}$ to $\vec{x}+z\hat{z}$.

We perform a calculation using gauge ensembles with clover valence fermions on a $48^3\times 144$ lattice with $2+1+1$ flavors (degenerate up and down, strange, and charm degrees of freedom) of highly improved staggered quarks (HISQ)~\cite{Follana:2006rc} generated by the MILC Collaboration~\cite{Bazavov:2012xda}. The lattice spacing $a\approx 0.06$~fm, and $m^\text{sea}_\pi=310$~MeV. Hypercubic (HYP) smearing~\cite{Hasenfratz:2001hp} is applied to the configurations. The bare quark masses and clover parameters are tuned to recover the lowest pion mass of the staggered quarks in the sea. Correlators are calculated from momentum-smearing sources~\cite{Bali:2016lva} using 20 source locations on each of the 95 configurations (1900 measurements in total).

We make two predictions using the ML algorithm. One is to predict the correlators at larger link length $z_\text{pred}$ from the correlators at $z_\text{in}<z_\text{pred}$. The other is to predict the correlators of larger momentum $p_\text{pred}$ from the correlators of $p_\text{in}<p_\text{pred}$.

\begin{figure}[htb]
	\centering
	\includegraphics[width=0.49\textwidth]{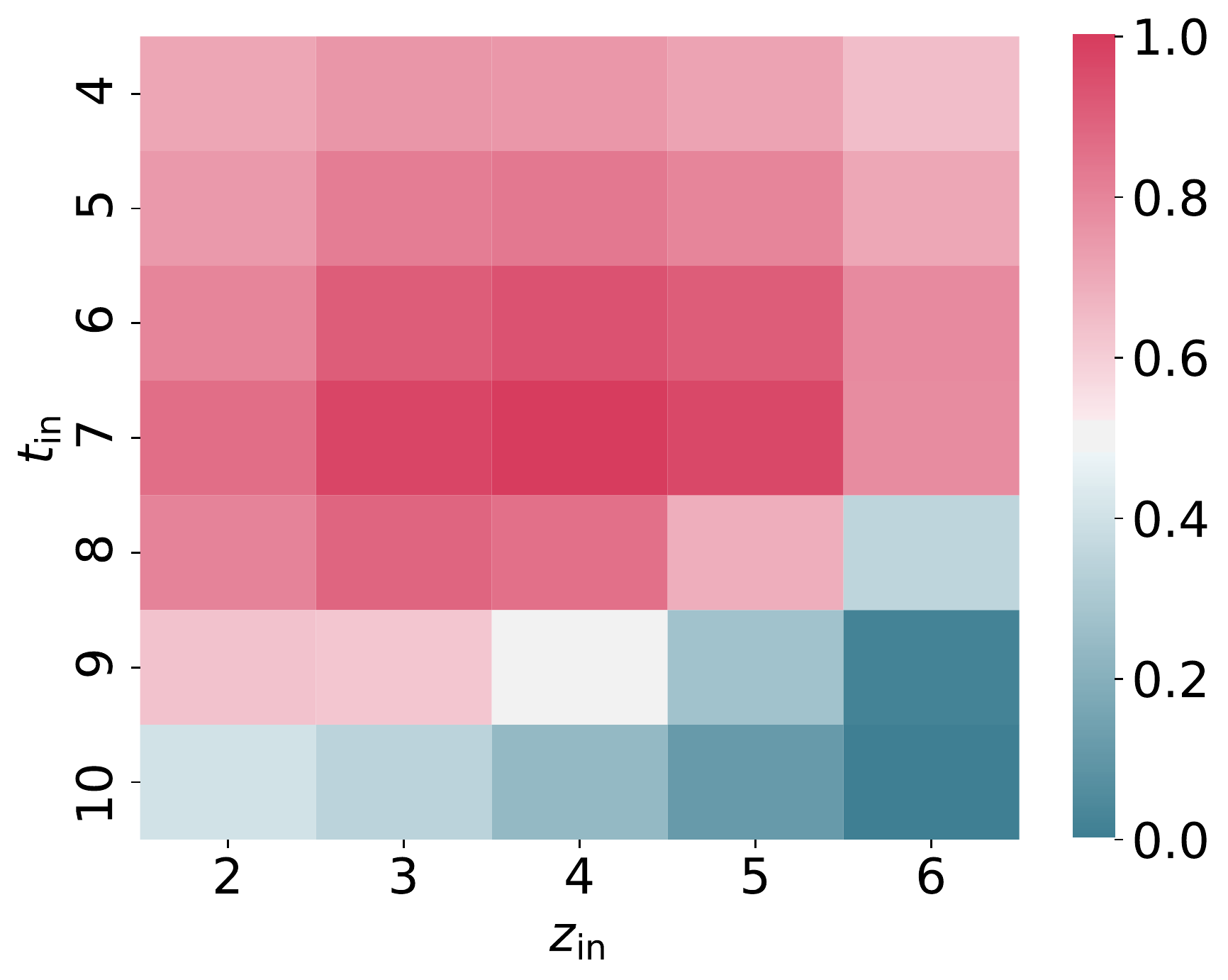}
	\includegraphics[width=0.49\textwidth]{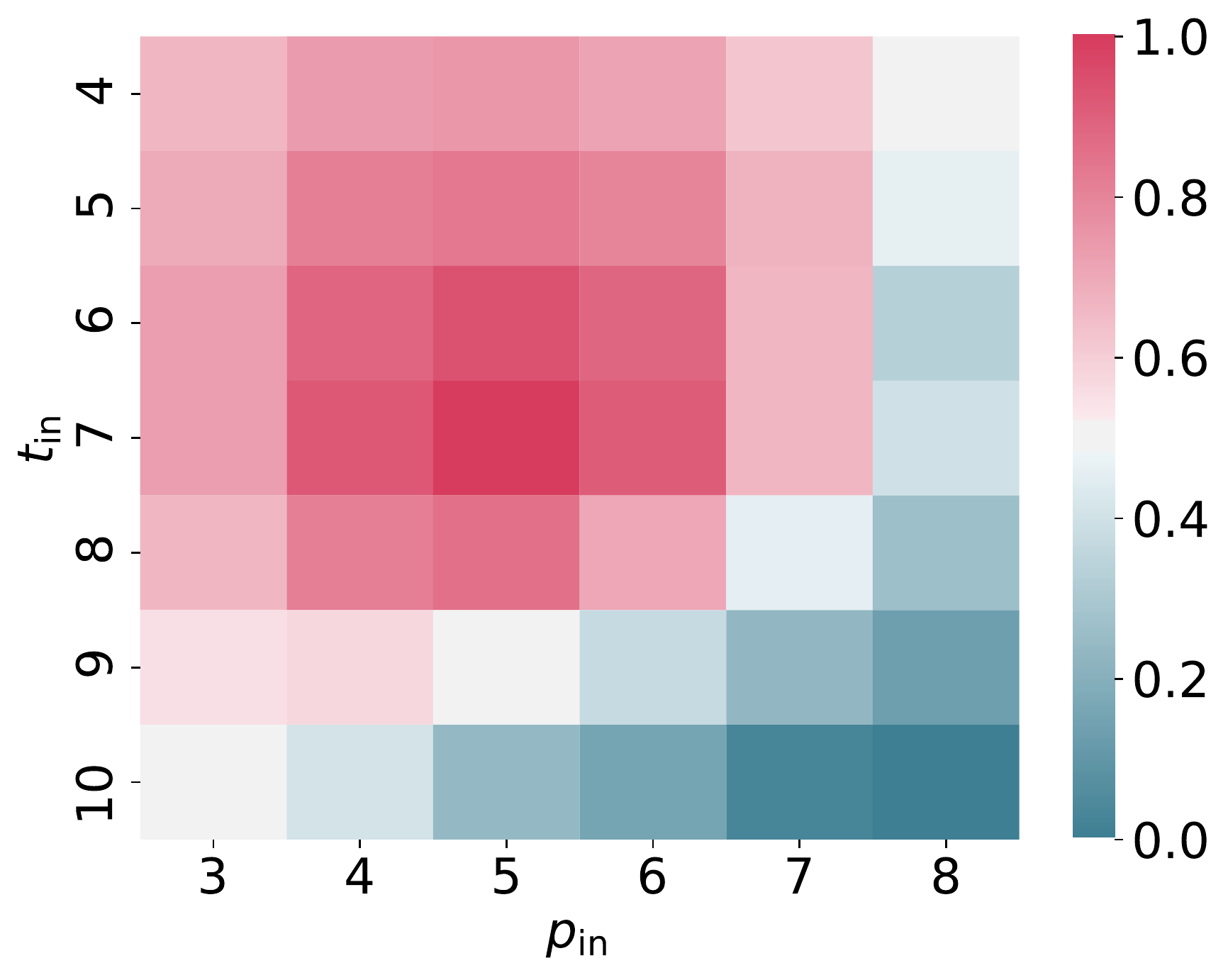}
	\caption{Correlations between target $\eta_s$ DA $C_\text{2pt}$ data at $z_\text{pred}=4$, $p_\text{pred}=5$, $t_\text{pred}=7$ with input data at a different link length (momentum) and timeslice for $z$-prediction (left) and $p$-prediction (right). The correlation decays quickly, especially at larger $t$.}
	\label{fig:pzcorr}
\end{figure}

To determine what input data to use for these predictions, we first check the correlations among datasets with different momenta, link lengths and timeslices. The results are shown in Fig.~\ref{fig:pzcorr}. Here, we set the target data to be the 2-point quasi-DA correlators at $p_\text{pred}=5$, $z_\text{pred}=4$ with input data $p_\text{in}=4$, $z_\text{in}=4$ for $p$-prediction and $p_\text{in}=5$, $z_\text{in}<4$ for $z$-prediction. We select the timeslice $t_\text{pred}=7$ to check the correlations.

Despite the larger error, larger timeslices have a weaker correlation with the target data. This suggests that we should use input data close to the timeslice of the target data. On the other hand, we should be able to extend the range of momentum or links of the input.

\begin{figure}[htbp]
	\centering
	\includegraphics[width=0.49\textwidth]{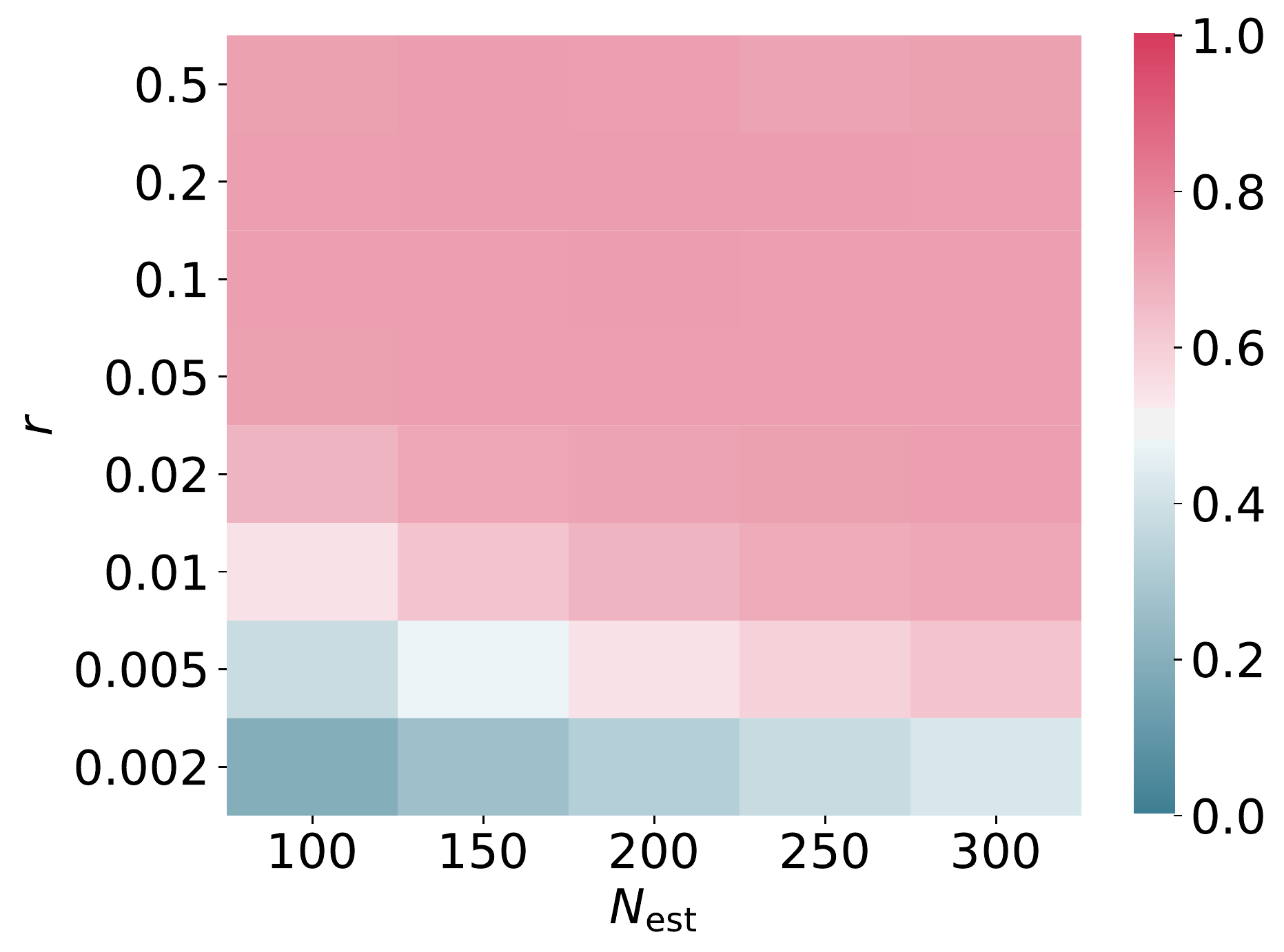}
	\includegraphics[width=0.49\textwidth]{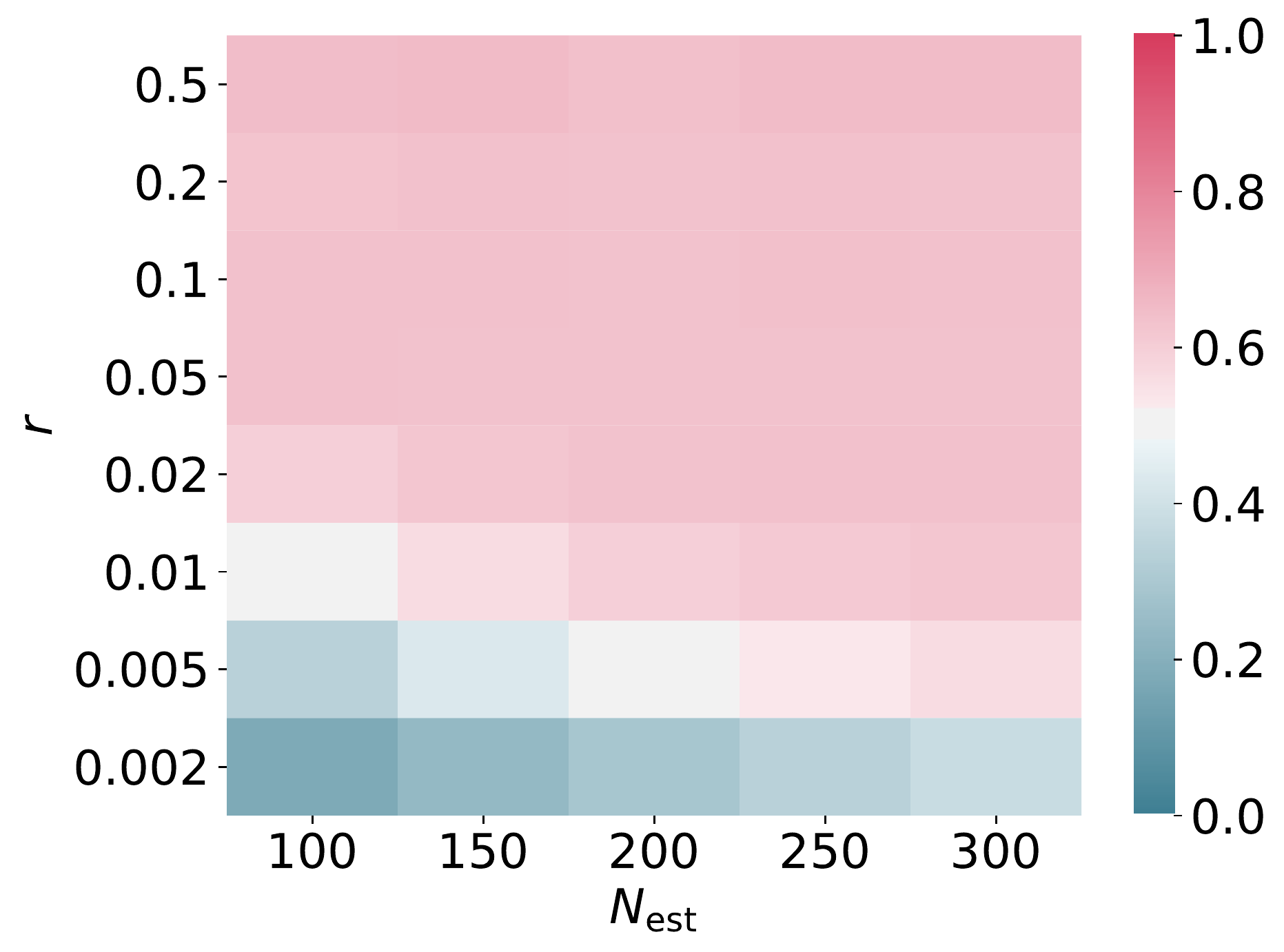}
	\caption{Fit variance $F_v$ of the unlabeled $\eta_s$ DA data for the $p_\text{pred}=5$, $z_\text{pred}=4$ prediction at $t_\text{pred}=4$ from $z_\text{in}=3$ (left) or $p_\text{in}\in[3,4]$ (right). $N_\text{tr}=400$, $N_\text{ul}=1000$. It is clear that more estimators are needed for smaller learning rate. Increasing $N_\text{est}$ without worsening the prediction indicates that the model is robust to overfitting.}
	\label{fig:mda_fq}
\end{figure}

In the training process, we tried different parameters for learning rate in $\{0.5, 0.2, 0.1, 0.02, 0.01, 0.005, 0.002\}$ and the number of estimators in $\{100,150,200,250,300\}$. The corresponding fit variance are plotted in a heatmap with range $[0,1]$, as shown in Fig.~\ref{fig:mda_fq}. Considering the fit quality for both $p$-predictions and $z$-predictions, we selected parameters $r=0.1$, $N_\text{est}=150$ as having highest fit quality in both cases; these will be used for further meson-DA predictions.

\begin{figure}[htbp]
	\centering
	\includegraphics[width=0.49\textwidth]{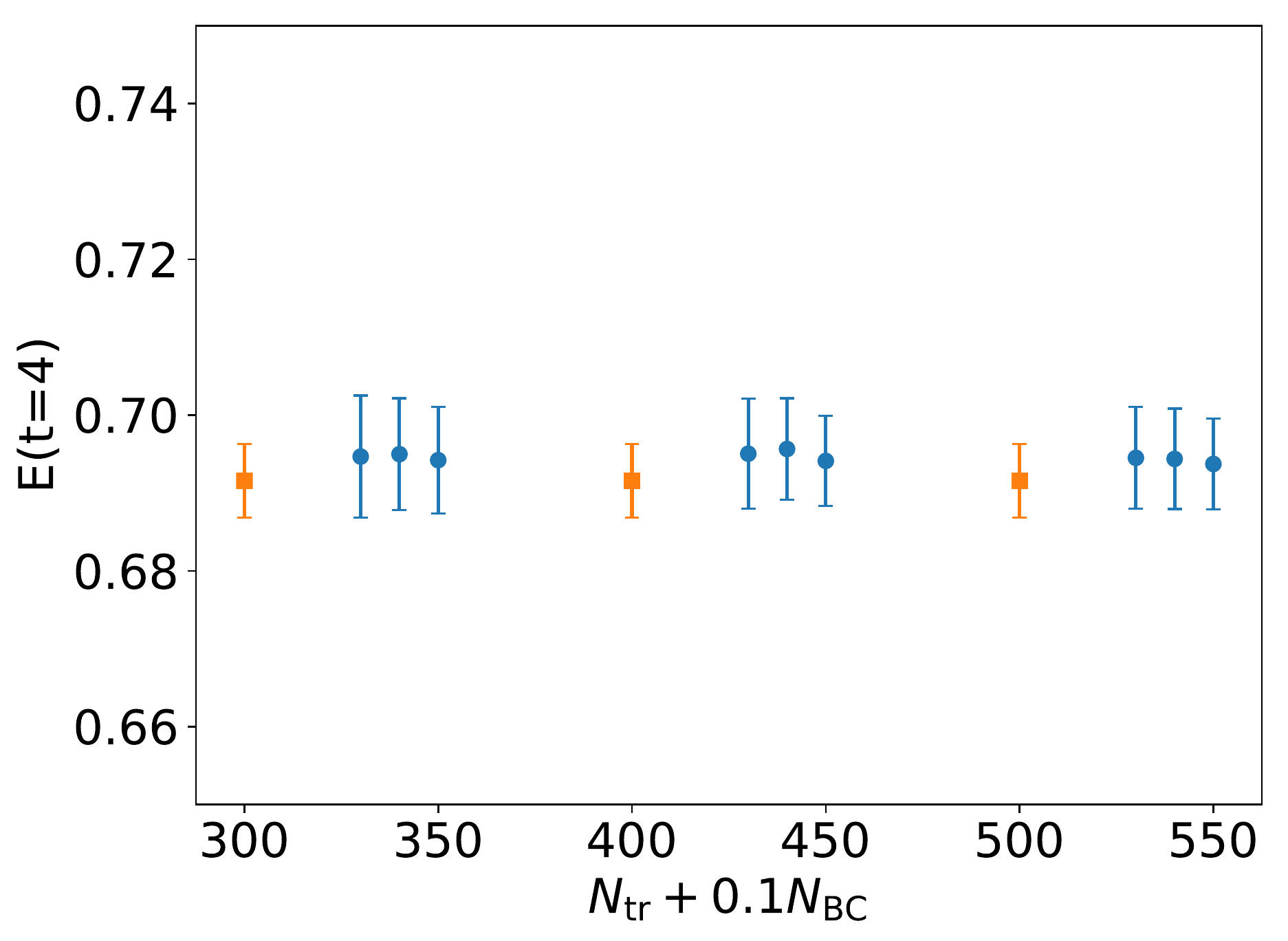}
	\includegraphics[width=0.49\textwidth]{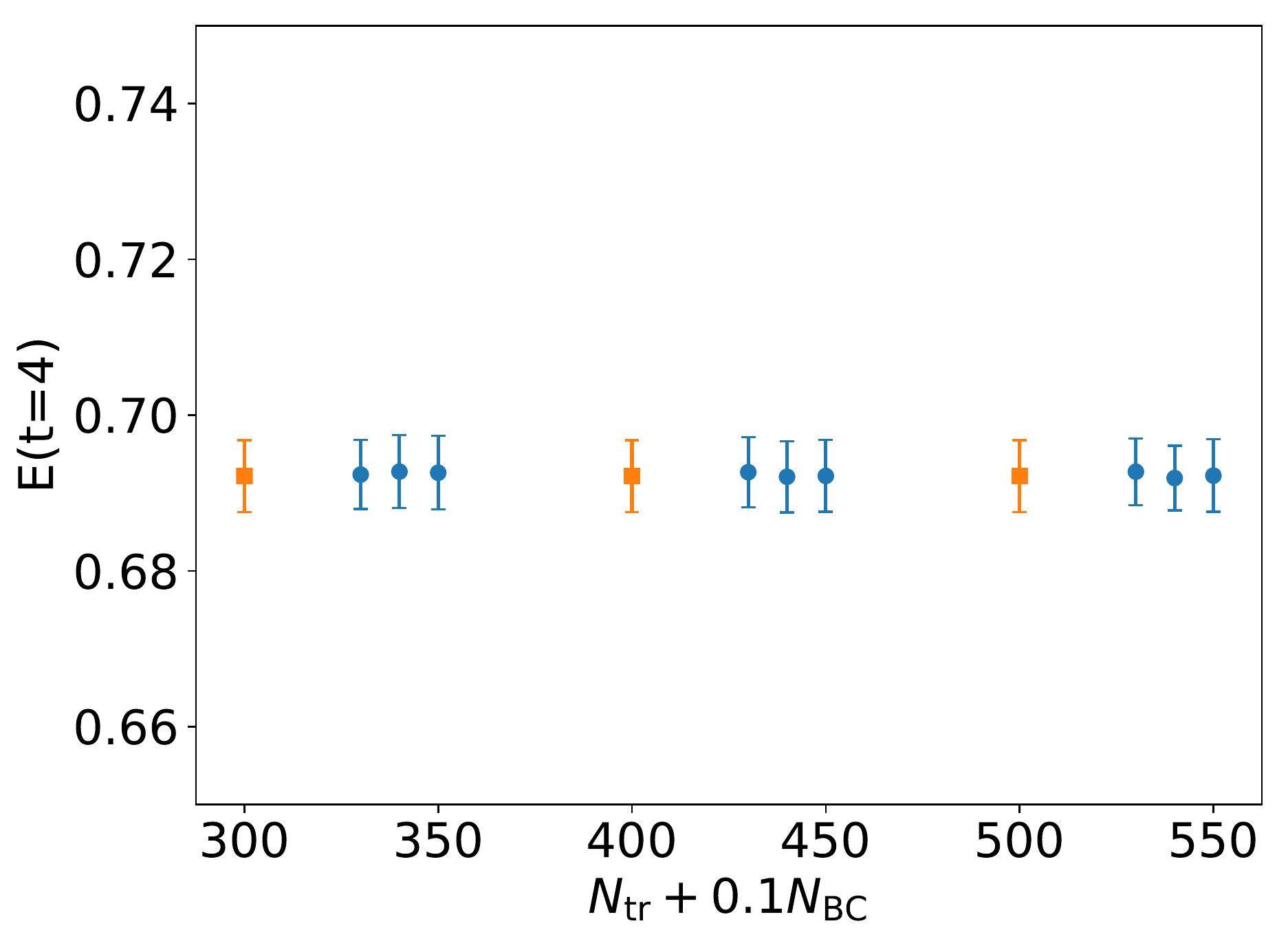}
	\caption{The observed and $z$-predicted $\eta_s$ DA effective mass of $p_\text{pred}=5$, $z_\text{pred}=4$ at $t_\text{pred}=4$ with input $p_\text{pred}=5$, $z_\text{pred}\in[0,3]$, $t_\text{in}\in[3,5]$ for different choices of training data counts and bias-correction data counts. The left (right) plot is the prediction of GBT (linear) model. The horizontal axis is $N_\text{tr}+0.1N_\text{BC}$, with $N_\text{ul}=900$ fixed. The GBT parameters are $N_\text{est}=150$, $r=0.1$. The blue points are predictions with bias correction for the unlabeled test data, and the brown points are observations for unlabeled test data.}
	\label{fig:mda_bc}
\end{figure}

The datasets were evenly distributed into three parts: training data, bias-correction data, and unlabeled test data. 
In practice, we want to minimize the labeled data size without sacrificing much prediction quality. We varied the amount of training data and bias-correction data from 300 to 500, while keeping the number of unlabeled test data $N_\text{ul}=900$ fixed, to look for a best trade-off between reduced data size and prediction quality. The results are shown in Fig.~\ref{fig:mda_bc}. When correlation is obvious, small number of training and bias-correction datasets provides precise estimate that is very close to the true observations for the unlabeled dataset. When correlation is vague, the prediction becomes more precise as one increases the size of the training or the bias-correction datasets. Based on the plot, we picked $N_\text{tr}=400$, $N_\text{BC}=500$ for further estimations.

To further check the consistency of our predictions with the observations, we calculate the effective mass from $C_\text{2pt}$ and compare the results. The effective mass is defined as
\begin{equation}
E(t)=\ln\frac{C_\text{2pt}(t)}{C_\text{2pt}(t+1)}
\end{equation}

Then, we compared different input data to be used for $z$-prediction in Table~\ref{tbl:mda_effmass_pred}. The bias correction makes the prediction noisier by converting the systematic error into statistical error, which improves the accuracy of the prediction for most cases.

\begin{table}[!htbp]
	\centering
	\begin{tabular}{|c|c|c|c|c|c|c|c|}
		\hline
		Type& Input & Method & $E_\text{tr}$& $E_\text{pred}$& $E_\text{pred,BC}$&$E_\text{ul}$&$F_v$ \\
		\hline
		\multirow{2}{*}{$p$-pred}& \multirow{2}{*}{$p_\text{in}\in[3,4]$, $z_\text{in}=4$, $t_\text{in}=7$}&GBT & 0.679(11) & 0.684(13) & 0.683(14) & 0.6923(80) & 0.50(13) \\
		\cline{3-8}
		& &linear & 0.679(11) & 0.6960(86) & 0.6961(91) & 0.6920(74) & 0.911(43) \\
		\hline
		\multirow{2}{*}{$z$-pred}& \multirow{2}{*}{$p_\text{in}=5$, $z_\text{in}\in[0,3]$, $t_\text{in}=7$}&GBT & 0.679(11) & 0.694(13) & 0.692(12) & 0.6923(80) & 0.62(14) \\
		\cline{3-8}
		& &linear & 0.679(11) & 0.6913(76) & 0.6912(75) & 0.6920(74) & 0.99935(40) \\
		\hline
	\end{tabular}
	\caption{Effective mass calculated from the prediction of $\eta_s$ DA $C_\text{2pt}$ at $p_\text{pred}=5$, $z_\text{pred}=4$, $t_\text{pred}=7$ with different models and different inputs. Models are trained with $N_\text{tr}=400$, $N_\text{BC}=500$, $N_\text{ul}=900$, $N_\text{est}=150$, $r=0.1$. The linear model is more accurate than GBT. Both models have better performance for $z$-prediction than $p$-prediction.}
	\label{tbl:mda_effmass_pred}
\end{table}

For small datasets, such as what we have for the quasi-DA data, it can be difficult for the GBT model to extract the nonlinear pattern of the training dataset. As a consequence, the fit quality of the GBT model for the test data is poor. Instead, the simpler linear regression shows better performance. Sometimes, however, when input data when the dataset is noisy (e.g., larger-$t$ data), the linear regression fails with poor prediction quality, as shown in Table~\ref{tbl:model}, while GBT was able to capture the correlation and make predictions. Using cleaner and more correlated data like the closest timeslice, momentum and link can significantly improve the fit quality for linear regression.

\begin{table}[!htbp]
	\centering
	\begin{tabular}{|c|c|c|c|c|c|c|c|}
		\hline
		Type& Input & Method &$E_\text{tr}$& $E_\text{pred}$& $E_\text{pred,BC}$&$E_\text{ul}$&$F_v$ \\
		\hline
		\multirow{2}{*}{$p$-pred}& \multirow{2}{*}{$p_\text{in}\in[3,4]$, $z_\text{in}=4$, $t_\text{in}=10$}&GBT &0.686(43) & 0.678(45) & 0.683(40) & 0.675(26) & 0.36(19) \\
		\cline{3-8}
		& &linear & 0.683(37) & 0.692(39) & 0.695(39) & 0.676(27) & 0.72(13) \\
		\hline
		\multirow{2}{*}{$p$-pred}& \multirow{2}{*}{$p_\text{in}\in[3,4]$, $z_\text{in}=4$, $t_\text{in}\in[7,13]$}&GBT & 0.686(43) & 0.677(51) & 0.676(43) & 0.675(26) & 0.25(27) \\
		\cline{3-8}
		& &linear & 0.683(37) & 0.675(88) & 0.677(77) & 0.676(27) & -0.13(85) \\
		\hline
	\end{tabular}
	
	\caption{Effective mass calculated from the prediction of $\eta_s$ DA $C_\text{2pt}$ at $p_\text{pred}=5$, $z_\text{pred}=4$, $t_\text{pred}=10$ with different models and different input timeslices. Models are trained with $N_\text{tr}=400$, $N_\text{BC}=500$, $N_\text{ul}=900$, $N_\text{est}=150$, $r=0.1$. The linear model has better performance on correlated cleaner data but fails when more uncorrelated noisy data input are included. The GBT is more stable and less sensitive to these inputs.}
	\label{tbl:model}
\end{table}

\begin{figure}[htbp]
	\centering
	\includegraphics[width=0.49\textwidth]{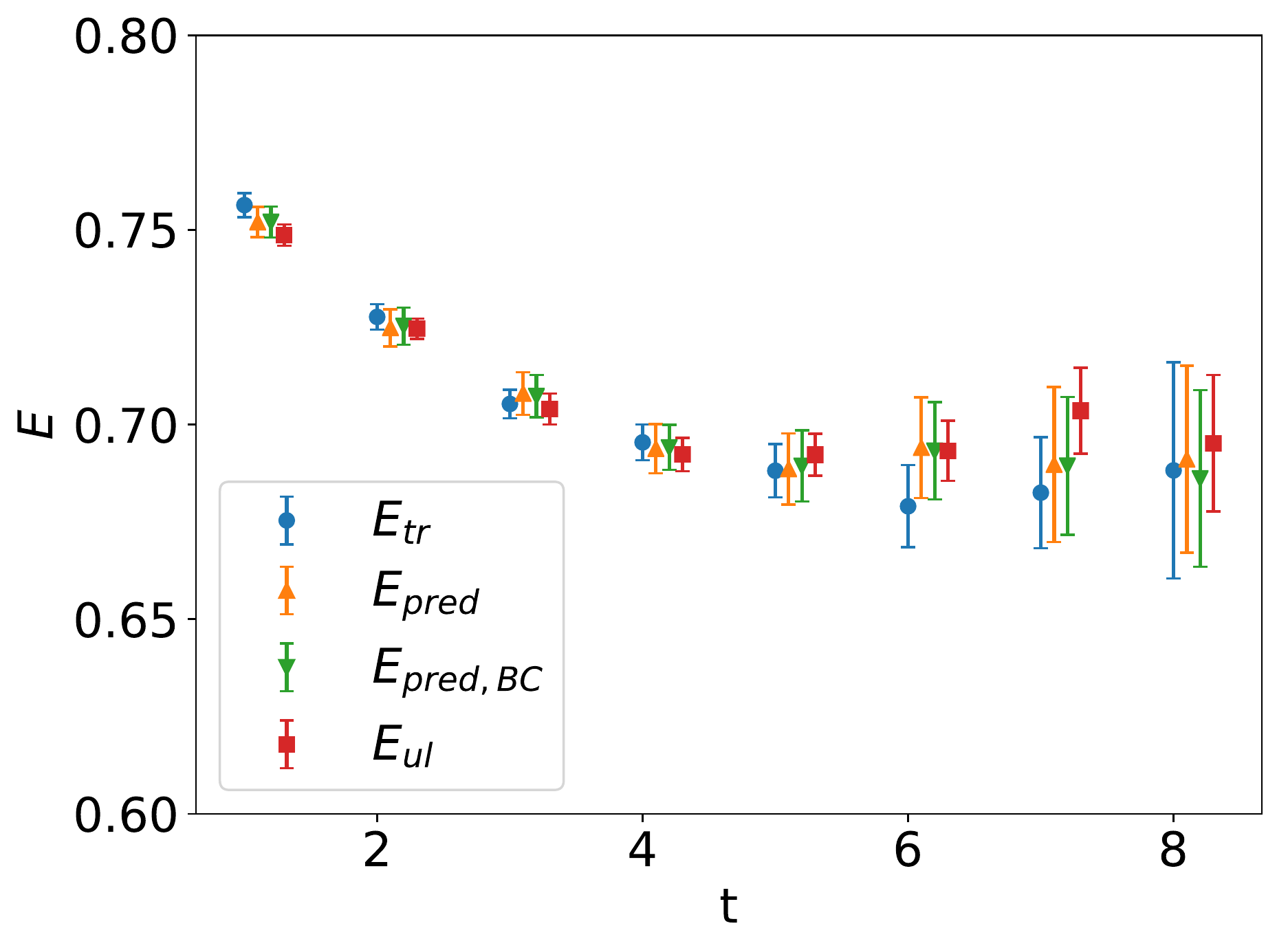}
	\includegraphics[width=0.49\textwidth]{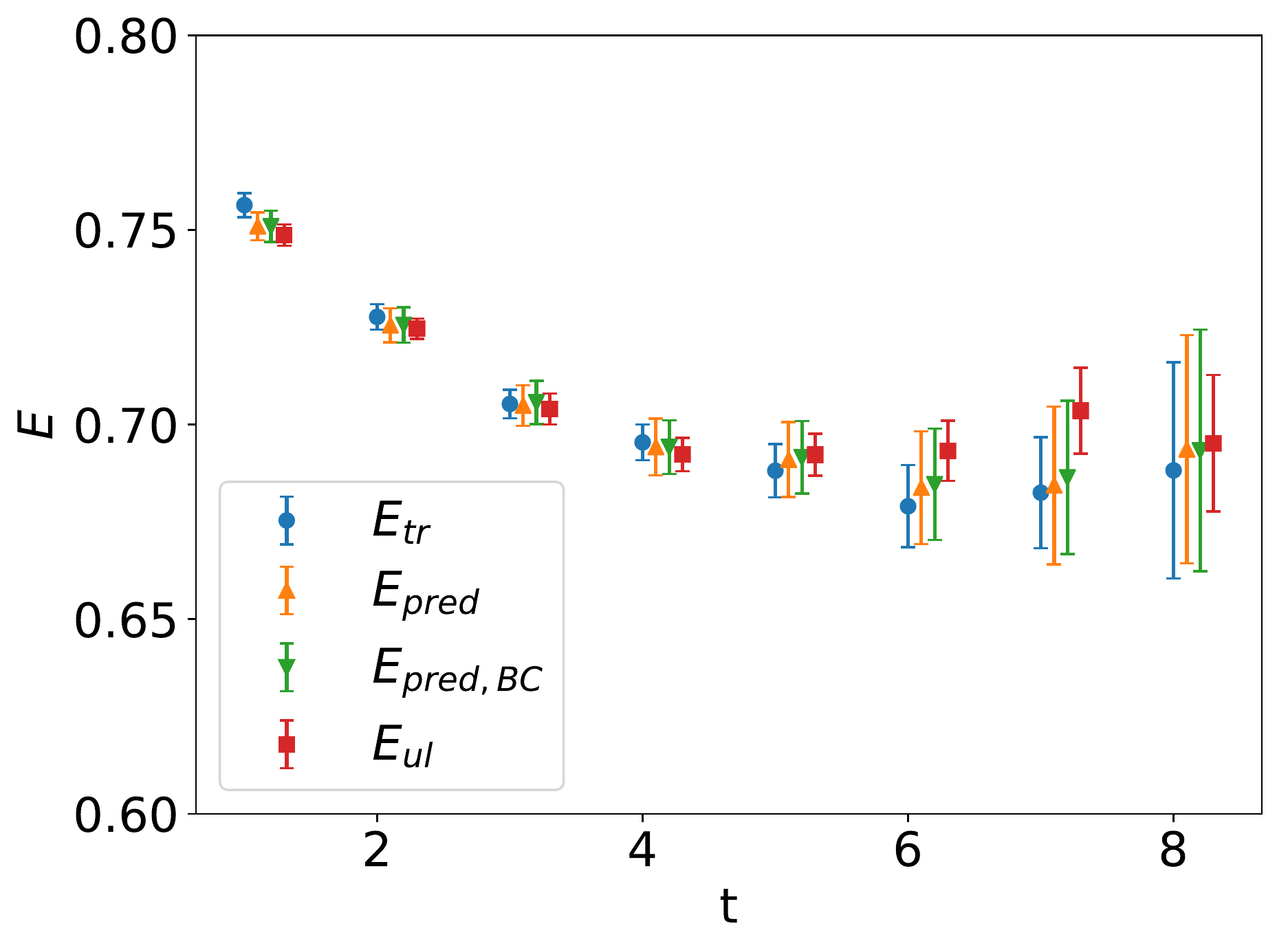}
	\includegraphics[width=0.49\textwidth]{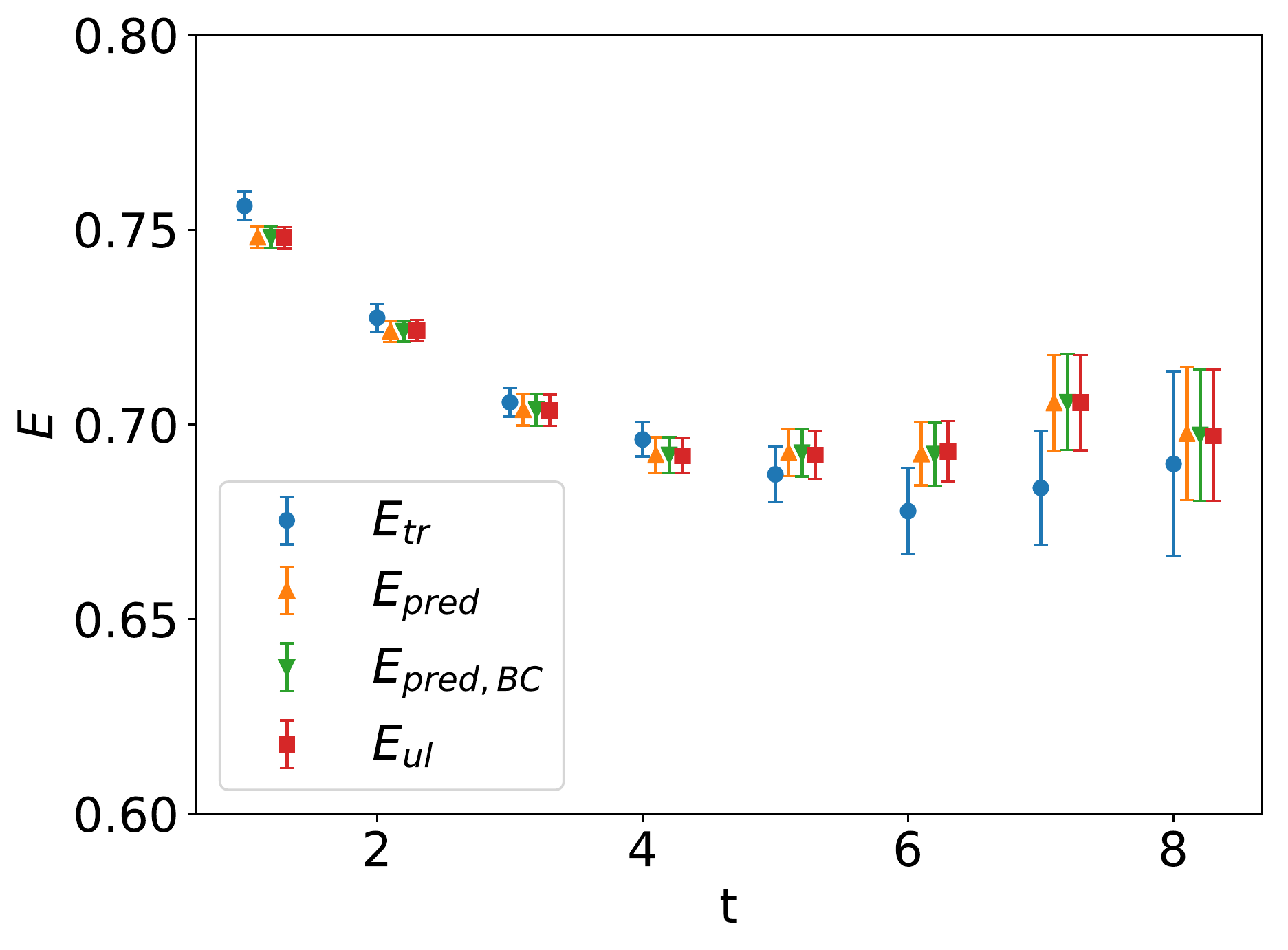}
	\includegraphics[width=0.49\textwidth]{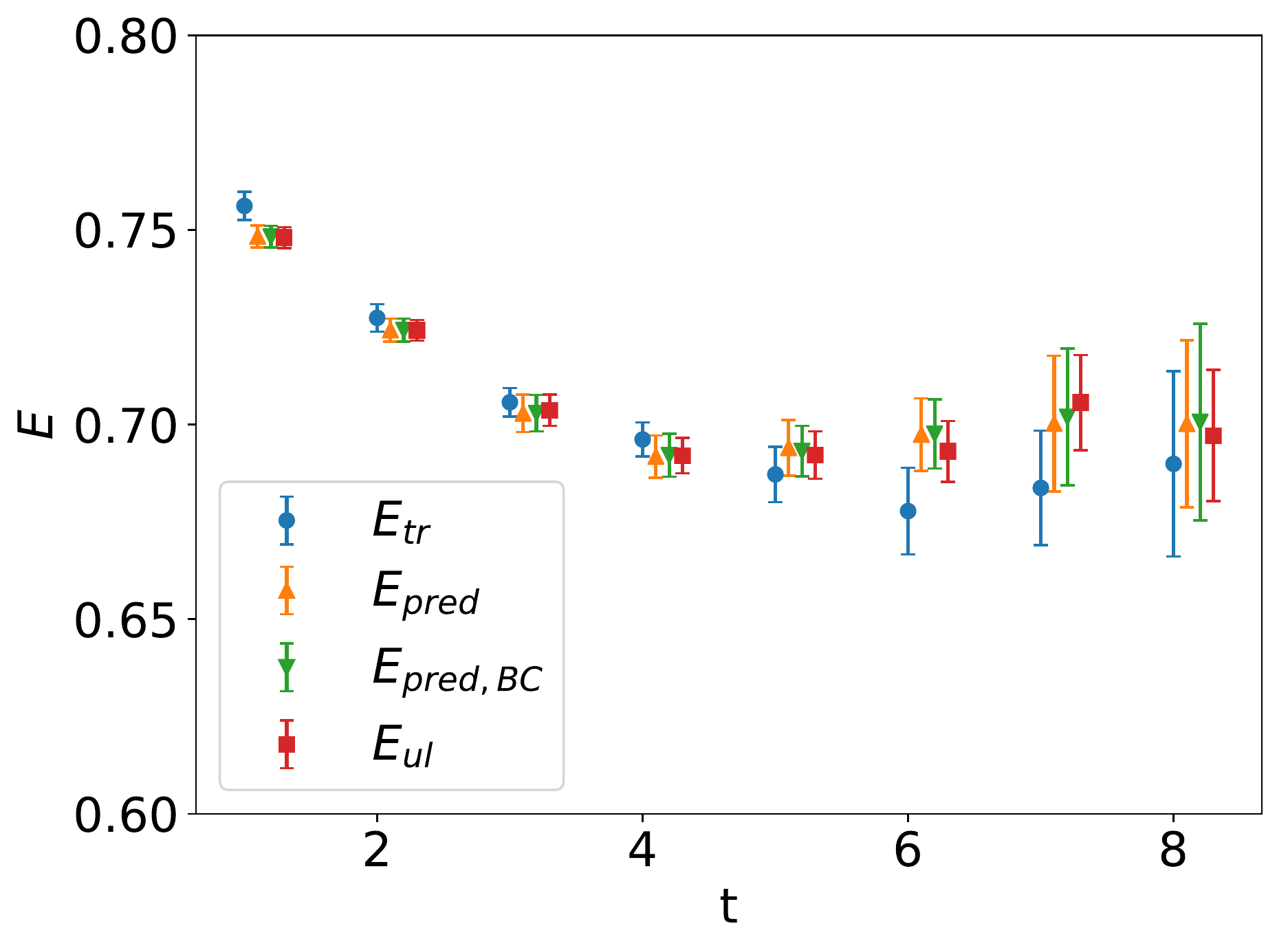}
	\caption{The observed/predicted $\eta_s$ DA effective mass at $p_\text{pred}=5$, $z_\text{pred}=4$ from $p_\text{in}=5$, $z_\text{in}\in[0,3]$(left) and $p_\text{in}\in[3,4]$, $z_\text{in}=4$ (right). The top (bottom) plots are obtained by using GBT (linear) model with $N_\text{tr}=400$, $N_\text{BC}=500$, $N_\text{ul}=900$.  The GBT parameters are $N_\text{est}=150$, $r=0.1$. The linear model shows better consistency with the unlabeled data, while at some timeslices the GBT model fails to give a good prediction.}
	\label{fig:mda_pred}
\end{figure}

After determining the parameters, we run the ML program and show the effective mass of our predictions along with the observed datasets for both $p_\text{pred}$ and $z_\text{pred}$ predictions in Fig.~\ref{fig:mda_pred}. The linear model works well for $z$-prediction, but the GBT model and $p$-predictions still need to be improved.

\subsection{Predictions of kaon quasi-PDFs} 

As Nambu-Goldstone bosons associated with dynamical chiral SU(3) symmetry breaking, 
the pion and kaon serve as a fundamental test ground for our understanding of QCD theory at the hadronic scale. 
The \textit{ab initio} calculation of hadron PDFs from lattice QCD provides theoretical background for particle-discovery experiments and Standard-Model (SM) tests at colliders~\cite{Aguilar:2019teb}.
After decades of theoretical and experimental efforts, 
the precision required in PDFs for more stringent tests of the SM has increased significantly.
In Ref.~\cite{Chen:2018fwa}, we presented the first direct lattice calculation of the valence-quark distribution in the pion, using the MILC HISQ coarse ensemble with $M_\pi \approx 330$~MeV.
Since the computational cost of quasi-PDF measurements on an ensemble at lighter pion mass or reduced lattice spacing would increase significantly, 
in this work we investigate a ML algorithm to reduce the computational cost.

We test on the meson unpolarized quasi-PDF measurements on the lattice:
\begin{align}
C_\text{3pt}(z, t) &= \langle 0 |\int d^3y\, e^{-i y\cdot P} M_\text{ps}(\vec{y},t_\text{sep}) \bar{s}(z, t) \gamma_4 \prod_{x=0}^{z-1} U_z(x, t) s(0, t)  \bar{M}_\text{ps}(\vec{0},0) | 0 \rangle, \\
C_\text{2pt}(t_\text{sep}) &= \langle 0 |\int d^3y\, e^{-i y\cdot P} M_\text{ps}(\vec{y},t_\text{sep}) \bar{M}_\text{ps}(\vec{0},0) | 0 \rangle,
\label{eq:kaon_3pt}
\end{align}
where $C_\text{3pt}$ is the three-point correlator, $C_\text{2pt}$ is the two-point correlator, 
$M_\text{ps}= \bar{q} \gamma_5 q$ is the pseudoscalar meson operator, 
$z$ is the length of the Wilson link, $U_\mu(x, t)$ is the gauge link, 
and $\gamma_i$ are Dirac spinor matrices. 
For this study we use Wilson clover valence quarks on a MILC HISQ ensemble. 
The lattice spacing is $a \approx 0.12$~fm,
the lattice volume $V = 40^3 \times 64$, 
and the pion mass $M^\text{sea}_\pi \approx 220$~MeV. 
The valence quark masses are tuned to match the valence pion to the sea pion mass. 
We adopt Gaussian momentum smearing~\cite{Bali:2016lva} to generate quark sources, 
to enhance the ground-state signal at nonzero momentum near 1.55~GeV. 
The Gaussian smearing width is chosen to be 3, with 50 iterations, and the momentum parameter $k = 4.82$.
Measurements are done on 495 configurations, using 4 quark-source locations per configuration, making 1960 measurements in total. 
Measurements are averaged over these quark sources before being passed to the ML algorithm, 
as this has shown to provide predictions with smaller statistical errors. 
The ratio of the three-point correlator ($C_\text{3pt}$) to the two-point correlator ($C_\text{2pt}$) is a useful way to extract the matrix elements:
\begin{equation}
R(t) = C_\text{3pt}(t) / C_\text{2pt}(t_\text{sep})\label{eq:ratio}
\end{equation}
where $R(t)$ is the ratio at the operator insertion time $t$, 
and $t_\text{sep}$ is the meson source and sink temporal separation.

\subsubsection{Kaon quasi-PDF results}

\begin{figure}[htb]
	\centering
	\includegraphics[width=0.49\textwidth]{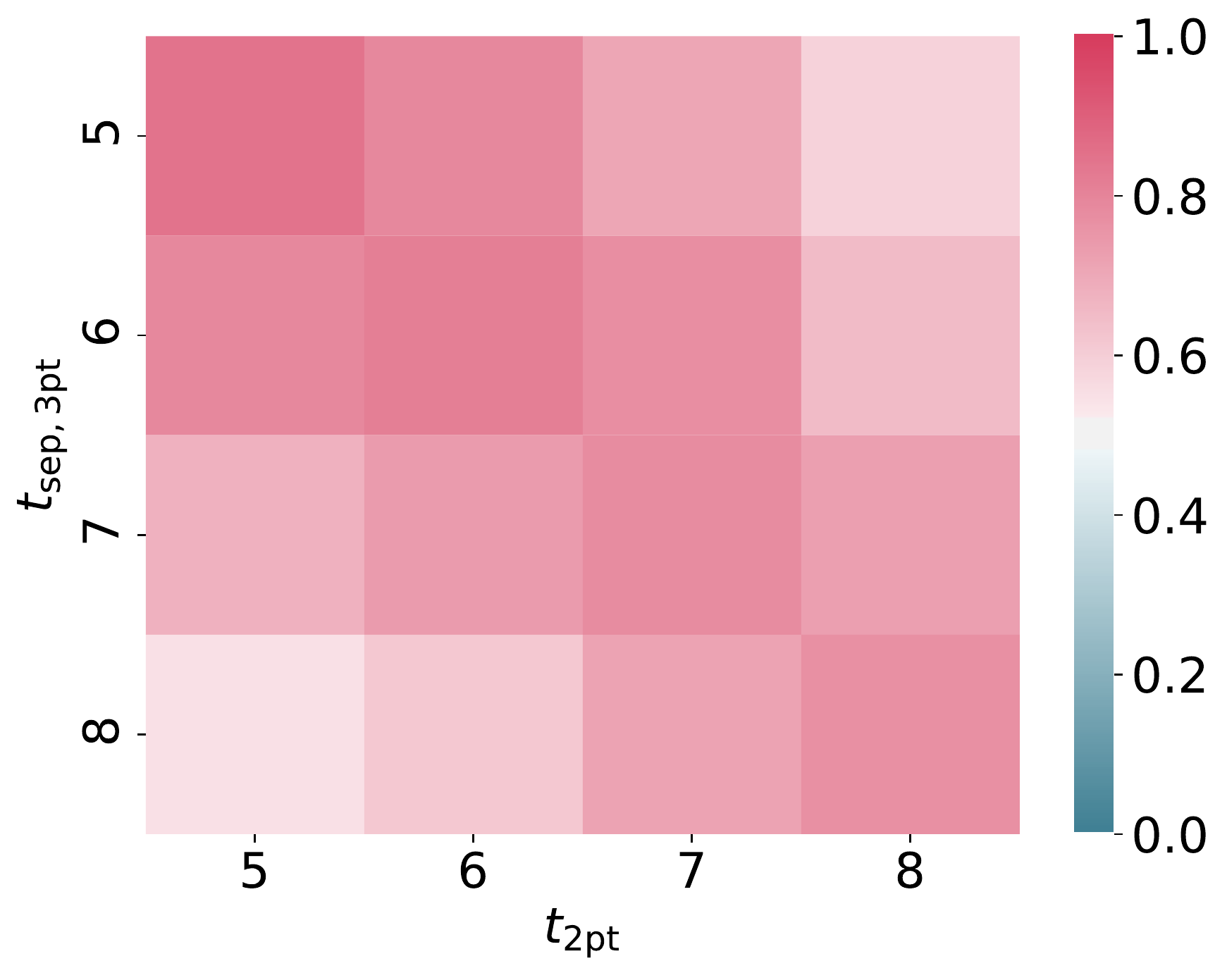}
	\includegraphics[width=0.49\textwidth]{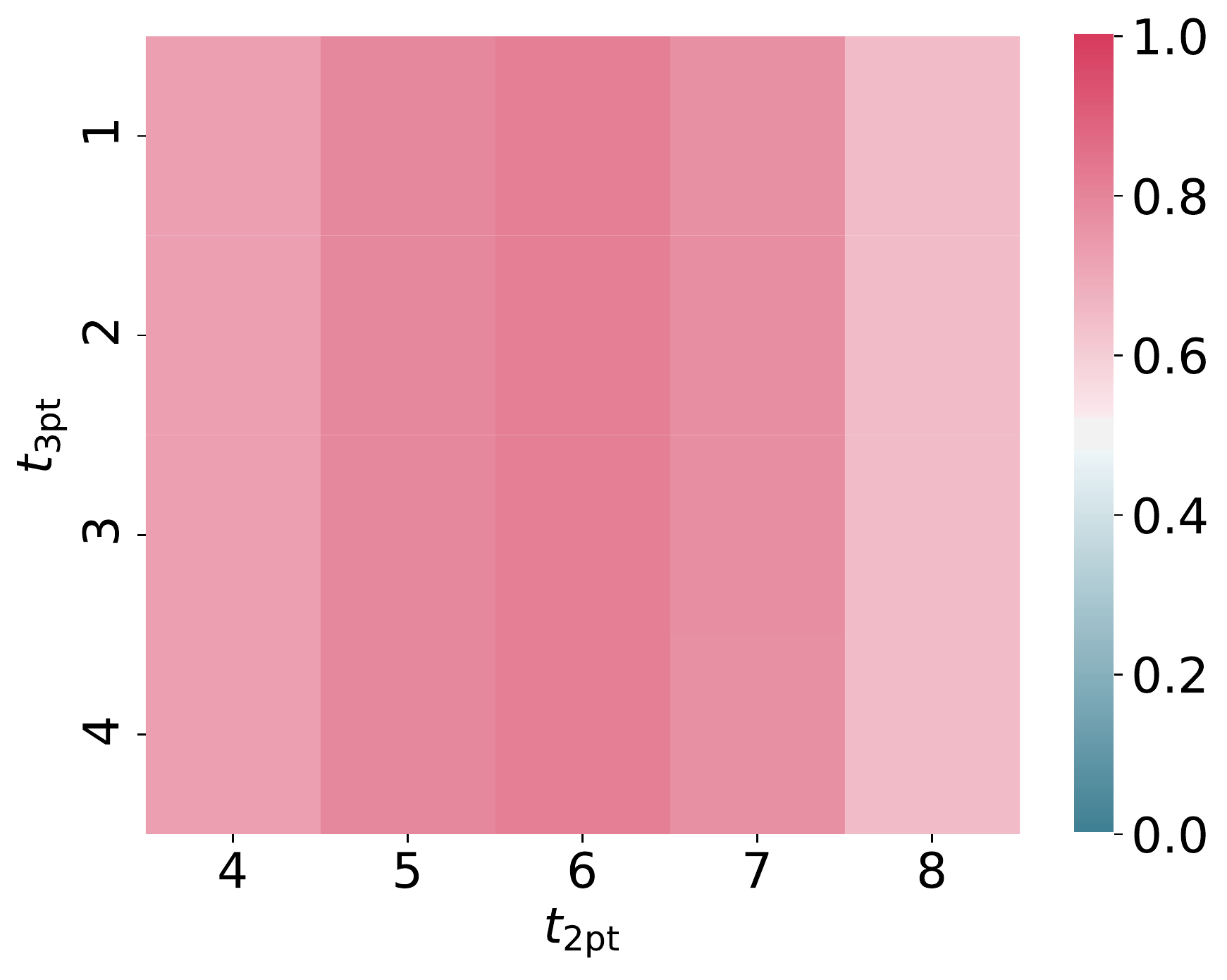}
	\caption{Correlations between $C_\text{2pt}$ and $C_\text{3pt}$ of kaon quasi-PDF at different time separations (left, with insertion time $t_\text{3pt} = t_\text{sep}/2$) and at different insertion times (right, with $t_\text{sep}=6$). The correlation is insensitive to the insertion time.
	}
	\label{fig:tcorr}
\end{figure}

\begin{figure}[htb]
	\centering
	\includegraphics[width=0.49\textwidth]{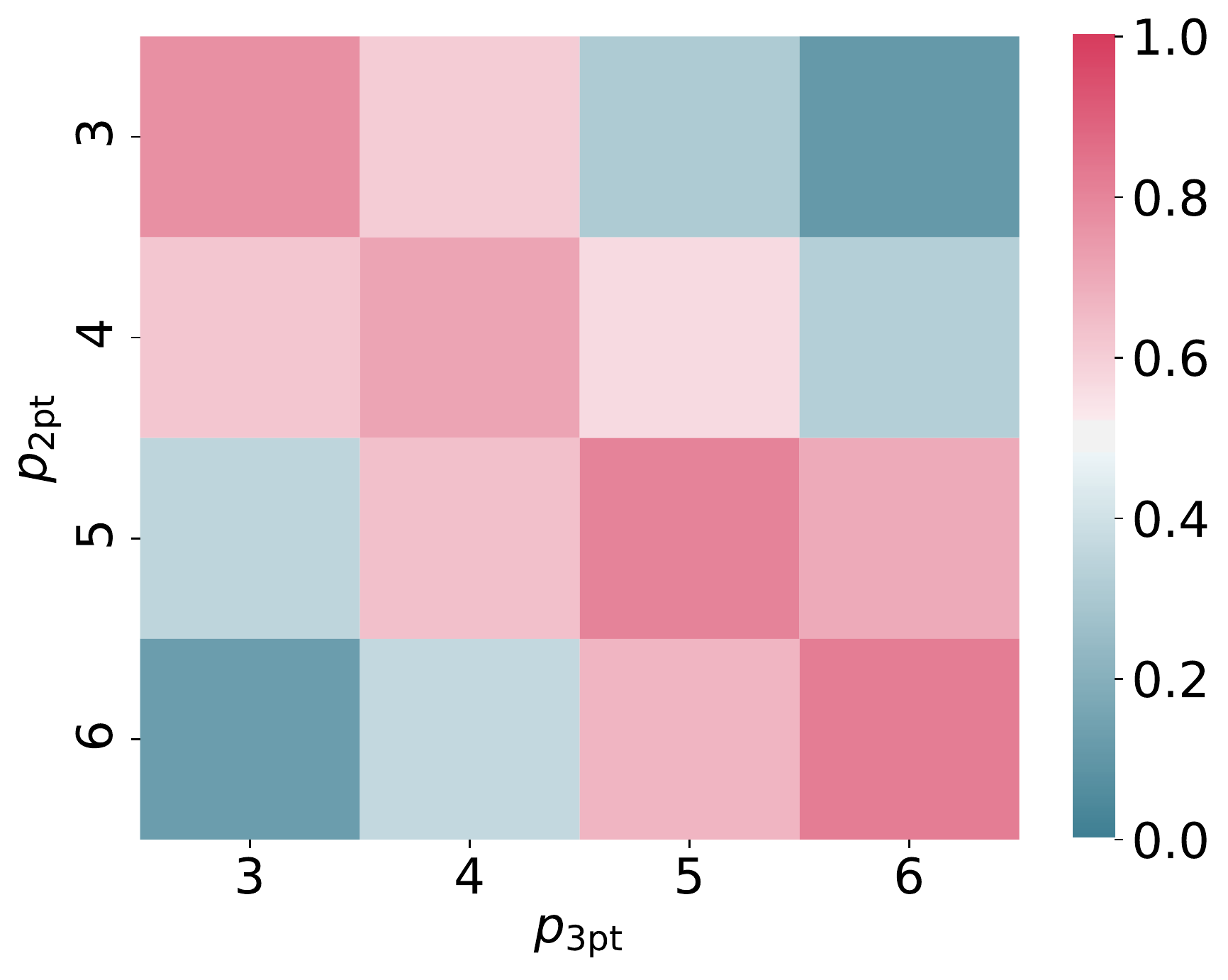}
	\includegraphics[width=0.49\textwidth]{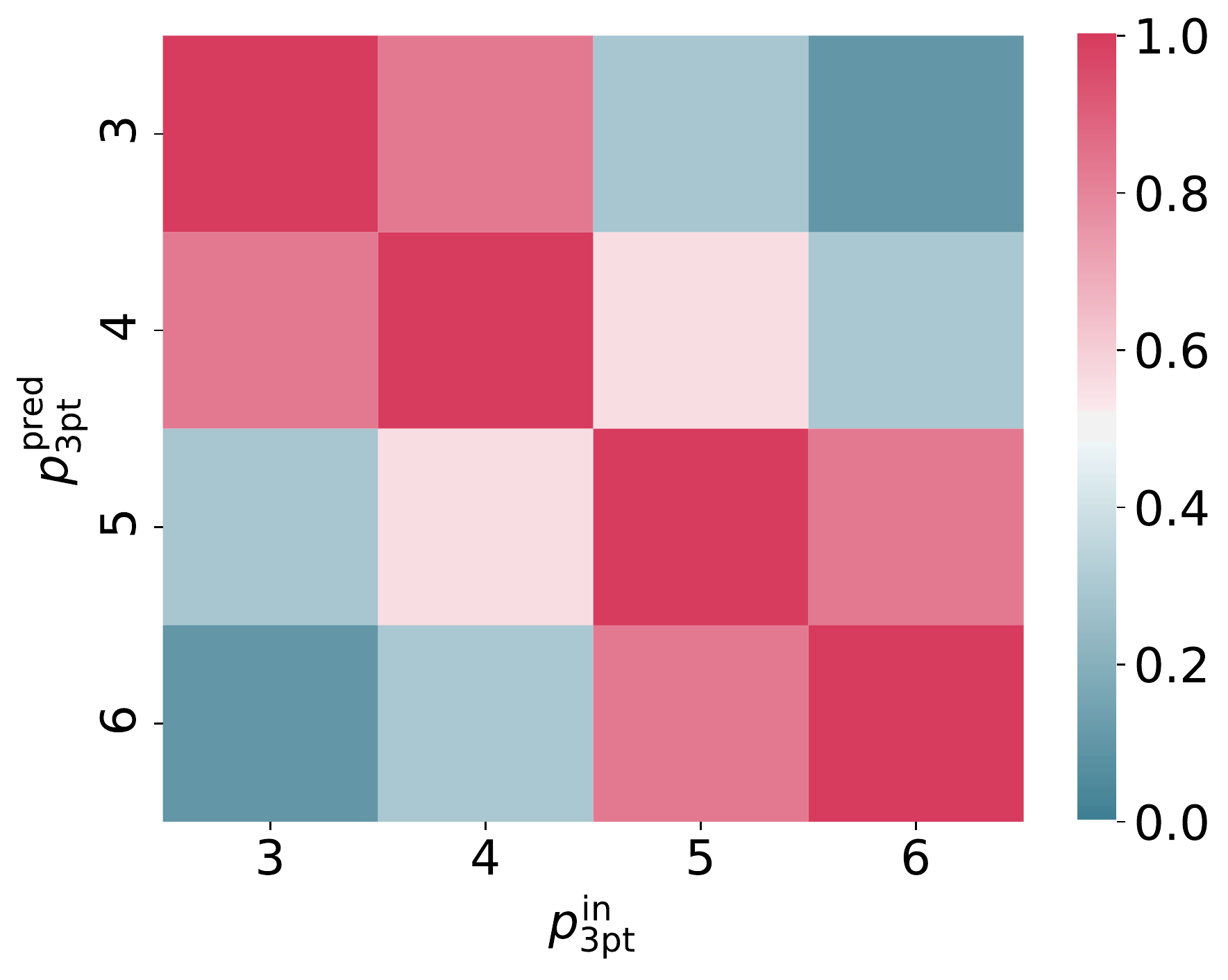}
	\caption{Correlations between the kaon quasi-PDF three-point correlators and two-point correlators (left) or three-point correlators (right) at different momenta. The three-point correlation seem to be clustered; $p\in\{3,4\}$ and $p\in\{5,6\}$ are correlated separately. Thus, the prediction of $p_\text{pred}=5$ from smaller momentum has bad quality.
	}
	\label{fig:p3corr}
\end{figure}

\begin{figure}[htb]
	\centering
	\includegraphics[width=0.49\textwidth]{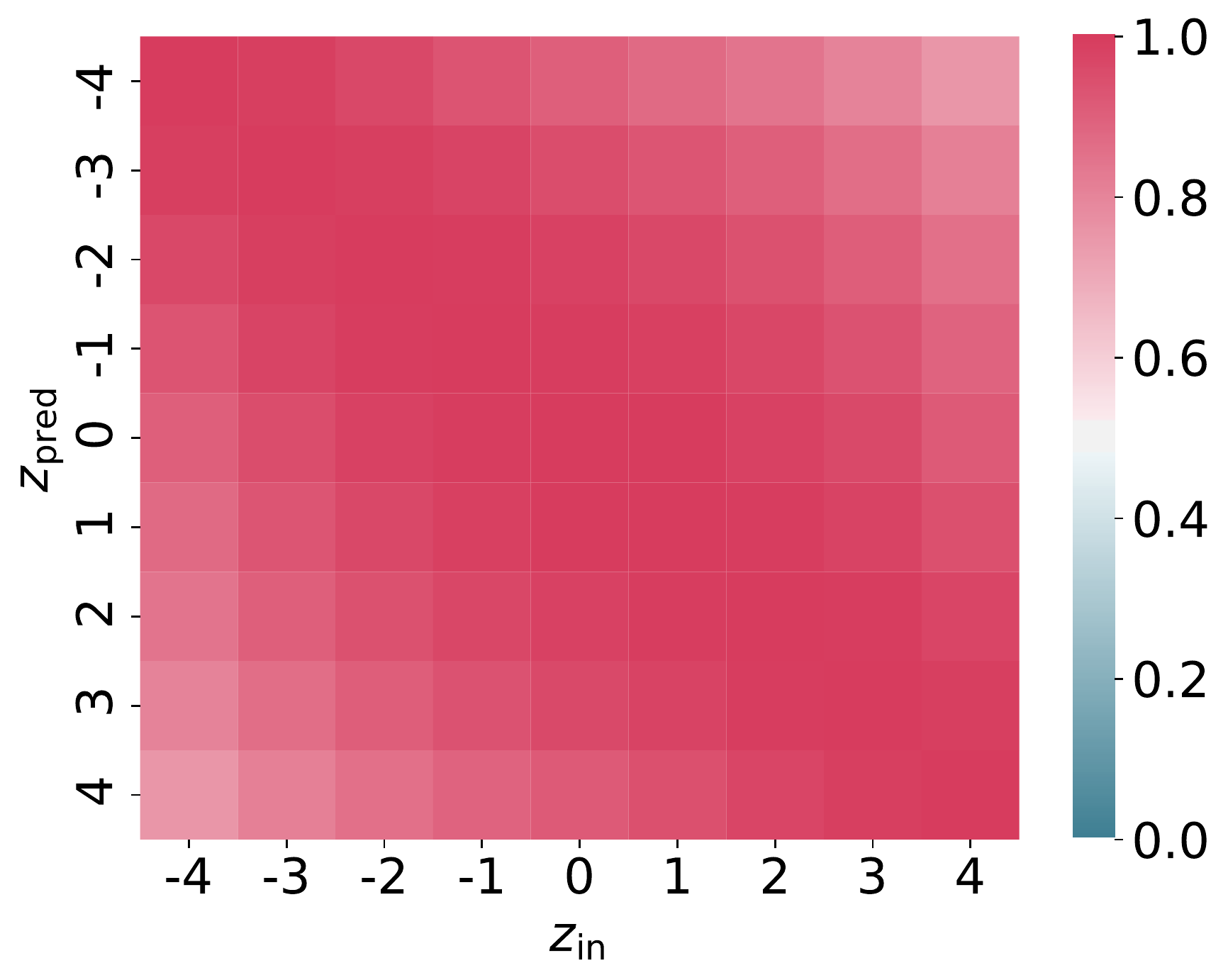}
	\caption{Correlations between the kaon quasi-PDF three-point correlators with $p_\text{in}=p_\text{pred}=3$ at different link lengths. Shorter link lengths have better correlation, and the correlation decay in the $z$ direction is slow.}
	\label{fig:z3corr}
\end{figure}

For the kaon quasi-PDF, the meson operator is $K= \bar{u} \gamma_5 s$.
We first check the correlation for three-point correlators with insertion operator $\gamma_4$. Generally, the correlations are better than for the DA case. The correlations between different time  are shown in Fig.~\ref{fig:tcorr}. The correlation is insensitive to insertion time, but sensitive to the difference between two-point timeslice and three-point source-sink time separation. Because the correlators are similar for different insertion times, we can use all the insertion timeslices as input in the same procedure. An anomaly is observed in the momentum correlation in Fig.~\ref{fig:p3corr}, which may due to the different number of measurements for $p\in\{3,4\}$ and $p\in\{5,6\}$, since we had an extra run for $p\in\{5,6\}$ with different source locations. The link correlation is then displayed in Fig.~\ref{fig:z3corr}. The correlation decays slowly in the $z$-direction, suggesting that we may use more data at different links as inputs.

\begin{figure}[htbp]
	\begin{minipage}{0.5\textwidth}
		\includegraphics[width=\linewidth]{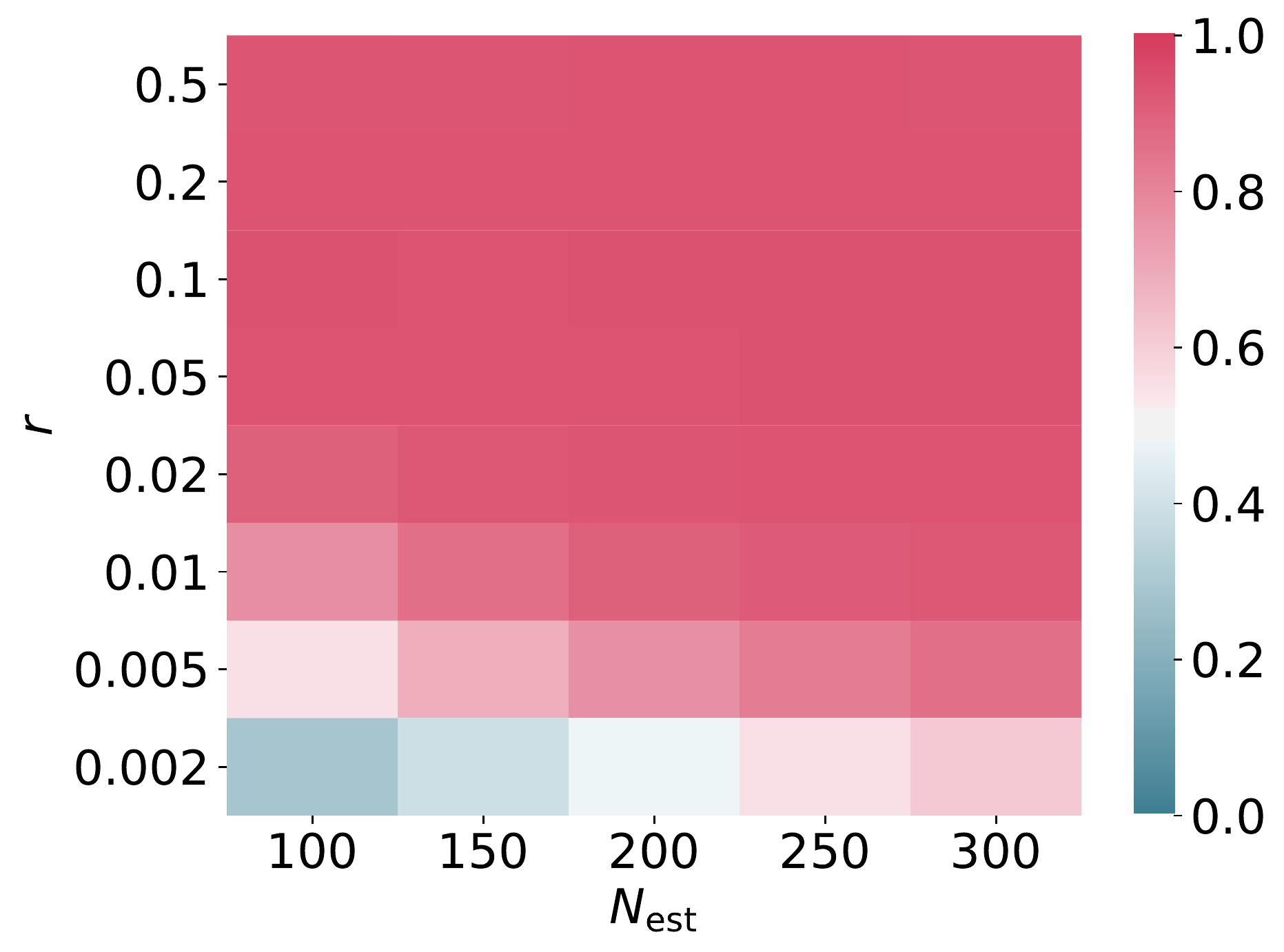}
	\end{minipage}\hfill
	\begin{minipage}{0.5\textwidth}
		\includegraphics[width=\linewidth]{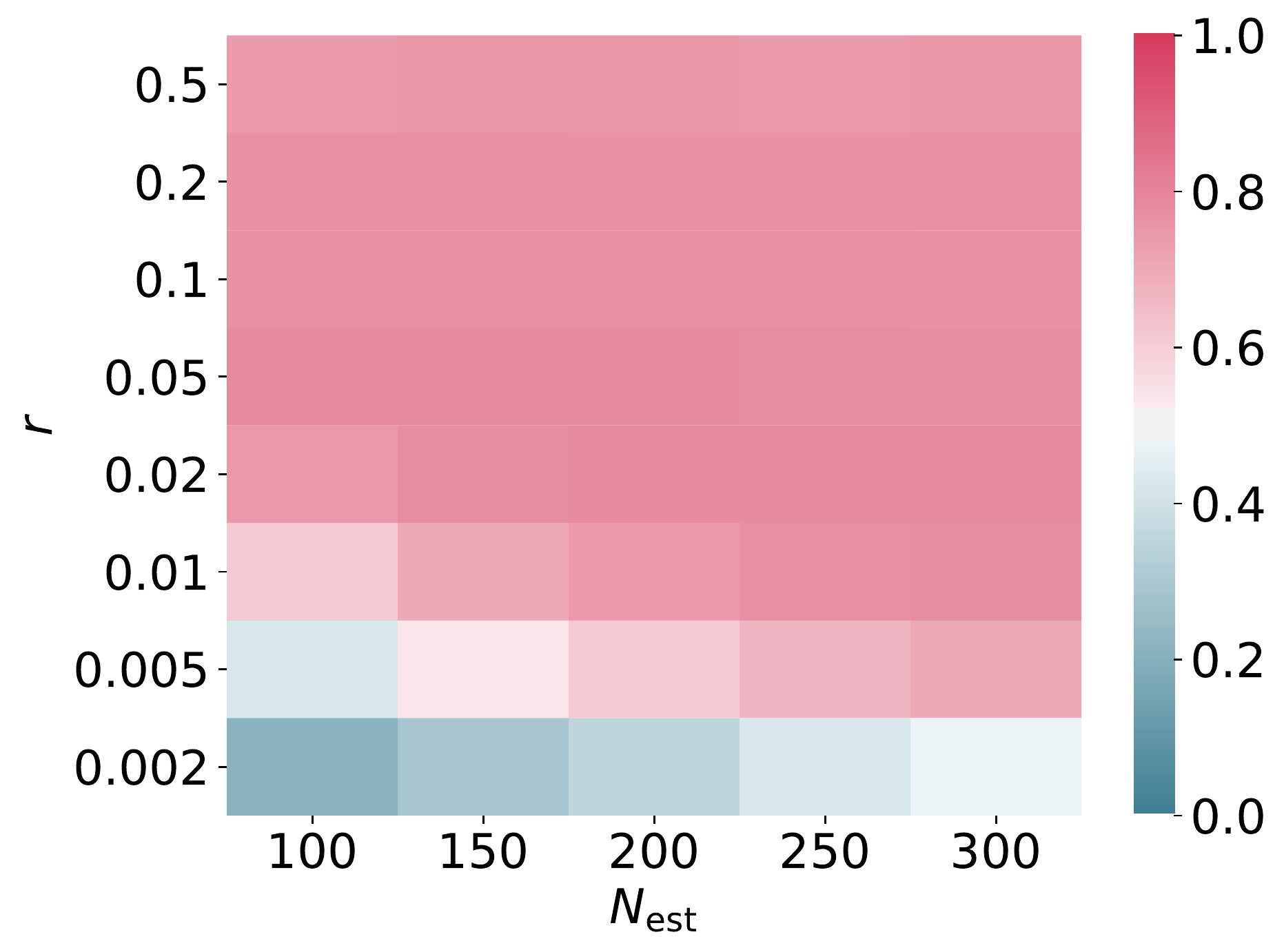}
	\end{minipage}
	\caption{Estimates of the fit variance $F_v(t=2)$ as a function of learning rate $r$ and number of estimators $N_\text{est}$ from the kaon quasi-PDF measurements at $p_\text{pred} = 4$, $z_\text{pred} = 4$, $t_\text{sep} = 5$. $N_\text{tr}=400$ and $N_\text{ul}=1180$ are used.
		The left (right) plot shows the results from the $z$-prediction ($p$-prediction). The $z$-prediction has a much better fit variance because of the good correlations between close links.
	}
	\label{fig:kaon_fv}
\end{figure}

Again, we compare the parameters used for the GBT model.
Fig.~\ref{fig:kaon_fv} shows the fit variance estimate $F_v$ from both the $z$-prediction (with $p_\text{in} = p_\text{pred}$ and $z_\text{in} < z_\text{pred}$), 
and the $p$-prediction (with $p_\text{in} < p_\text{pred}$ and $z_\text{in} = z_\text{pred}$) using the GBT model trained on 400 measurements. 
The horizontal axis shows the number of estimators $N_\text{est}$, and the vertical axis shows the learning rate $r$. 
The target measurement is at $p_\text{pred}=4$, $z_\text{pred} = 4$, $t_\text{sep}= 5$, and $t = 2$. 
For each prediction we used both $C_\text{3pt}$ and the $C_\text{2pt}$. 
Thus, in either case a set of fit parameters can be chosen as, e.g., $N_\text{est} = 150$, $ r = 0.1$. 
As expected, with reduced learning rate, one needs more estimators to achieve a similar fit variance. With fixed learning rate, the fit variance becomes stable when we keep increasing $N_\text{est}$, indicating that the model is robust to overfitting.

\begin{figure}[htbp]
	\begin{minipage}{0.5\textwidth}
		\includegraphics[width=\linewidth]{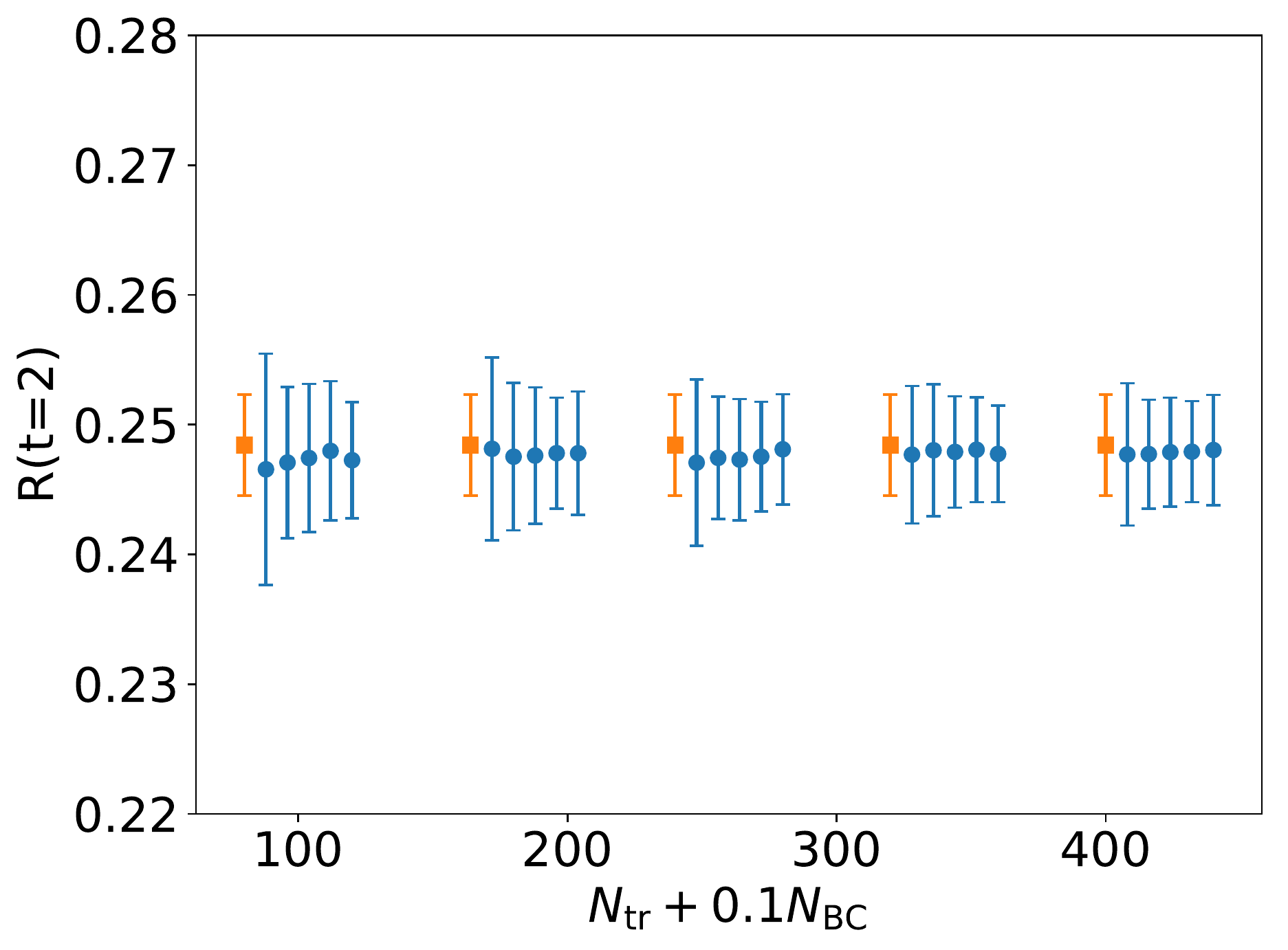}
	\end{minipage}\hfill
	\begin{minipage}{0.5\textwidth}
		\includegraphics[width=\linewidth]{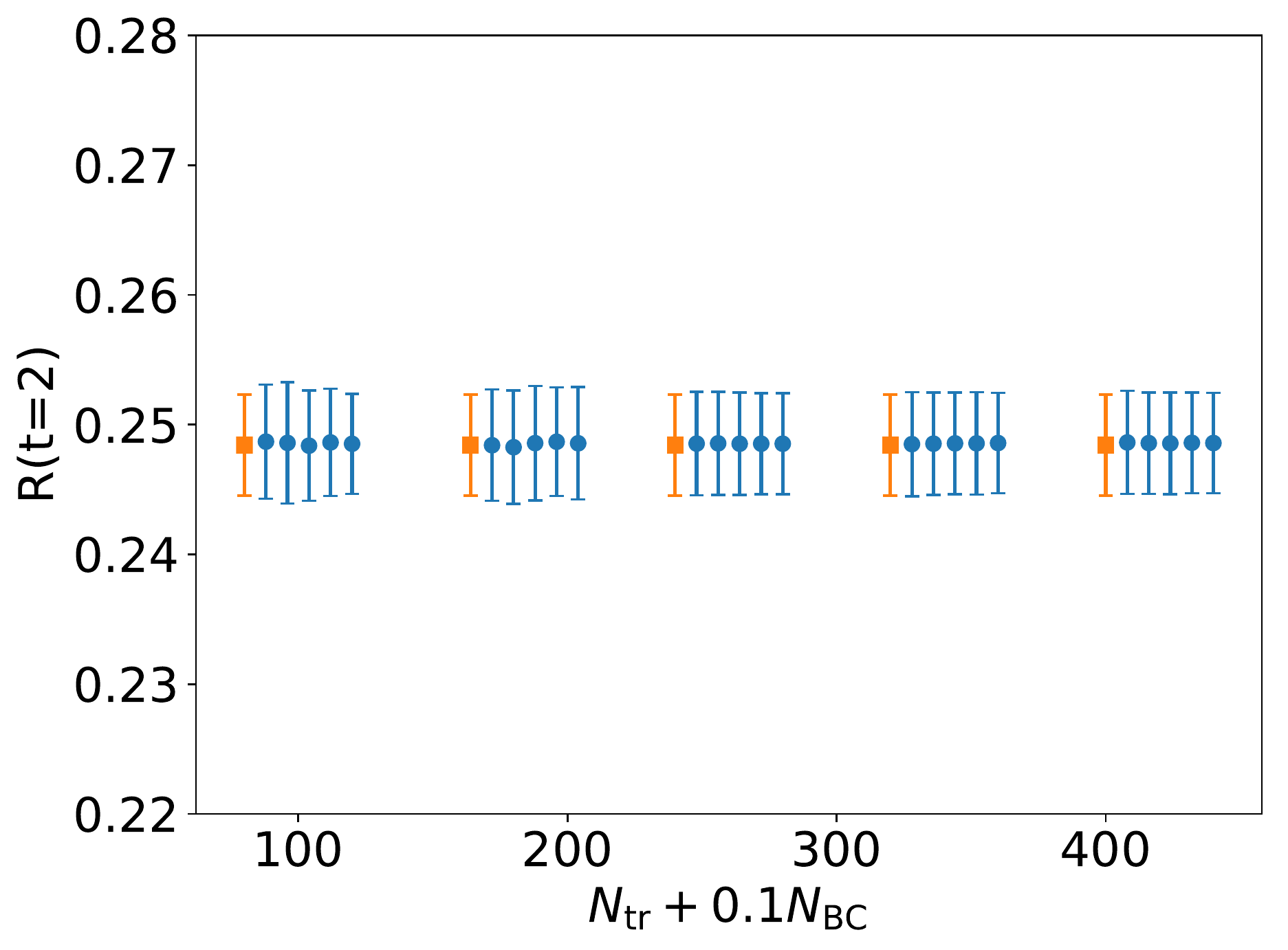}
	\end{minipage}
	\caption{Observations and predictions of the ratio $R(t=2)$ of kaon quasi-PDF correlators at $p_\text{pred} = 4$, $z_\text{pred}= 4$, $t_\text{sep} = 5$ from input data at $p_\text{in}=4$, $z_\text{in}\in[0,3]$, $t_\text{sep}=5$. 
		The left and right sides show the results from using the GBT and linear models, respectively. We use $N_\text{est} = 150$, $r = 0.1$ for the GBT model.
		The horizontal axis shows $N_\text{tr} + 0.1 N_\text{BC}$, and the number of unlabeled measurements is fixed to 1180. 
		Points in blue are for predictions with bias correction, and orange for observations. 
	}
	\label{fig:kaon_bc}
\end{figure}

Fig.~\ref{fig:kaon_bc} compares the final predictions among various training and bias-correction measurements: 
$N_\text{tr}$ and $N_\text{BC}$ are selected from $\{80, 160, 240, 320, 400\}$, 
and the number of unlabeled measurements is fixed to $N_\text{ul} = 1180$. 
The fit parameters are adopted as above. 
We observe a reduced error size of final predictions with increased $N_\text{tr}$ and $N_\text{BC}$.

\begin{figure}[htbp]
	\begin{minipage}{0.5\textwidth}
		\includegraphics[width=\linewidth]{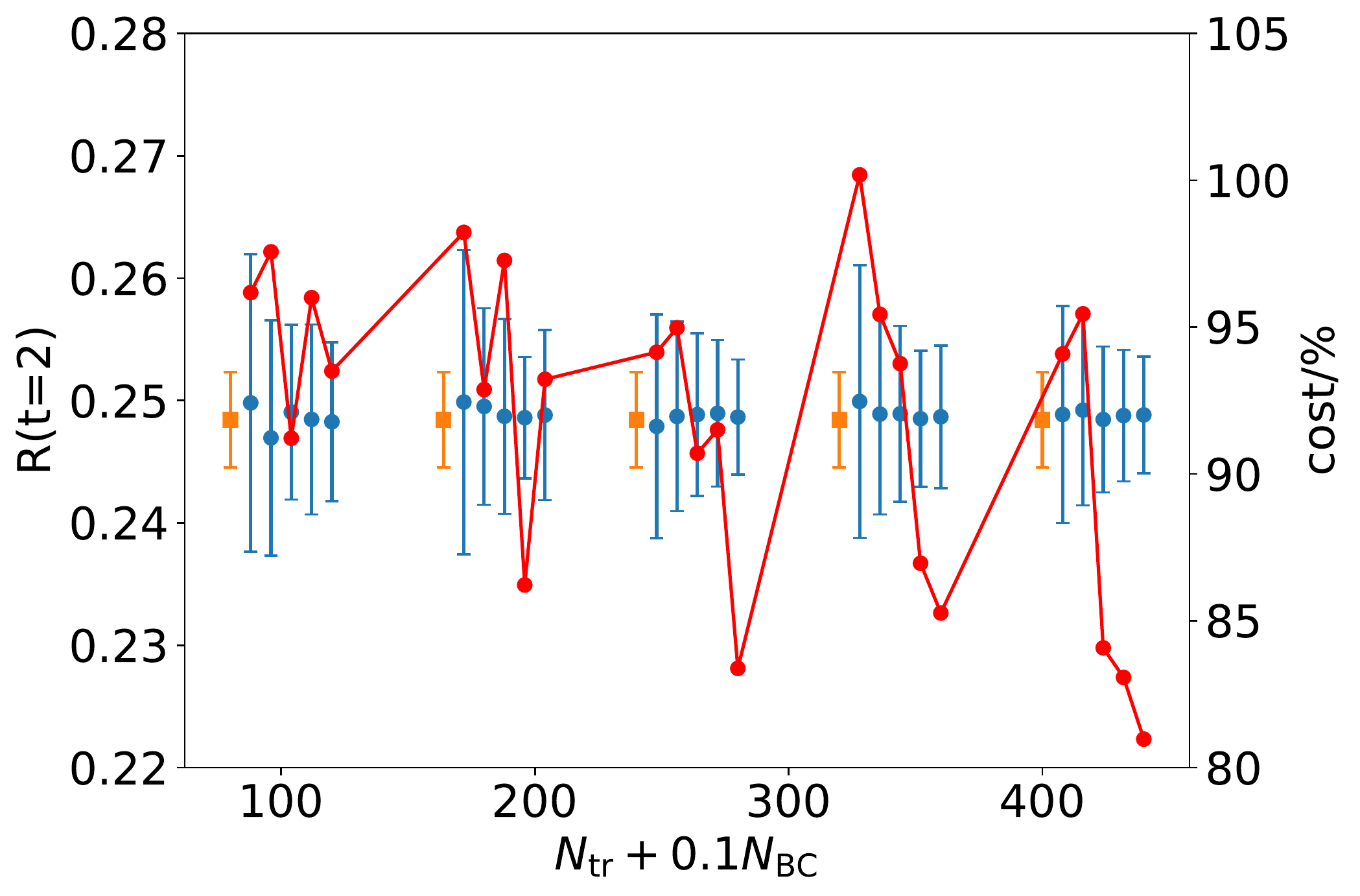}
	\end{minipage}\hfill
	\begin{minipage}{0.5\textwidth}
		\includegraphics[width=\linewidth]{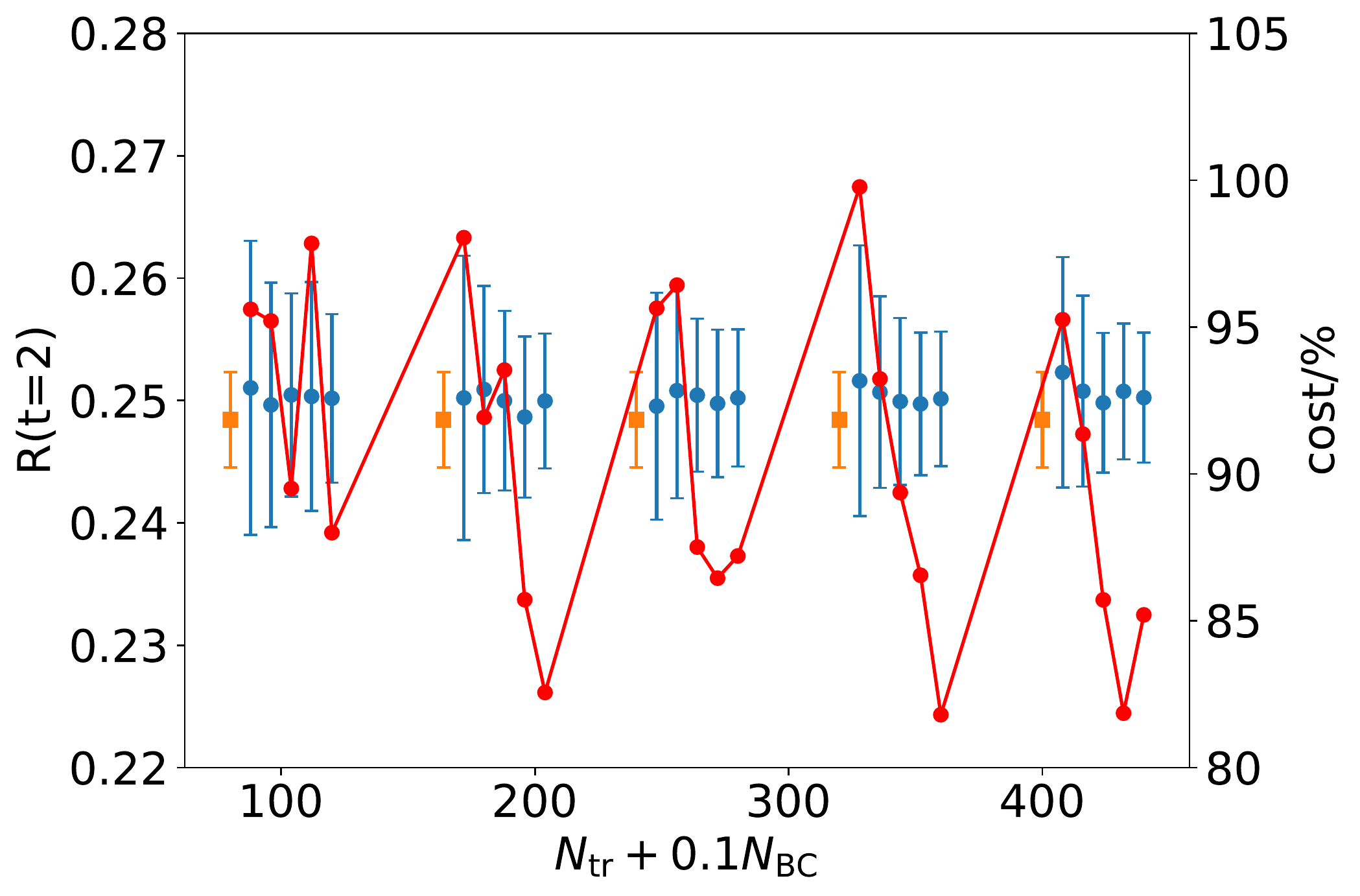}
	\end{minipage}
	\caption{Observations and predictions of the ratio $R(t=2)$ of kaon quasi-PDF correlators at $p_\text{pred} = 4$, $z_\text{pred}= 4$, $t_\text{sep} = 5$ from input data at $p_\text{in}=3$, $z_\text{in}=4$, $t_\text{sep}=5$. The red line shows the effective cost averaged on $R(t\in[1,4])$.
		The left and right sides show the results from using the GBT and linear models, respectively. We use $N_\text{est} = 150$, $r = 0.1$ for the GBT model.
		The horizontal axis shows $N_\text{tr} + 0.1 N_\text{BC}$, and the number of unlabeled measurements is fixed to 1180. 
		Points in blue are for predictions with bias correction, and orange for observations.
	}
	\label{fig:kaon_bc_cost}
\end{figure}

Using $p$-prediction on kaon quasi-PDFs can reduce the computational cost, because calculating $C_\text{3pt}$ at different momenta requires the calculation of different propagators from different sequential sources. The effective computational savings of the ML calculation can be derived by considering the number of propagators needed to achieve the same precision as in a calculation without ML. In our case, to use $p_\text{in}=3$ to predict $p_\text{pred}=4$, we need to calculate $N_\text{in}$ propagators at $p_\text{in}=3$ and $N_\text{BC}+N_\text{tr}$ propagators at $p_\text{pred}=4$ for the ML setup. Then, we can use the model to obtain the $N_\text{ul}$ predictions at $p_\text{pred}=4$. This amount of data is equivalent to a non-ML calculation with $N_\text{in}$ propagators at $p_\text{in}=3$ and $N_\text{ul}\times\sigma^2(R_\text{ul})/\sigma^2(R_\text{comb})$ propagators at $p_\text{pred}=4$. The cost with ML can be quantified by:
\begin{equation}
\text{Cost} = \frac{N_\text{in}+N_\text{BC}+N_\text{tr}}{N_\text{in}+N_\text{ul} \left\langle\frac{\sigma^2(R_\text{ul})}{\sigma^2(R_\text{comb})}\right\rangle_t}
\end{equation}
where $N_\text{BC}, N_\text{tr}$ and $N_\text{ul}$ are the numbers of propagator calculations (which represent the computational cost) needed to obtain the corresponding datasets (bias-correction, training and unlabeled), and $N_\text{in}$ is that of the input data. The ratio $\sigma^2(R_\text{ul})/\sigma^2(R_\text{comb})$ is the scaling factor of the effective number of measurements we can obtain by employing the ML prediction, accounting for the increase of statistical error due to prediction error. We assume that the errors of observables scale as $1/\sqrt{N}$ as the number of measurements increases. For the cost estimate, we use an average value of the ratios over different insertion timeslices. We calculate the $R_\text{comb}=C_\text{3pt}^\text{comb}/C_\text{2pt}$ here from each bootstrap sample by taking the weighted average of the measurements on labeled data and BC predictions on unlabeled data in each sample:
\begin{equation}
C^\text{comb}_\text{3pt} = \frac{\bar{C}^\text{pred,BC}_\text{3pt}/\sigma^2(C^\text{pred,BC}_\text{3pt})+\bar{C}^\text{lb}_\text{3pt}/\sigma^2(C^\text{lb}_\text{3pt})}
{1/\sigma^2(C^\text{pred,BC}_\text{3pt})+1/\sigma^2(C^\text{lb}_\text{3pt})}
\end{equation}
while the error $\sigma(R_\text{comb})$ is estimated from all bootstrap results. A smaller cost indicates higher prediction efficiency, so we vary $N_\text{tr}$ and $N_\text{BC}$ to find the optimal cost reduction, as shown in Fig.~\ref{fig:kaon_bc_cost}. By choosing optimal $N_\text{tr}$ and $N_\text{BC}$, we can obtain about $20\%$ reduction in computational cost.

\begin{figure}[htbp]
	\centering 
	\includegraphics[width=0.49\linewidth]{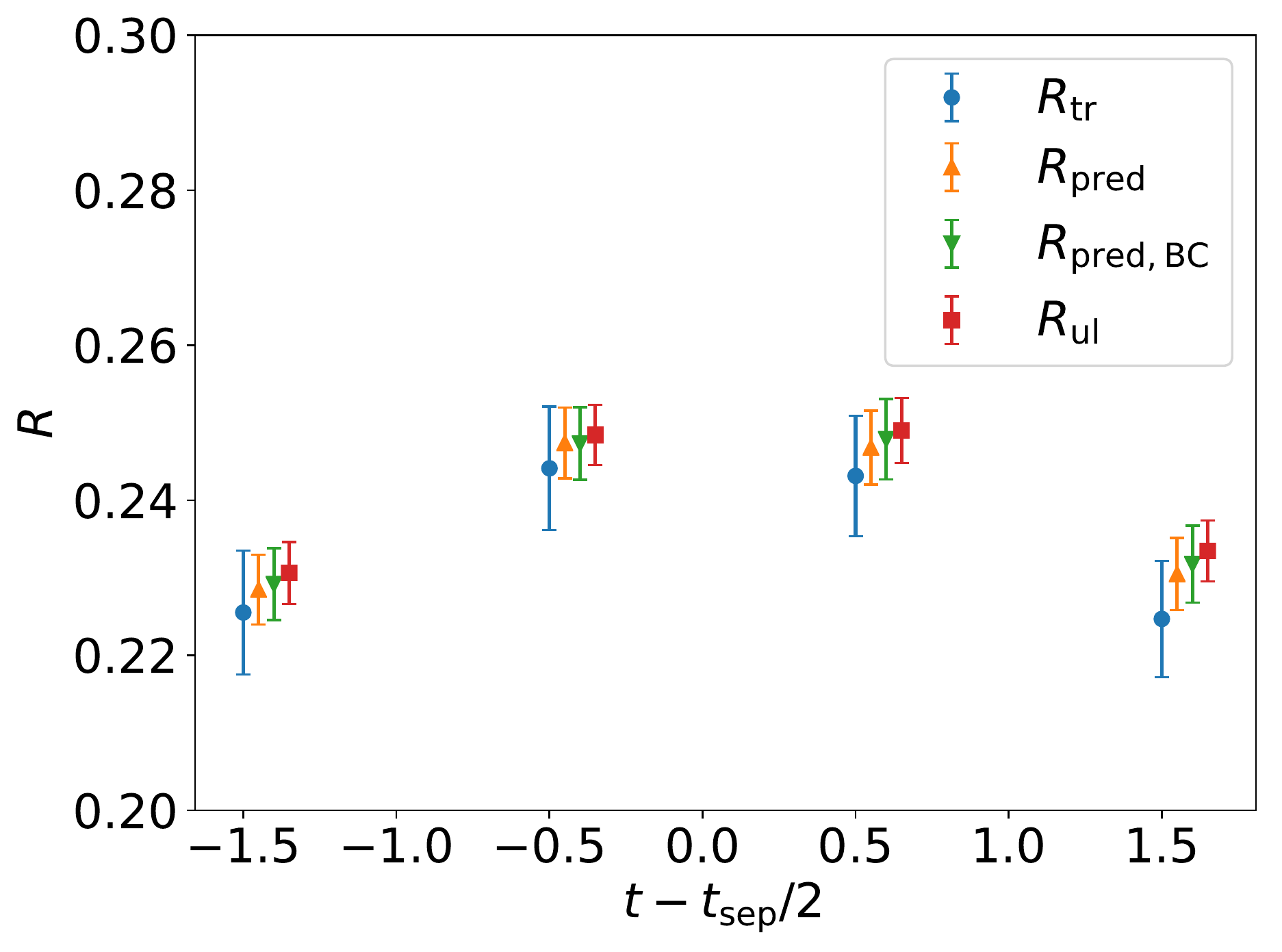}
	\includegraphics[width=0.49\linewidth]{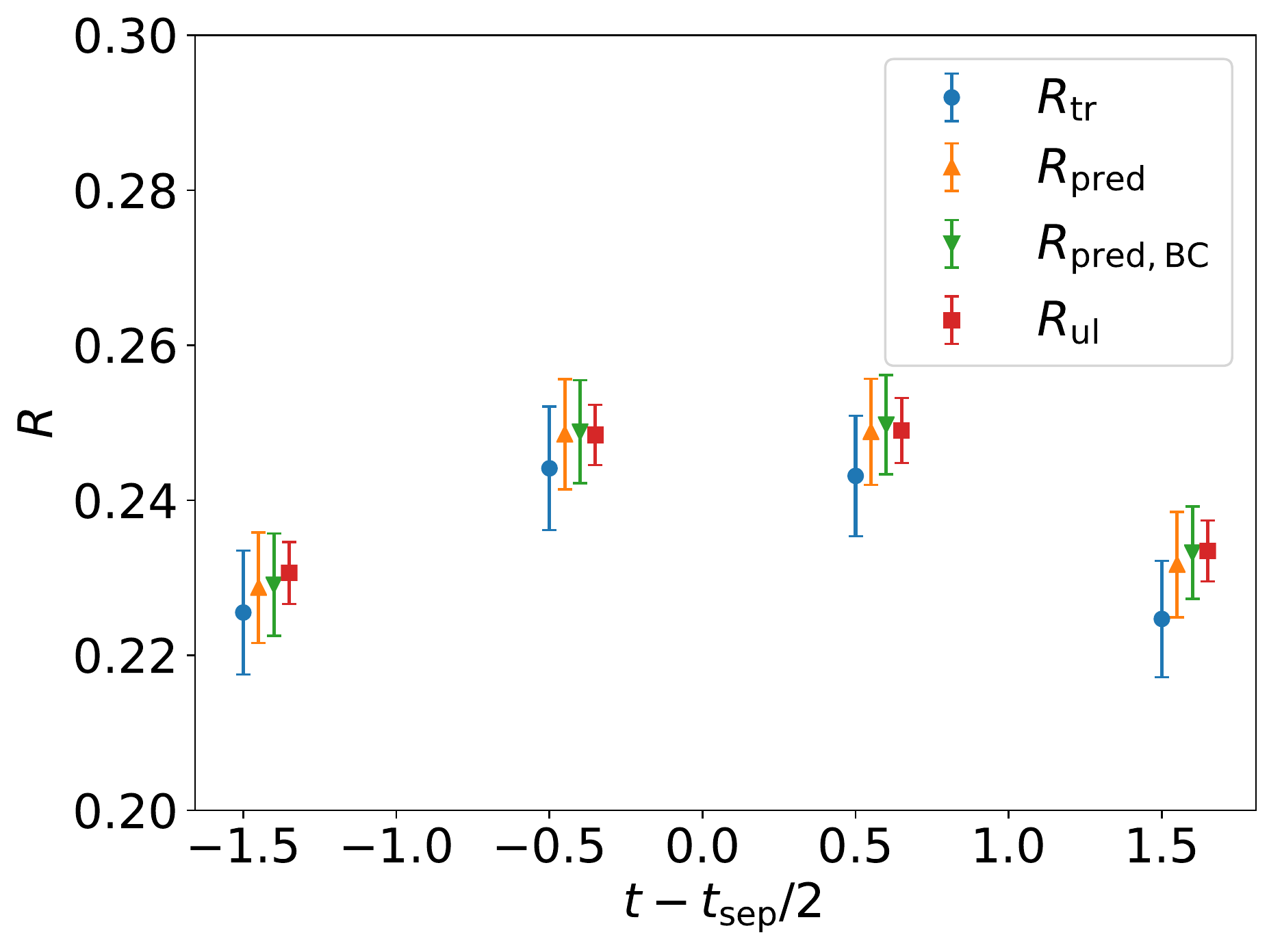}
	\includegraphics[width=0.49\linewidth]{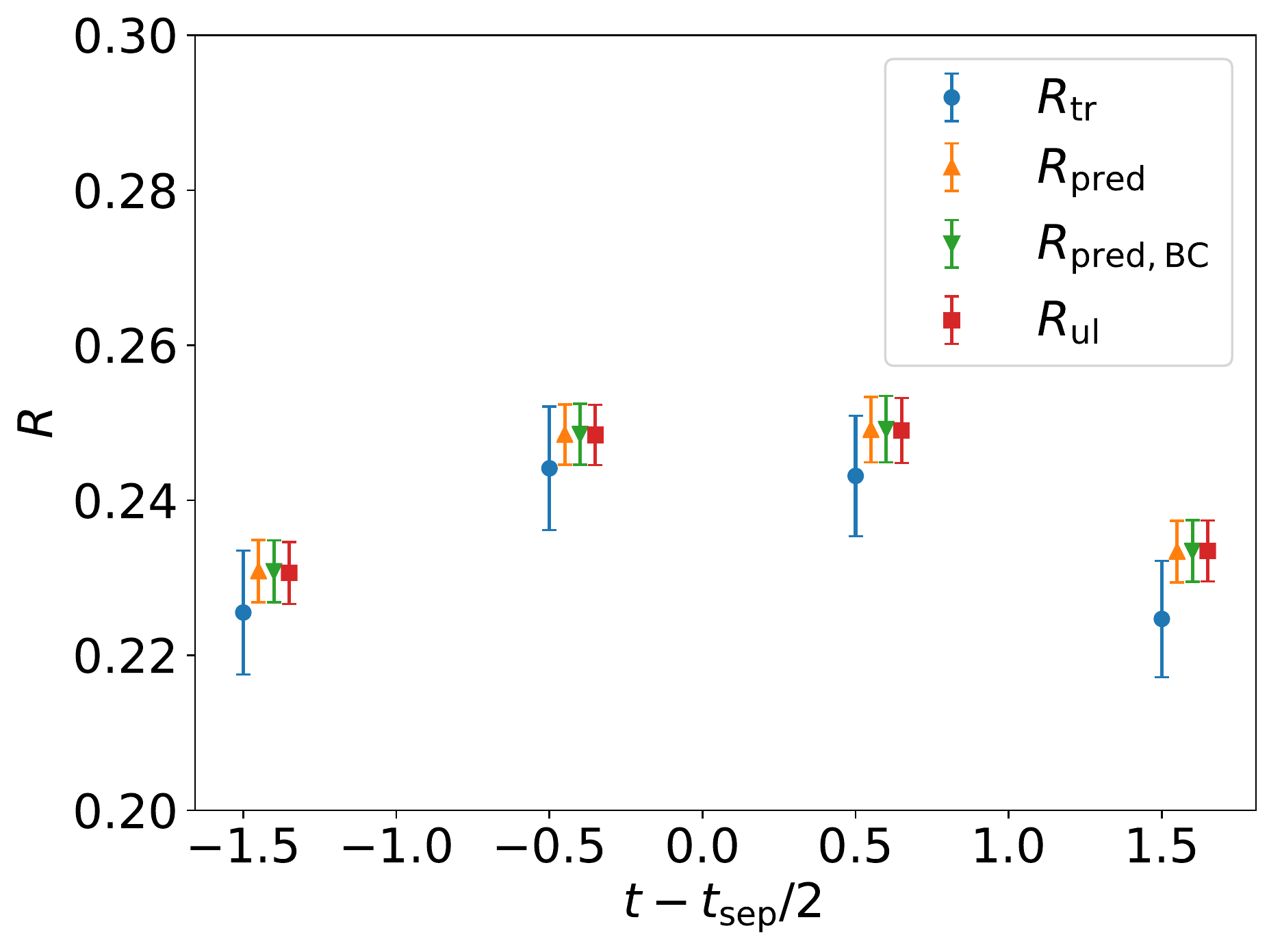}
	\includegraphics[width=0.49\linewidth]{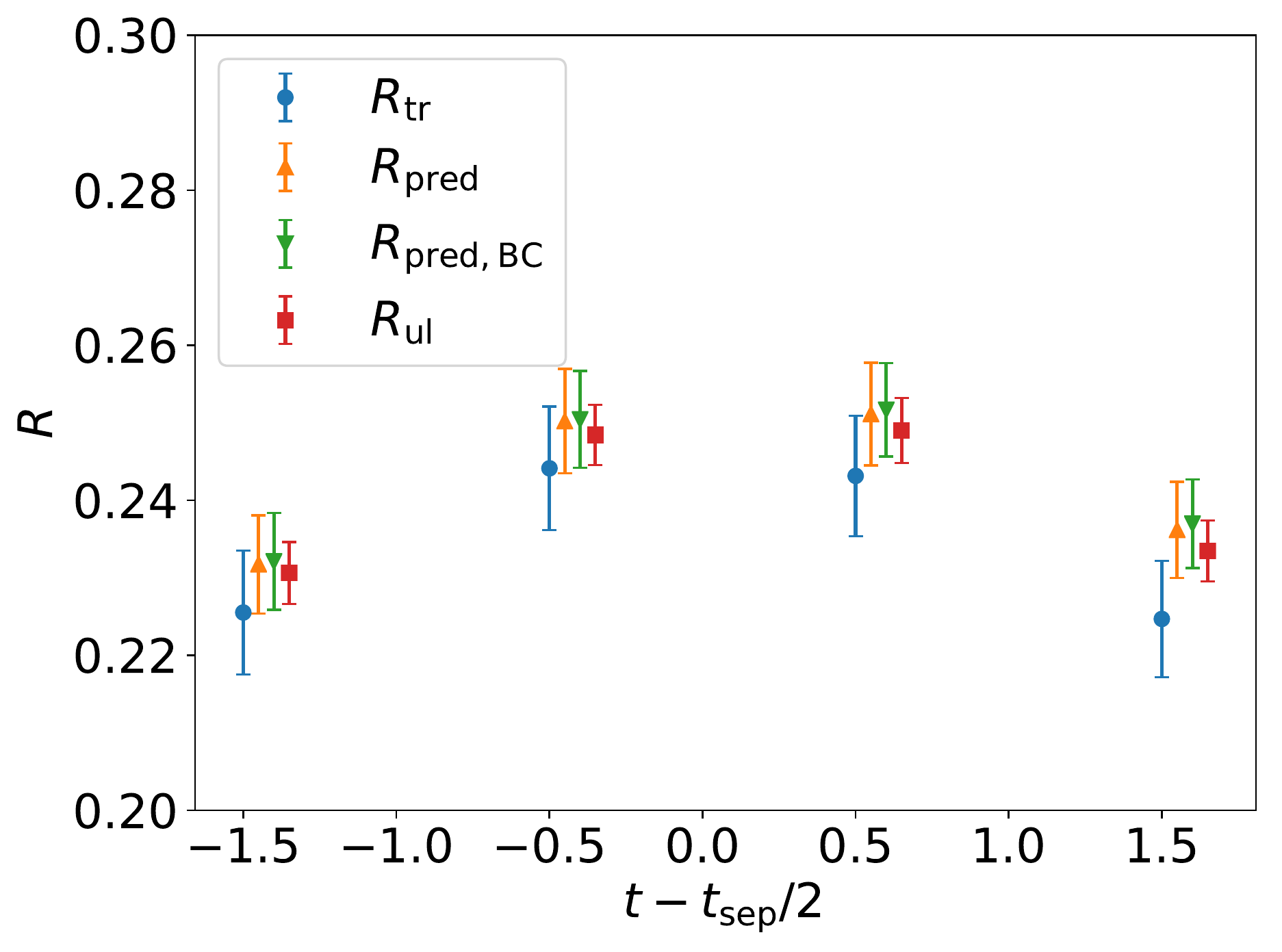}
	\caption{The ratio $R(t)$ of the kaon quasi-PDF correlators at $z_\text{pred}=4$, $p_\text{pred}=4$ from direct measurements and the predictions of the three models. The top (bottom) row is GBT (linear) model with $N_\text{est} = 150$, $r = 0.1$. $N_\text{tr}=N_\text{BC}=240$ and $N_\text{ul}=1180$ are used. The left column uses $z_\text{in}\in[0,3]$, $p_\text{in}=4$ as inputs, while the right column uses $z_\text{in}=4$, $p_\text{in}=3$ as inputs. $z$-predictions are better than $p$-predictions.}
	\label{fig:kpdf_ratio}
\end{figure}

Figure~\ref{fig:kpdf_ratio} shows this set of fitted results from both the $z$-prediction and $p$-prediction at $N_\text{tr} = 240$, $N_\text{BC} = 240$, 
while Table~\ref{Tab2} compares several sets of $p$- and $z$-predictions and observations. 
The last column of the table shows the fit quality.

\begin{table}[h]
	\centering
	\begin{tabular}{@{} |c|c|c|c|c|c|c|c|c| @{}}
		\hline
		Type &Input& Method   & $R_\text{tr}$ & $R_\text{pred}$ & $R_{\text{pred},\text{BC}}$& $R_\text{comb}$ & $R_\text{ul}$  &$F_v$ \\
		\hline
		\multirow{2}{*}{$p$-pred}& \multirow{2}{*}{$p_\text{in}=3,z_\text{in}=4$}& GBT & 0.2441(70) & 0.2430(60) & 0.2439(56) & 0.2435(51) & 0.2471(35) & 0.692(41)\\
		\cline{3-9}
		& &linear & 0.2441(70) & 0.2479(63) & 0.2480(58) & 0.2472(54) & 0.2471(35) & 0.772(29)\\
		\hline
		\multirow{2}{*}{$z$-pred}& \multirow{2}{*}{$p_\text{in}=4,z_\text{in}\in[0,3]$}&GBT & 0.2441(70) & 0.2458(40) & 0.2455(41) & 0.2456(32) & 0.2471(35) & 0.890(26)\\
		\cline{3-9}
		& &linear & 0.2441(70) & 0.2470(36) & 0.2473(36) & 0.2466(32) & 0.2471(35) & 0.998(1)\\
		\hline
	\end{tabular}
	\caption{Observations and predictions of the ratio $R(t=2)$ of the kaon quasi-PDF correlators at $p_\text{pred}=4$, $z_\text{pred}=4$, $t_\text{sep}=5$ from different models and inputs. We use $N_\text{est} = 150$, $r = 0.1$ for the GBT model. The models are trained on 240 measurements with 240 bias-correction measurements, and then tested on 1180 unlabeled measurements. The linear model shows a better fit variance than GBT.}
	\label{Tab2}
\end{table}

We compare the predicted ratios for these models in Fig.~\ref{fig:kpdf_ratio}. The $z$-predictions are consistent with unlabeled data for both models, but the $p$-predictions still need to be improved.

\subsubsection{\texorpdfstring{$\eta_s$}{Eta-s} quasi-PDF results}

\begin{figure}[htbp]
	\begin{minipage}{0.5\textwidth}
		\includegraphics[width=\linewidth]{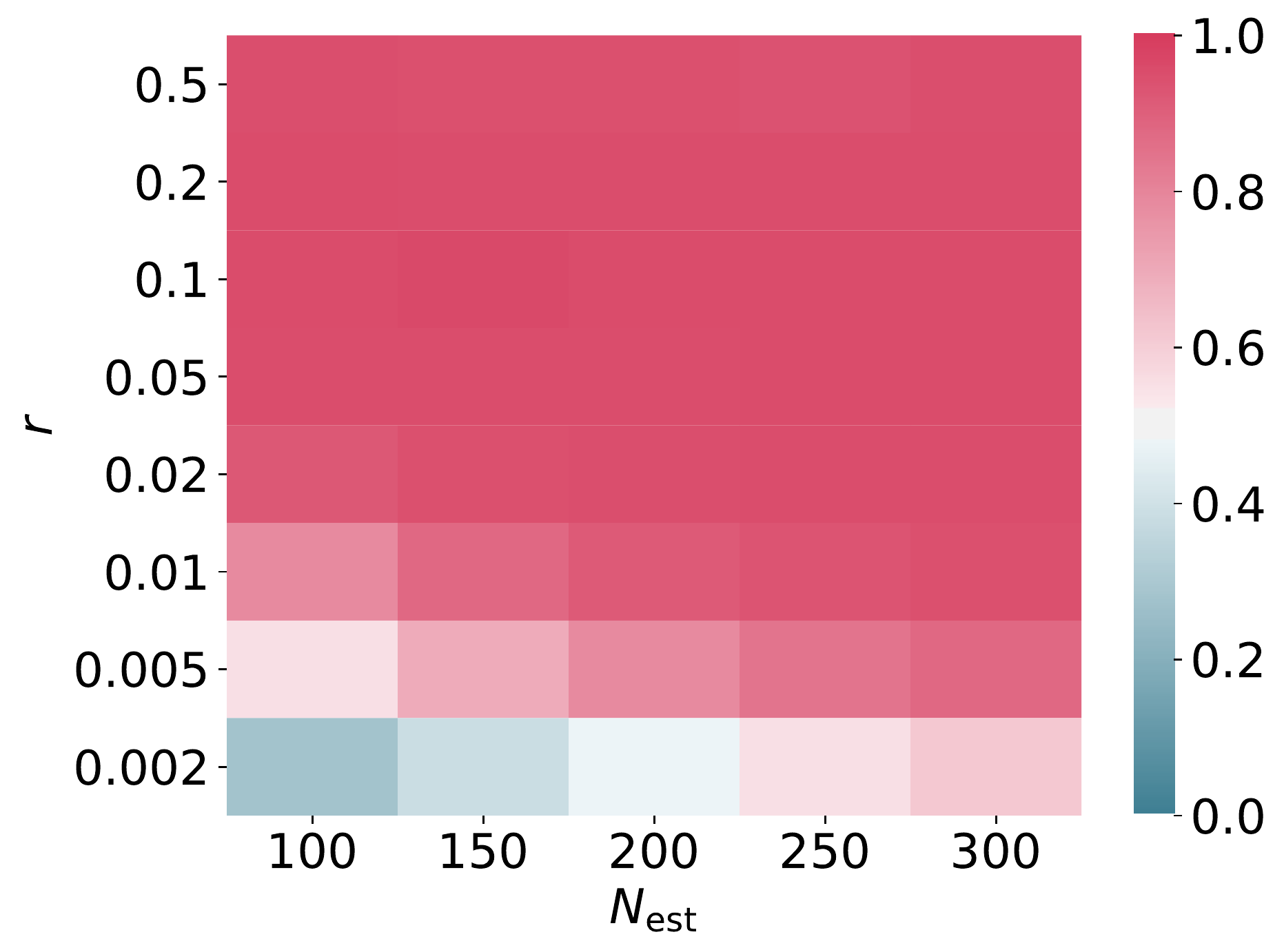}
	\end{minipage}\hfill
	\begin{minipage}{0.5\textwidth}
		\includegraphics[width=\linewidth]{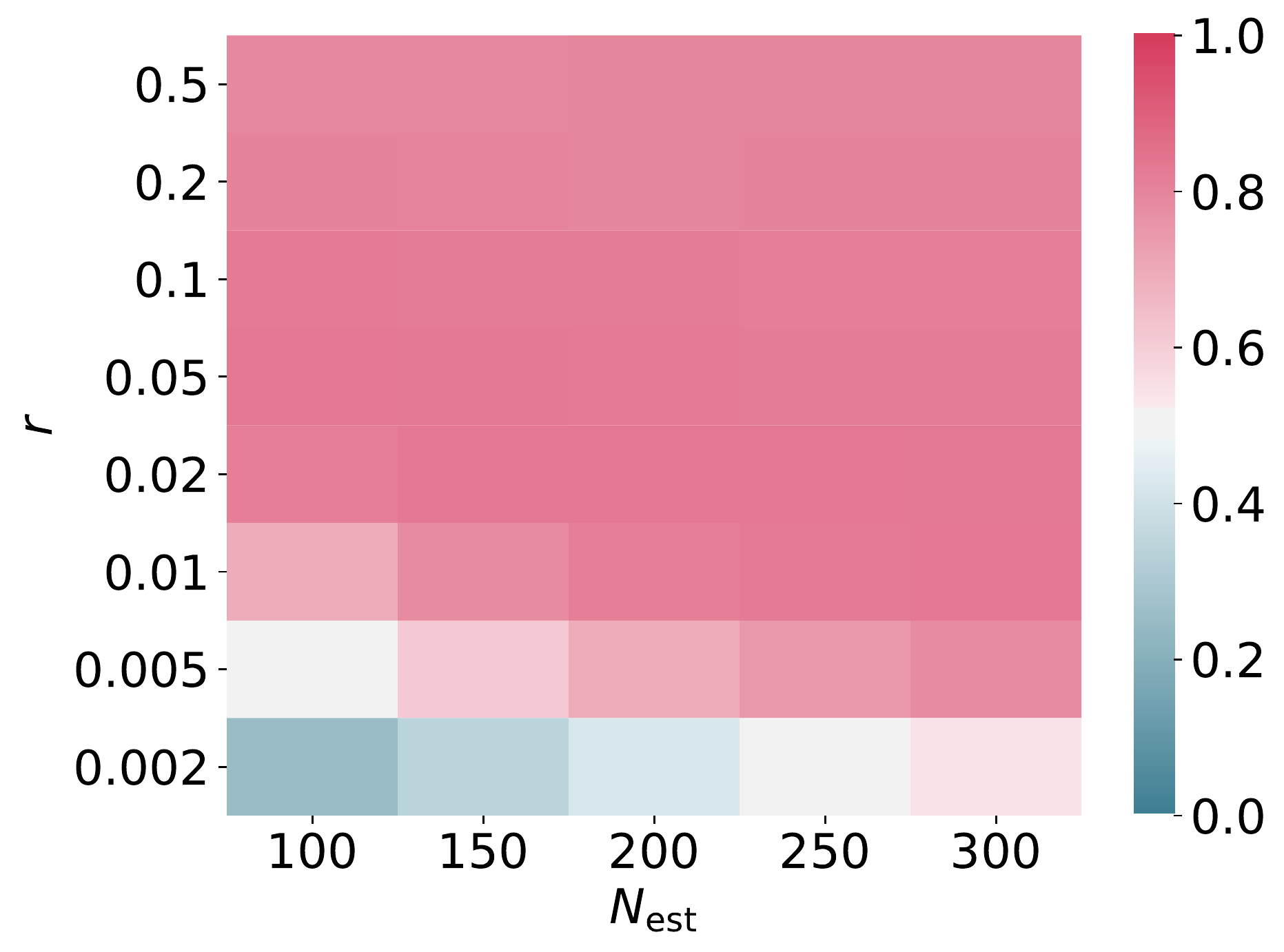}
	\end{minipage}
	\caption{Estimates of the fit variance $F_v$ as a function of learning rate $r$ and number of estimators $N_\text{est}$ from the $\eta_s$ quasi-PDF measurements at $p_\text{pred} = 4$, $z_\text{pred} = 4$, $t_\text{sep} = 5$, and $t_\text{pred} = 2$. $N_\text{tr}=400$ and $N_\text{ul}=1180$ are used.
		The left (right) side shows the results from the $z$-prediction ($p$-prediction). The performance is better than the model of kaon data.
	}
	\label{fig:eta_fv}
\end{figure}

\begin{figure}[htbp]
	\begin{minipage}{0.50\textwidth}
		\includegraphics[width=\linewidth]{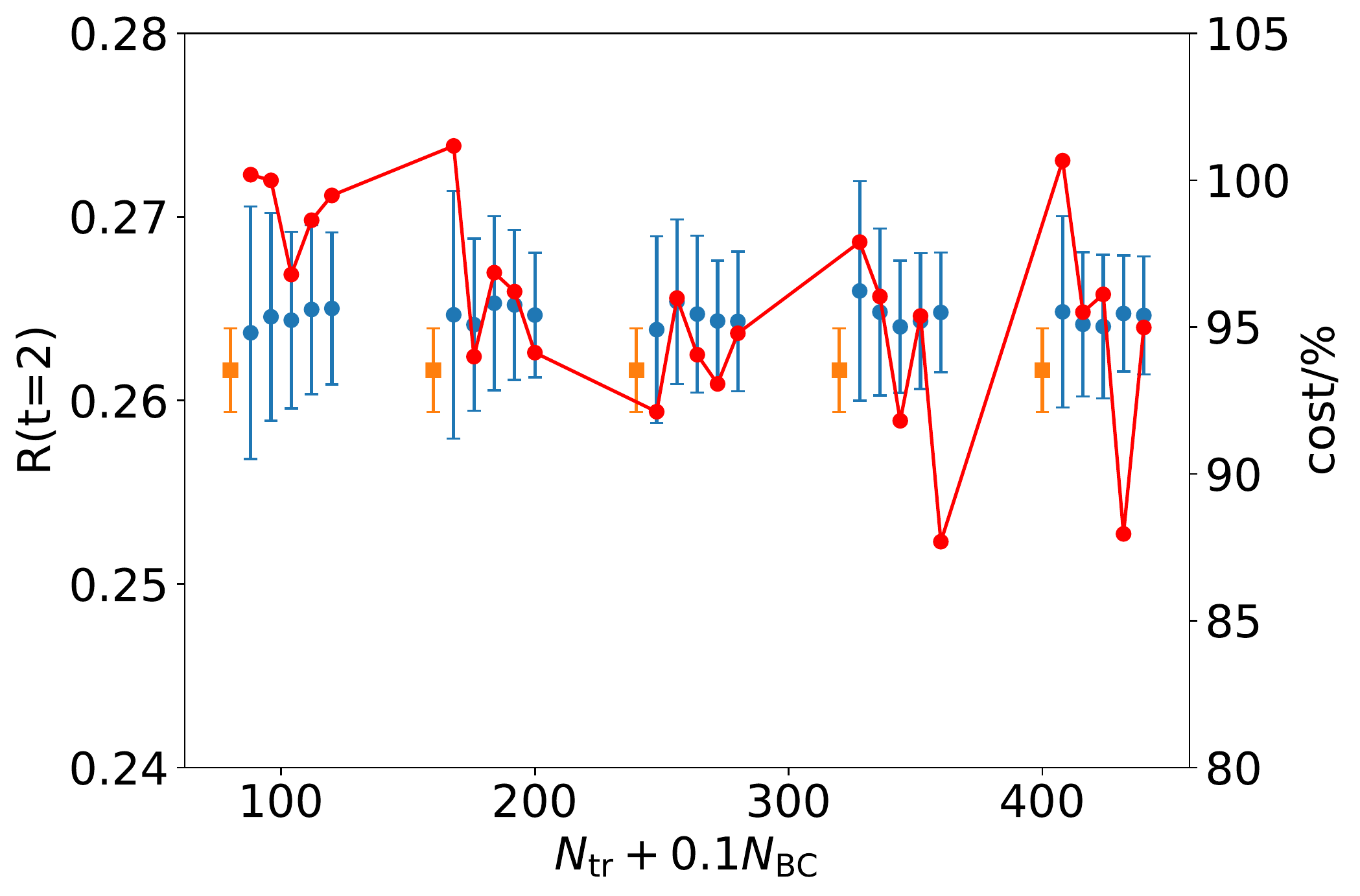}
	\end{minipage}\hfill
	\begin{minipage}{0.50\textwidth}
		\includegraphics[width=\linewidth]{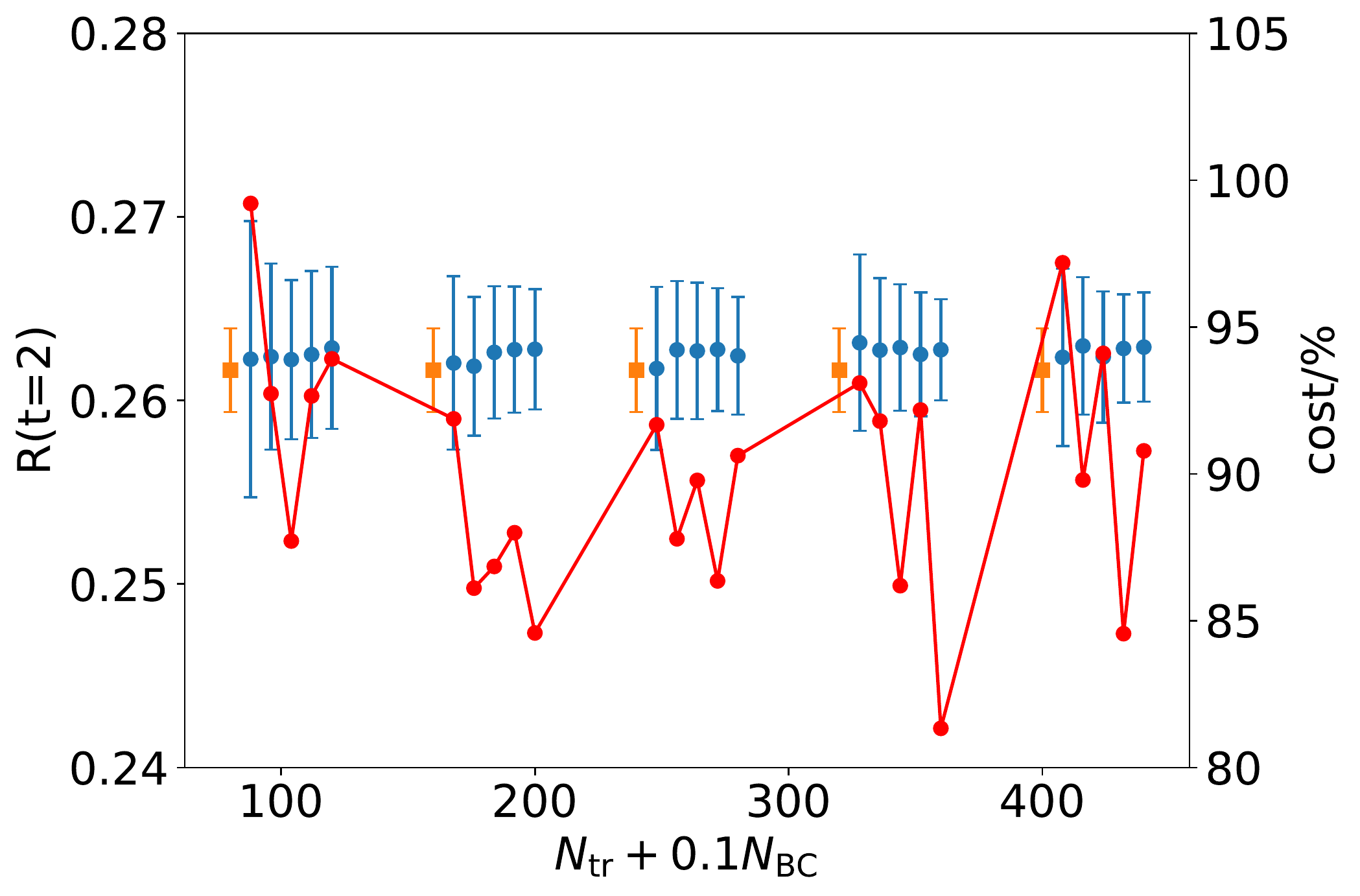}
	\end{minipage}
	\caption{Observations and predictions of the ratio $R(t=2)$ of $\eta_s$ quasi-PDF correlators at $p_\text{pred} = 4$, $z_\text{pred}= 4$, $t_\text{sep} = 5$ from input data at $p_\text{in}=3$, $z_\text{in}=4$, $t_\text{sep}=5$. Red line is the effective cost averaged on $R(t\in[1,4])$.
		The left and right sides show the results from using the GBT and linear models, respectively. We use $N_\text{est} = 150$, $r = 0.1$ for the GBT model.
		The horizontal axis shows $N_\text{tr} + 0.1 N_\text{BC}$, and the number of unlabeled measurements is fixed to 1180. 
		Points in blue are for predictions with bias correction, and orange for observations. 
	}
	\label{fig:eta_bc_cost}
\end{figure}

\begin{figure}[h]
	\centering 
	\includegraphics[width=0.49\linewidth]{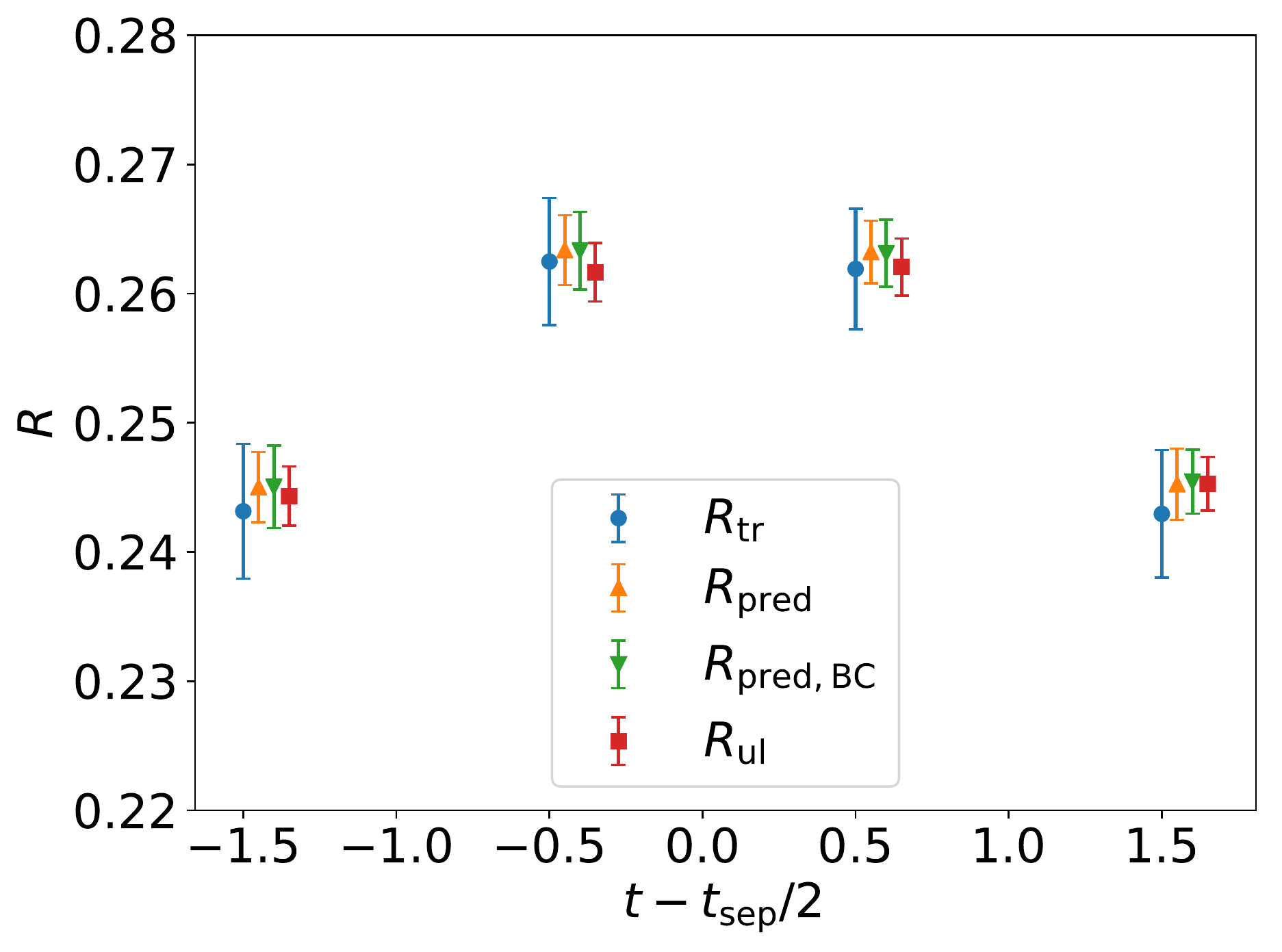}
	\includegraphics[width=0.49\linewidth]{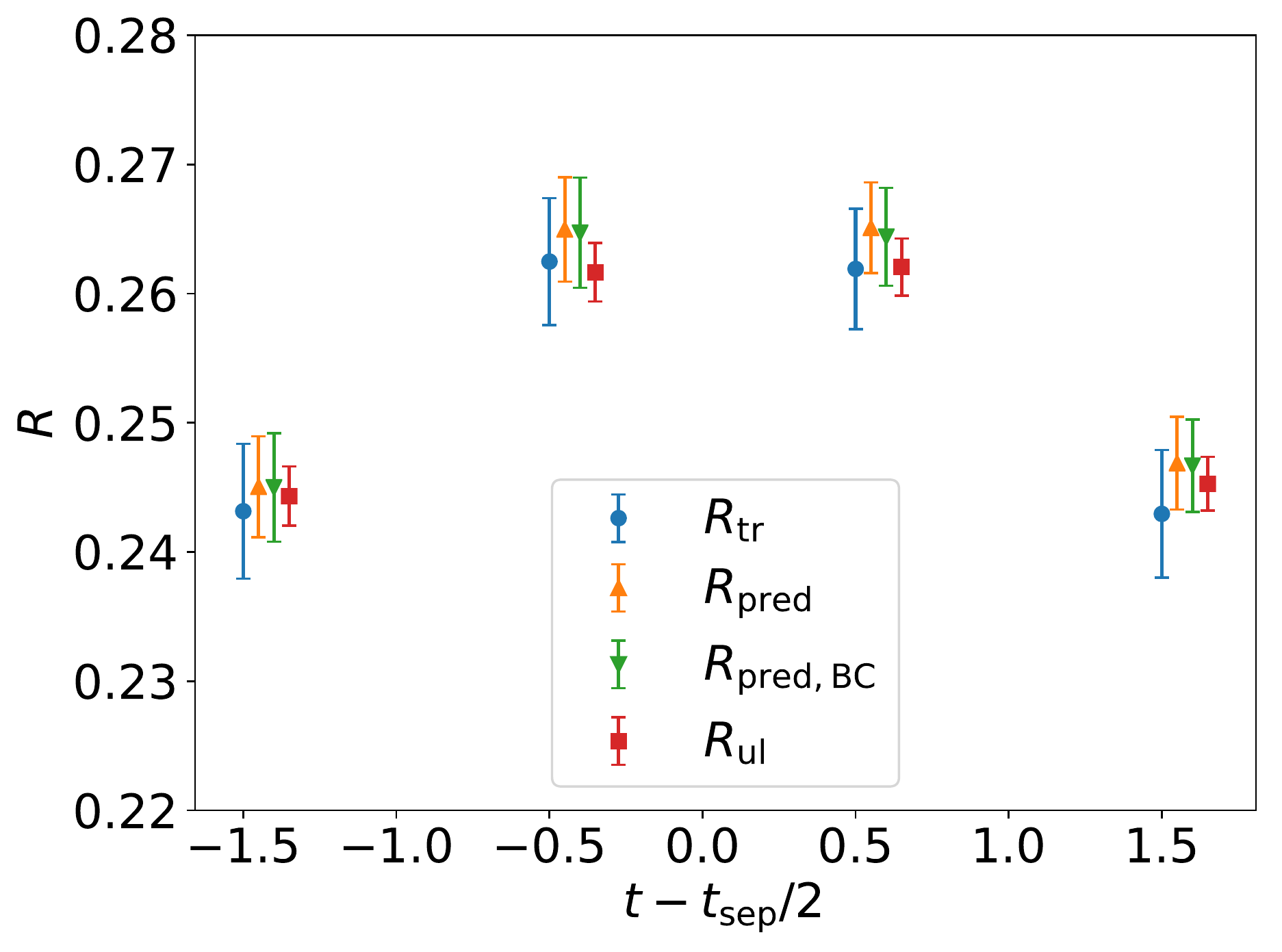}
	\includegraphics[width=0.49\linewidth]{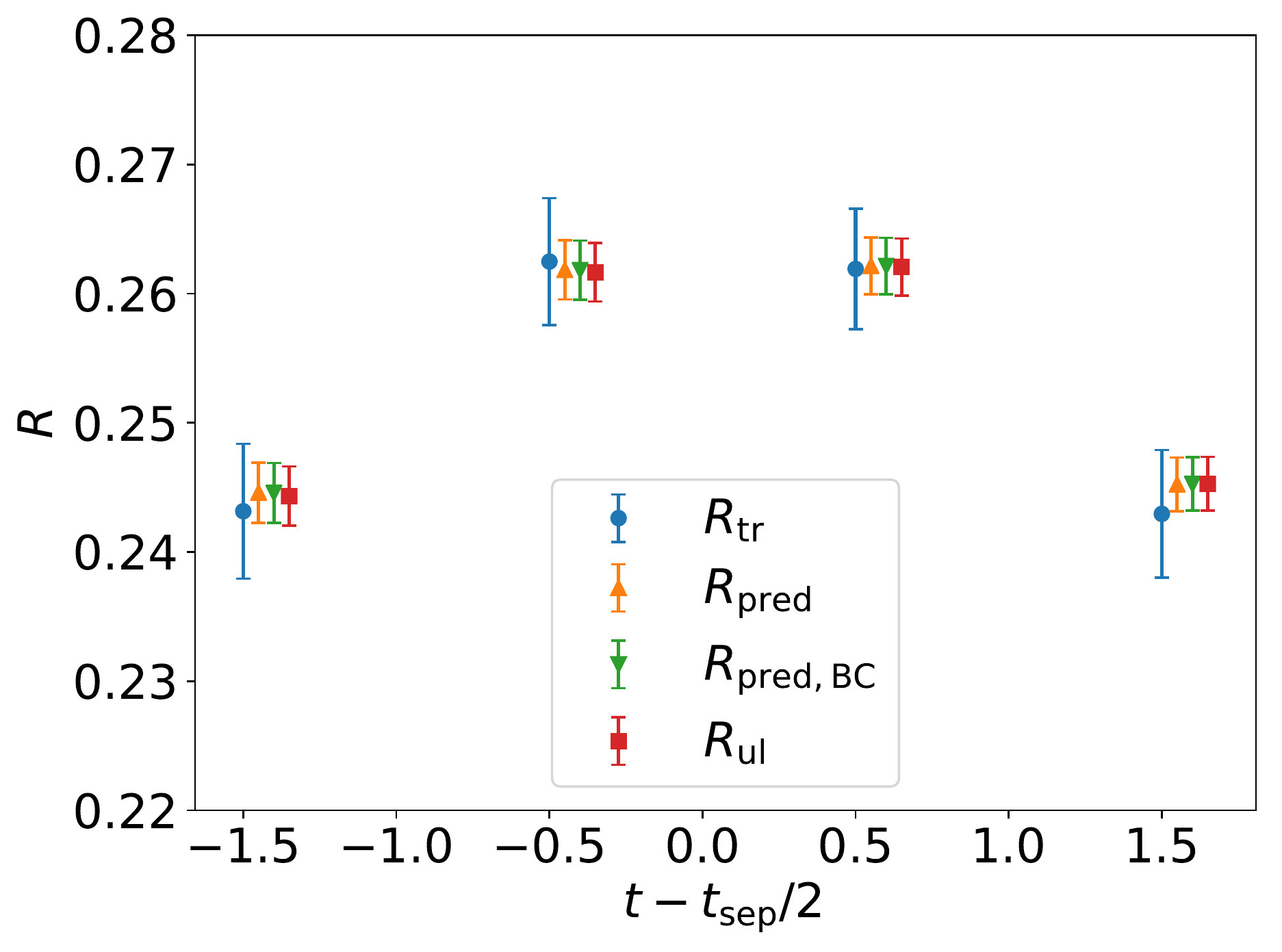}
	\includegraphics[width=0.49\linewidth]{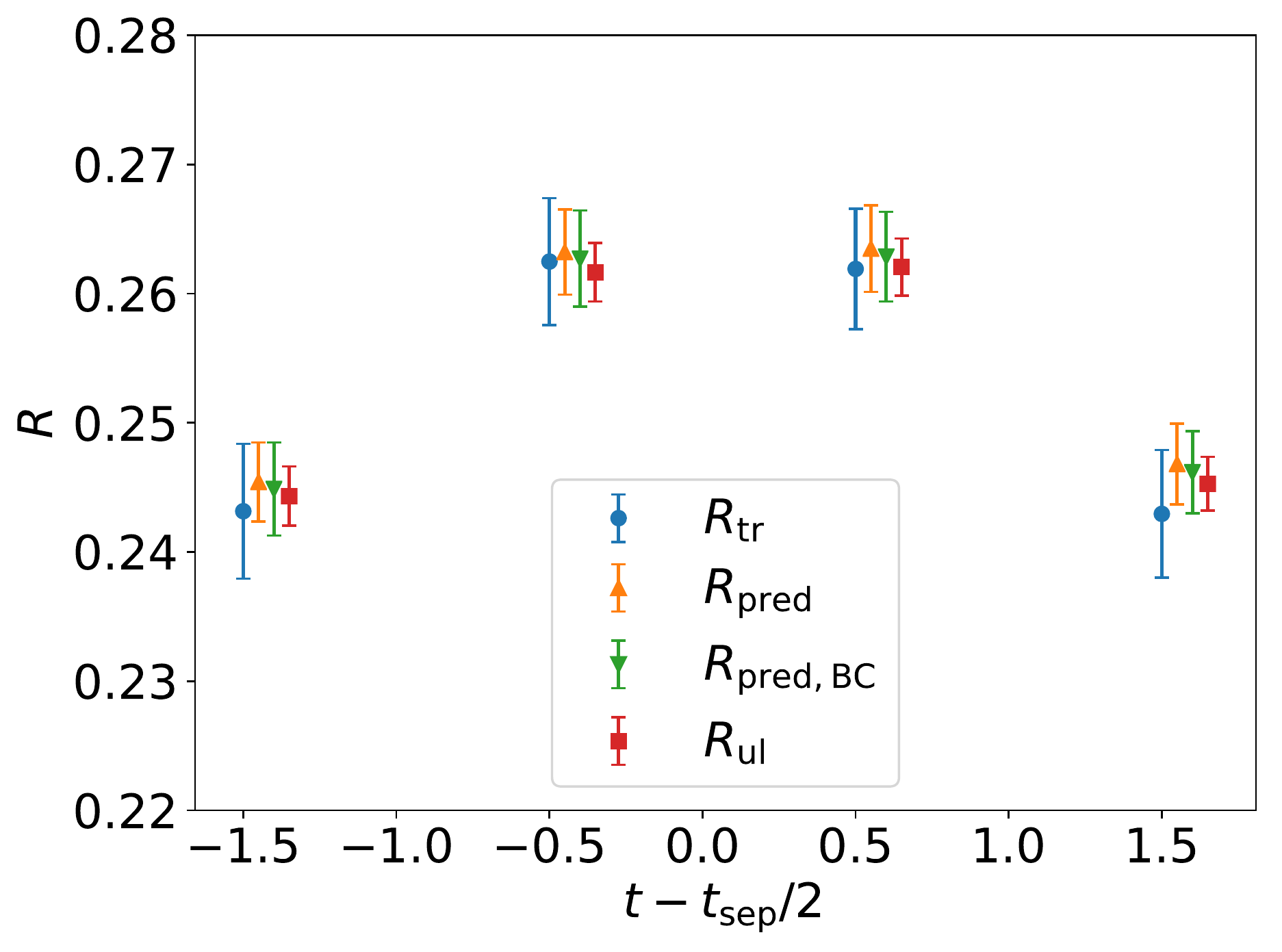}
	\caption{The ratio $R(t)$ of $\eta_s$ quasi-PDF correlators at $z_\text{pred}=4$, $p_\text{pred}=4$ from direct measurements and the predictions of the two models. The top (bottom) row is the GBT (linear) model. The left column uses $z_\text{in}\in[0,3]$, $p_\text{in}=4$, the right column uses $z_\text{in}=4$, $p_\text{in}=3$. The model performs better on these cleaner datasets. $N_\text{tr}=N_\text{BC}=240$, $N_\text{ul}=1180$ and $N_\text{est}=150$, $r=0.1$ are used in the model training.}
	\label{fig:epdf_ratio}
\end{figure}

For the $\eta_s$ quasi-PDF data, the meson operator is $\eta_\text{s}= \bar{s} \gamma_5 s$. The $\eta_s$ data have better signals, and the correlations among $\eta_s$ data show the same patterns as those of the kaon. Therefore, we select the same parameters for the model training, $N_\text{est}=150$, $r=0.1$, $N_\text{tr}=N_\text{BC}=240$, $N_\text{ul}=1180$. By comparing Fig.~\ref{fig:eta_fv} and Fig.~\ref{fig:kaon_fv}, we can see that the fit quality is slightly improved by the cleaner dataset. We infer that with more labeled kaon quasi-PDF data available for model training, the kaon model will show better performance as well. The predictions compared with observations are shown in Fig.~\ref{fig:epdf_ratio}. Both $z$-predictions and $p$-predictions are more precise compare to the kaon case. Figure~\ref{fig:eta_bc_cost} shows the cost on different $N_\text{tr}/N_\text{BC}$ set, the linear model shows a better optimal reduction than the kaon case. Overall, the cost reductions are $12\%$--$18\%$ at optimal choices of the sizes of the training and bias-correction datasets.

\subsection{Gluon Quasi-PDF Matrix Elements}

The gluon PDF contributes at next-to-leading order to deep inelastic scattering (DIS) cross sections, and it enters at leading order in jet production. Global fits have combined the data from both DIS and jet-production cross sections, and constraints on the gluon PDF from the experimental side are improving.  However, on the theoretical side the gluon PDF is poorly known. PDF cannot be calculated using perturbative QCD. Recently, it has been found that they can be calculated directly in lattice QCD using large-momentum effective field theory. The gluon unpolarized quasi-PDF matrix elements are computed on the lattice using
\begin{align}\label{}
C_\text{3pt}(z;t_\text{sep},t)&=
\langle 0|\Gamma\int d^3y\, e^{-iy\cdot P}\chi(\vec y,t_\text{sep})F_{\mu t}(z,t)[\prod_{x=0}^{z-1} U(x, t)]F^{z\mu}(0,t)\chi(\vec 0,0)|0\rangle, \\
C_\text{2pt}(z;t_\text{sep})&=
\langle 0|\Gamma\int d^3y\, e^{-iy\cdot P}\chi(\vec y,t_\text{sep})\chi(\vec 0,0)|0\rangle, 
\end{align}
where $C_\text{3pt}$ is the three-point correlator, $C_\text{2pt}$ is the two-point correlator, ${\cal O}(z,t)$ is the gluon operator, $\chi=\epsilon^{abc}[{u^a}^T(x)i\gamma_4\gamma_2\gamma_5 d^b(x)]u^c(x)$ is the nucleon interpolation field, \{a,b,c\} are color indices, $\Gamma=\frac{1}{2}(1+\gamma_4)$, and the field tensor $F_{\mu\nu}$ is defined by
\begin{equation}
F_{\mu\nu}=\frac{i}{8a^2g}(\mathcal{P}_{[\mu,\nu]}+\mathcal{P}_{[\nu,-\mu]}+\mathcal{P}_{[-\mu,-\nu]}+\mathcal{P}_{[-\nu,\mu]}),
\end{equation}
where the plaquette $\mathcal{P}_{\mu,\nu}=U_{\mu}(x)U_{\nu}(x+a\hat{\mu})U^{\dag}_{\mu}(x+a\hat{\nu})U^{\dag}_{\nu}(x)$ and $\mathcal{P}_{[\mu,\nu]}=\mathcal{P}_{\mu,\nu}-\mathcal{P}_{\nu,\mu}$. To improve the signal, we studied 1, 3, 5, 10 steps of hypercubic (HYP) smearing~\cite{Hasenfratz:2001hp} on the gluon momentum faction $\langle x\rangle_g$ in Eq.~(3) of Ref.~\cite{Fan:2018dxu}. After applying the renormalization to the bare matrix elements, the results from different numbers of HYP-smearing steps are consistent with each other and with phenomenology results 0.42(2) within the uncertainties~\cite{Fan:2018dxu} with the exception of the 10-step. Therefore, we apply 5 steps of HYP smearing to the gluon quasi-PDF operators in this work. The ratio $R$ of the three-point correlator to the two-point correlator follows the same definition as in Eq.~\eqref{eq:ratio}.

We use valence overlap fermions on RBC gauge configurations~\cite{blum2016domain} with $2+1$ flavors of domain-wall fermions (DWF), lattice volume $L^3\times T=24^3\times 64$, lattice spacing $a=0.1105(3)$~fm, and pion mass $m_\pi^\text{sea}=330$~MeV. We also compute clover valence quarks on the MILC $N_f = 2+1+1$ HISQ configurations~\cite{bazavov2013lattice} with $L^3\times T=32^3\times 96$, $a=0.0888(8)$~fm, and $m_\pi^\text{sea}=313$~MeV. For the nucleon two-point function, considering all timeslices and independent smeared point sources, the number of measurements for the two-point functions is $200\times 128\times 8=204,800$ on the RBC-24I lattices and $300\times 16\times 6=28,800$ on the MILC-a09m310 lattices.

\subsubsection{Predictions of the gluon correlators with the overlap valence fermions}

\begin{figure}[htb]
	\centering
	\includegraphics[width=0.49\textwidth]{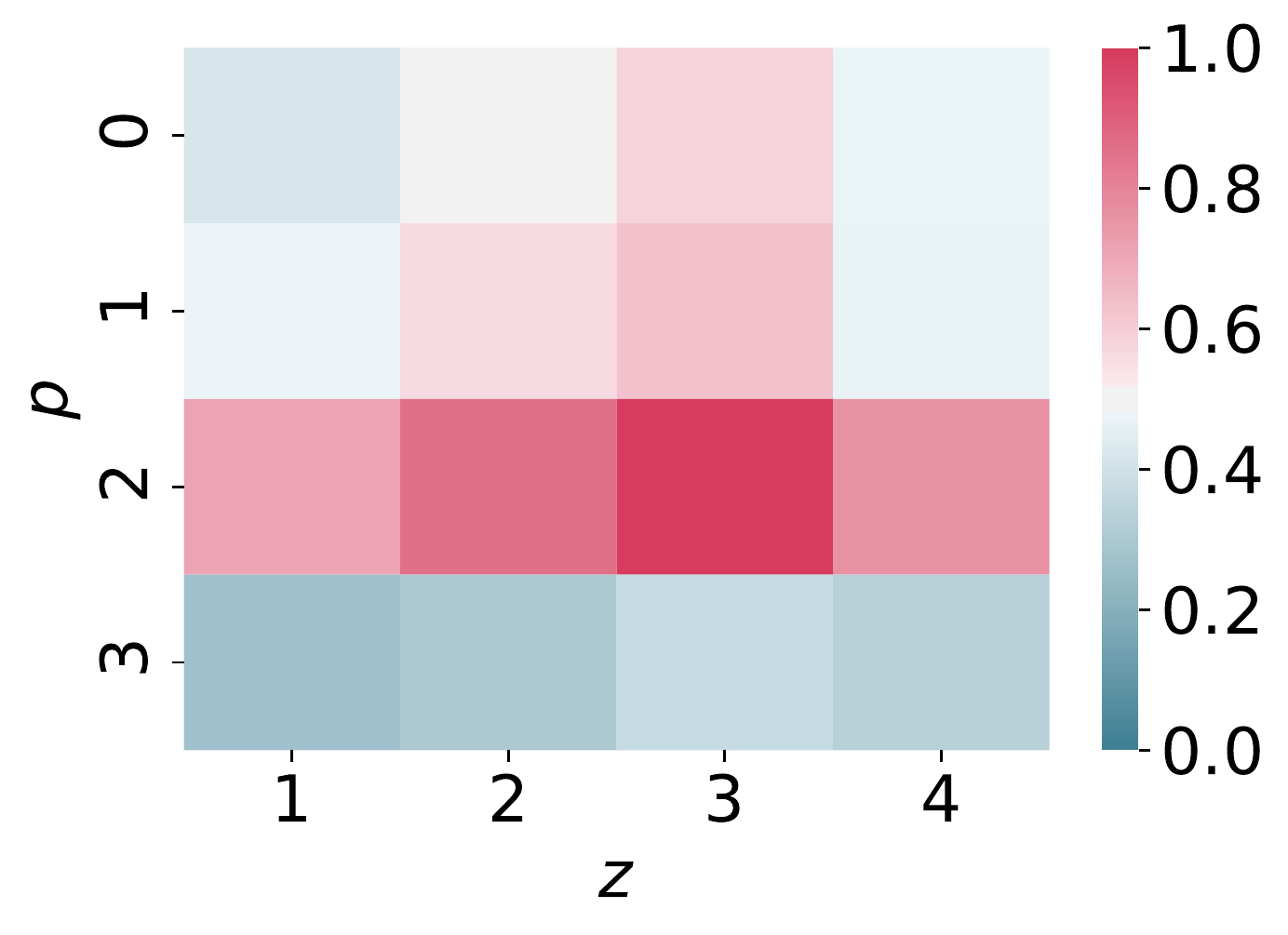}
	\caption{Correlation coefficient between the three-point correlation function at $p_\text{pred}=2$, $z_\text{pred}=3$ and at various choices of $p$ and $z$ calculated using the overlap valence fermions. Different $z$ at the same $p$ cases show higher correlation than different $p$ at the same $z$ cases.}
	\label{fig:p3corr_overlap}
\end{figure}

\begin{figure}[htbp]
	\centering
	\includegraphics[width=0.49\textwidth]{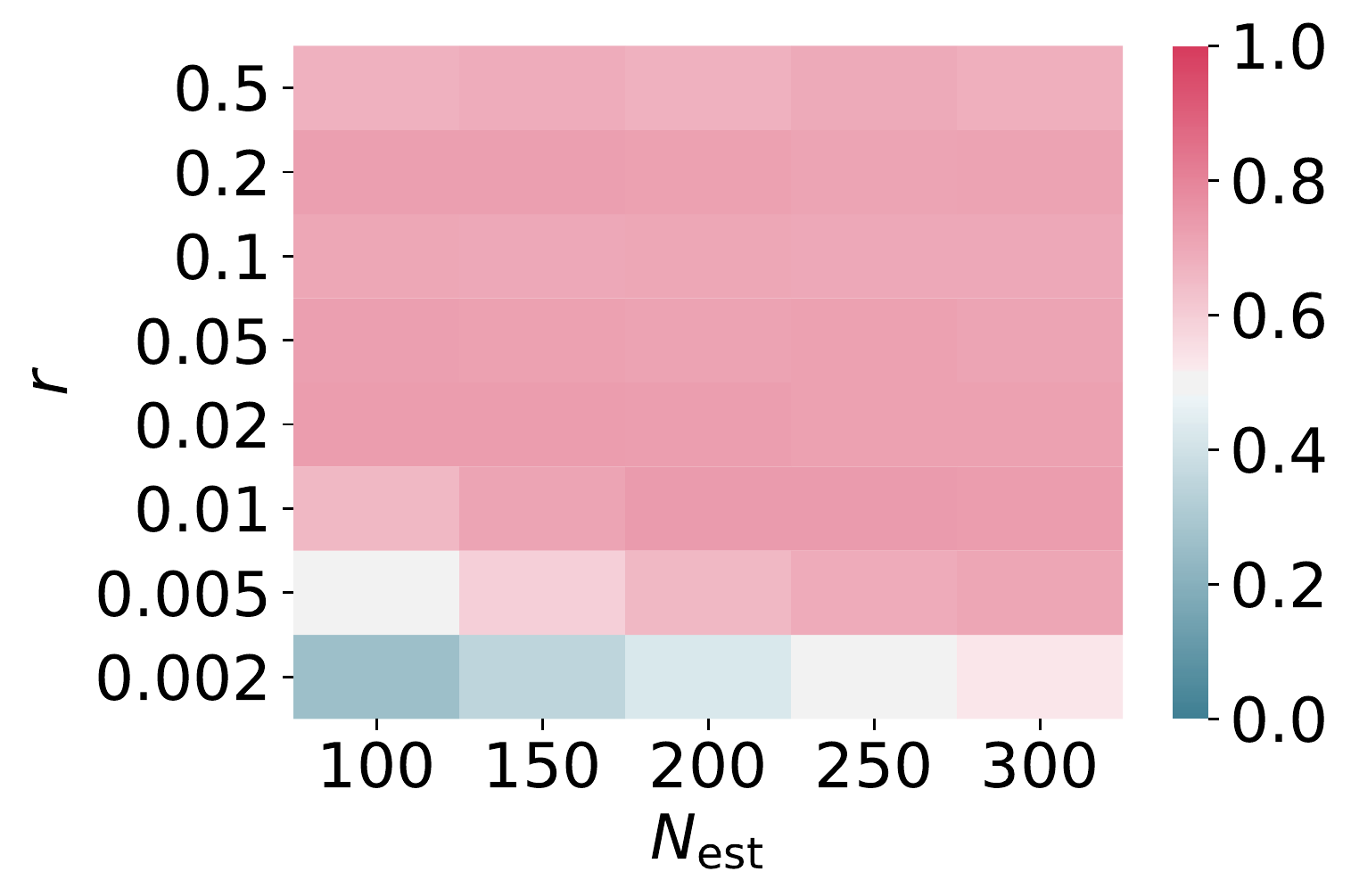}
	\includegraphics[width=0.49\textwidth]{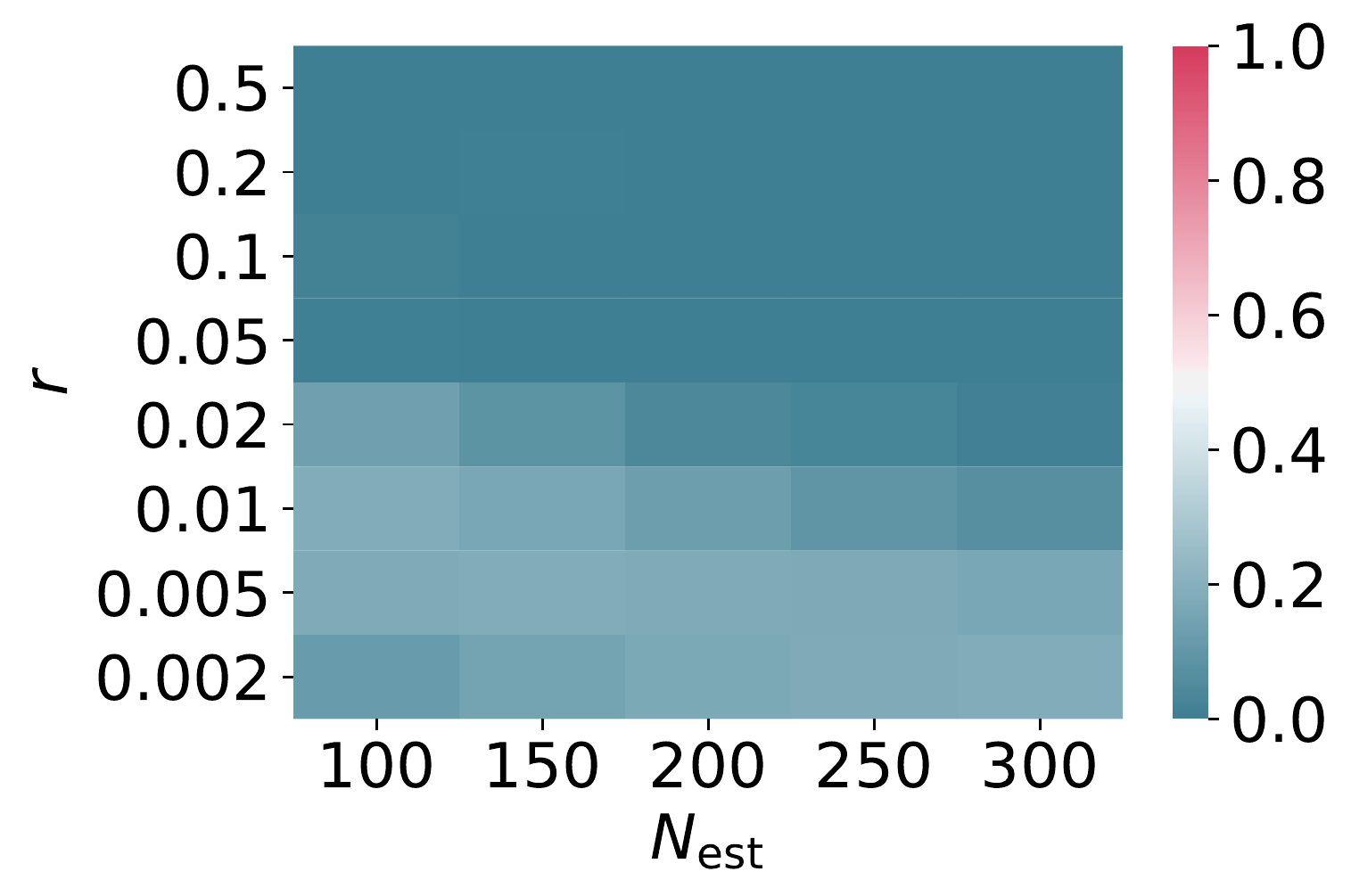}
	\caption{Gluon-correlator ratio fit variance for the $z$-prediction (left) and $p$-prediction (right) for the overlap valence fermions at $p_\text{pred}=5$, $z_\text{pred}=3$, $t_\text{sep}=8$ at $t=4$ from $p_\text{in}=5$, $z_\text{in}=2$, $t_\text{sep}=8$ and $p_\text{in}=4$, $z_\text{in}=3$, $t_\text{sep}=8$ using $N_\text{tr}=61440$, $N_\text{BC}=61440$, and $N_\text{ul}=81920$. Fit variance is closely related to the correlations between the input data and unlabeled data. $z$-prediction works much better than $p$-prediction.}
	\label{fig:gluon-fitvar-overlap}
\end{figure}

To make $z$/$p$-predictions on correlators based on smaller $z$/$p$ values, we should first check the correlations among correlators with different momenta and link lengths. In Fig.~\ref{fig:p3corr_overlap}, we show the correlations between the three-point correlation function at $p_\text{pred}=2$, $z_\text{pred}=3$ and the same three-point correlation functions at various choices of momenta $p_\text{in}=\{0,1,3\}$ and link lengths $z_\text{in}=\{1,2,3,4\}$. The source-sink time separation is fixed to $t_\text{sep}=8$. We notice that the correlations between different momenta are weaker than the correlations between different link lengths in this case, which will result in a relatively low $p$-prediction fit variance as shown in Fig.~\ref{fig:gluon-fitvar-overlap}.

The fit variances $F_v$ from the $p$-prediction and $z$-prediction are shown in Fig.~\ref{fig:gluon-fitvar-overlap} with different learning rates in $\{0.002, 0.005, 0.01, 0.02, 0.05, 0.1, 0.2, 0.5\}$ and different numbers of estimators in $\{100, 150, 200, 250, 300\}$.
The target measurement is with $p_\text{in}= [0, 1]$, $p_\text{pred} = 2$, $z_\text{in}= z_\text{pred} = 3$, $t_\text{sep} = 8$, and $t = 4$.
We used both $C_\text{3pt}$ and $C_\text{2pt}$ for prediction.
Thus, considering the $F_v$ for $z$/$p$-prediction shown in Fig.~\ref{fig:gluon-fitvar-overlap}, we choose $r=0.02$, $N_\text{est}=150$ as the parameter set we will use in further work.

\begin{figure}[htbp]
	\centering
	\includegraphics[width=0.49\textwidth]{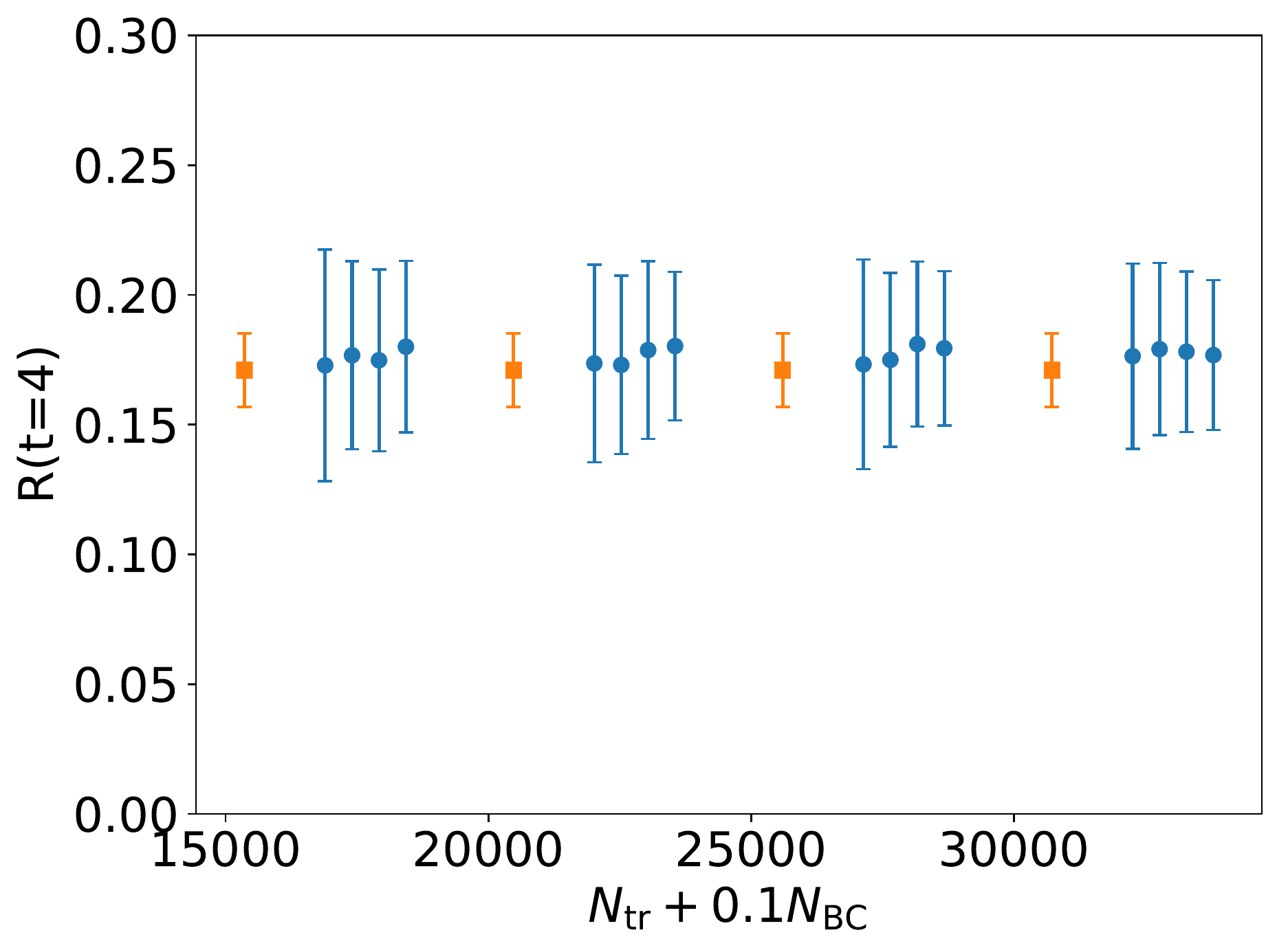}
	\includegraphics[width=0.49\textwidth]{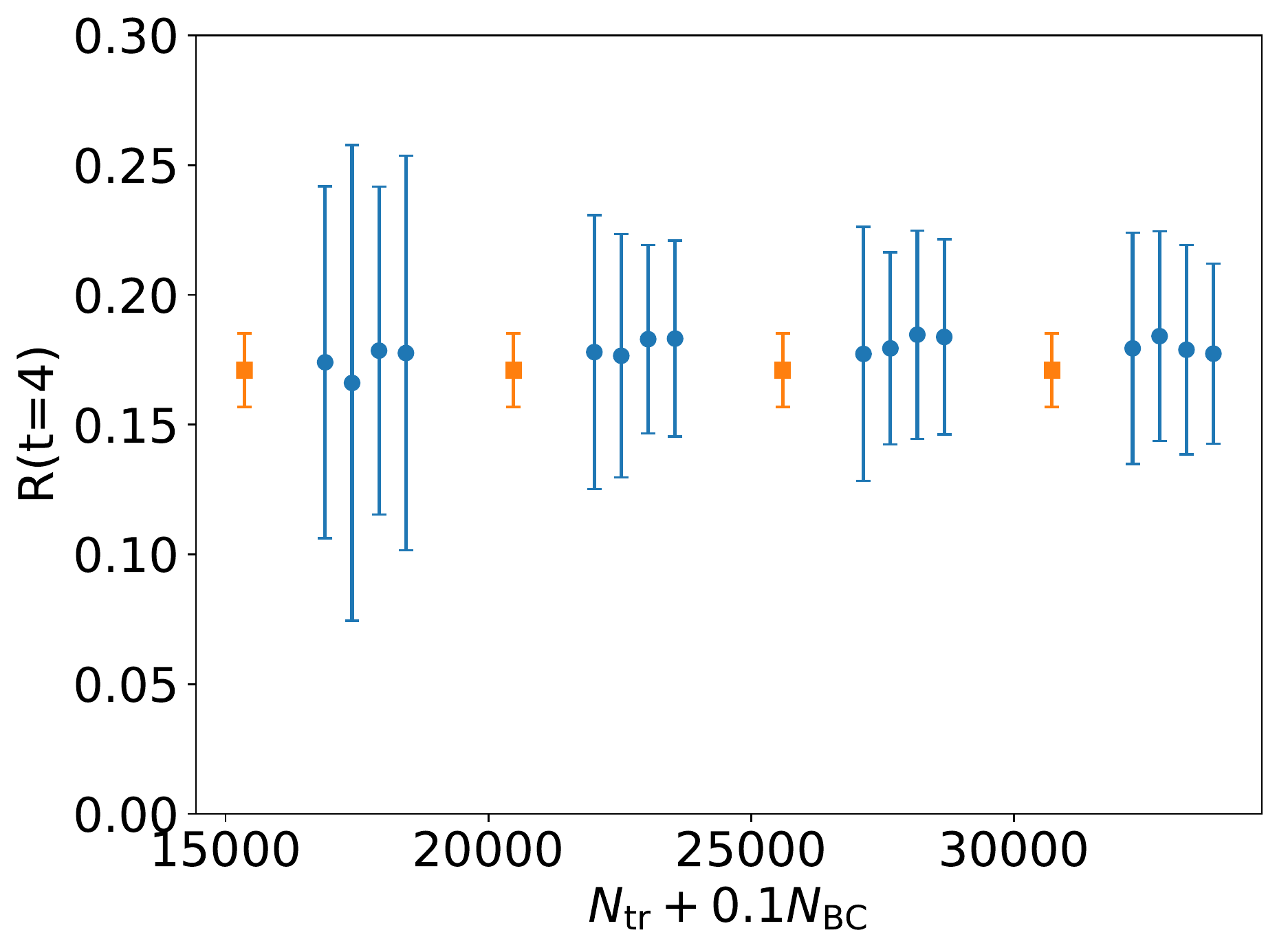}
	\includegraphics[width=0.49\textwidth]{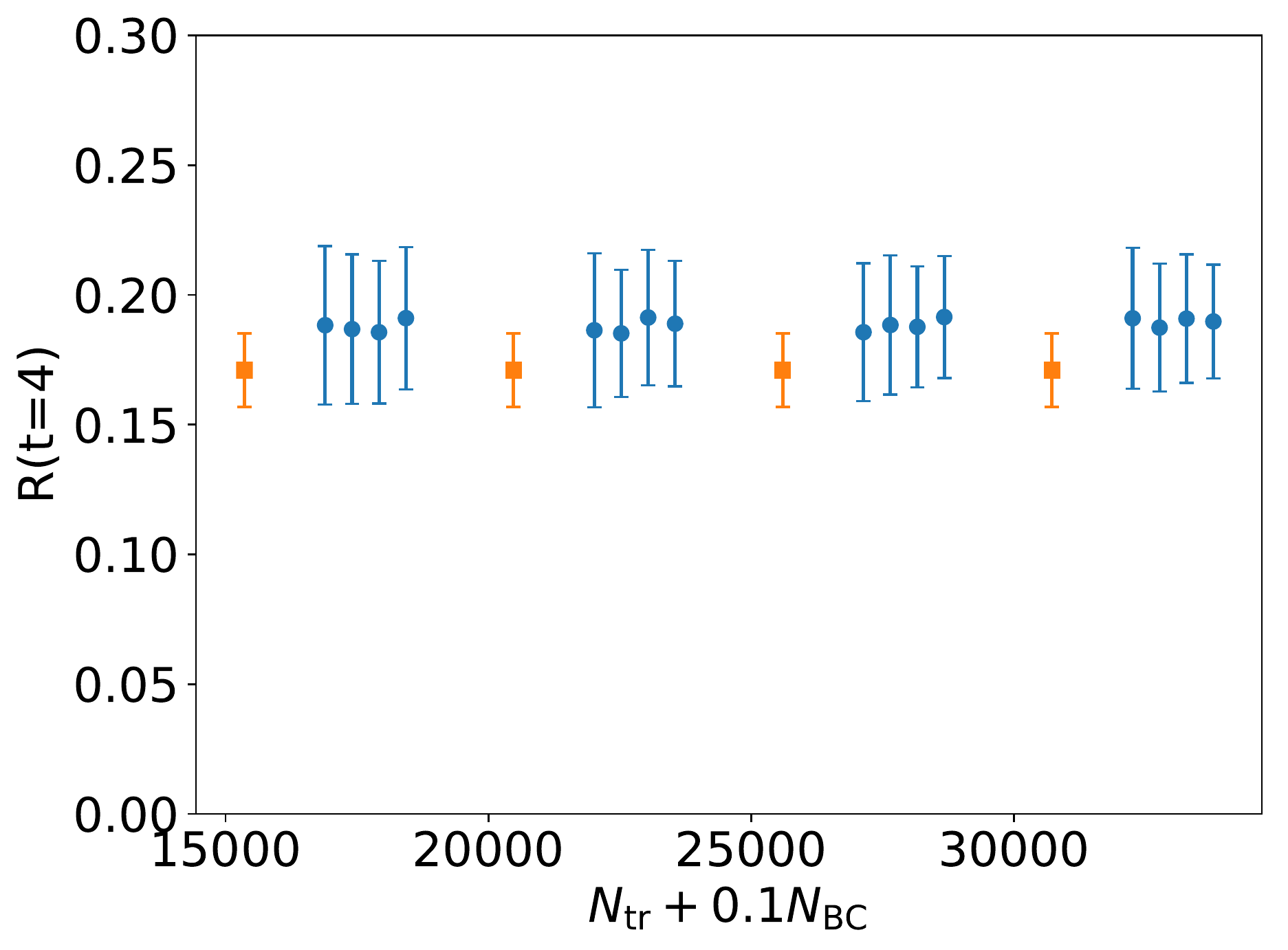}
	\includegraphics[width=0.49\textwidth]{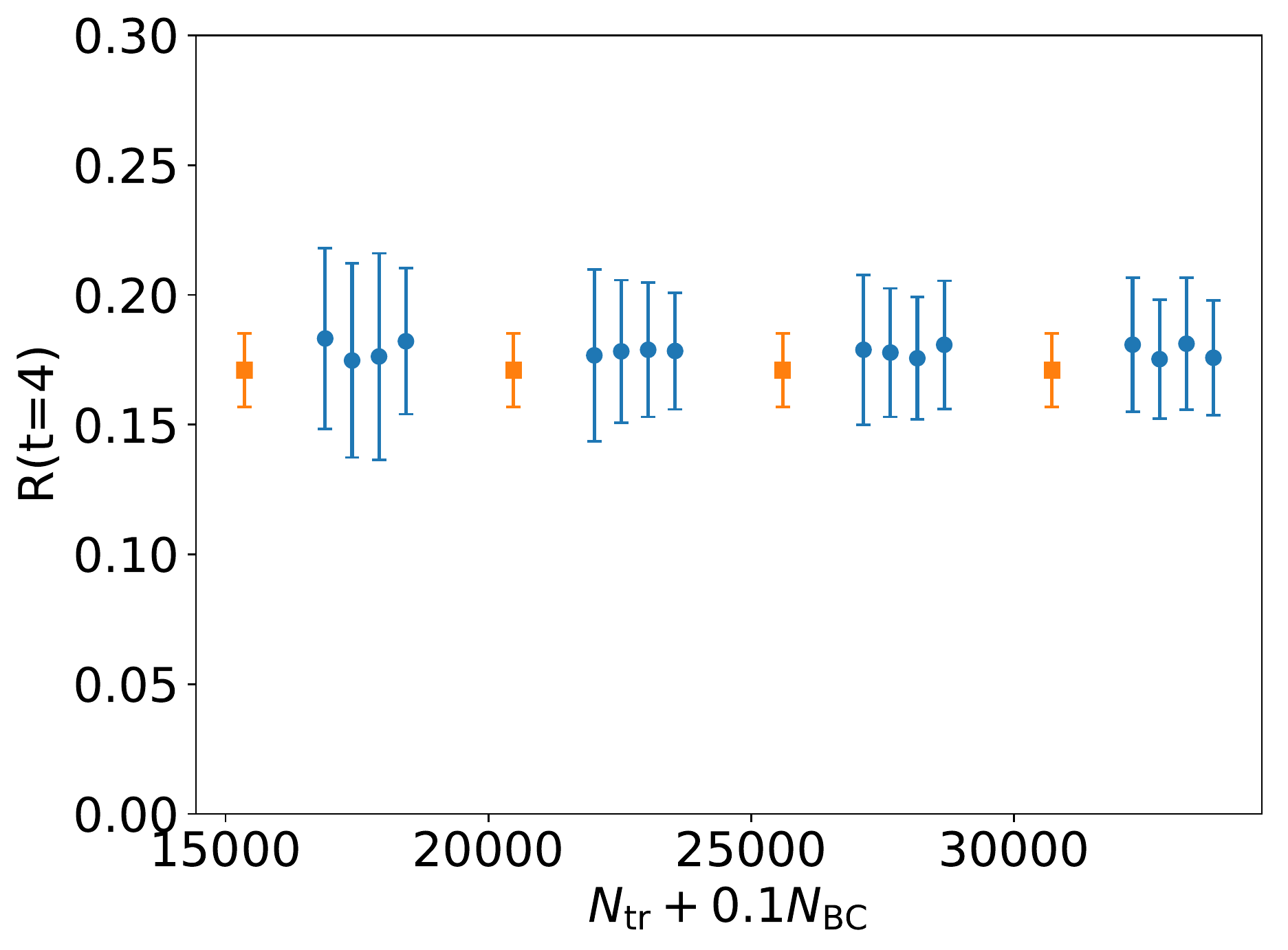}
	\caption{The GBT (left) and linear-regressor (right) results. The observed $p$- and $z$-predicted gluon-correlator ratios are for the overlap valence fermions at $p_\text{pred}=2$, $z_\text{pred}=3$ at $t_\text{sep}=8$ by using $r=0.02$ and $N_\text{est}=150$ for different counts of training data and bias -correction data. The horizontal axis is $N_\text{tr}+0.1N_\text{BC}$, with $N_\text{ul}=143360$ fixed. The blue points are predictions with bias correction for the unlabeled test data, and the orange points are observations for unlabeled test data.}
	\label{fig:gluon_overlap}
\end{figure}

For $p$-prediction, we varied the number of training data and bias-correction data from 15360 to 30720, while keeping the number of unlabeled test data $N_\text{ul}=143360$ fixed, to compare their performance. The results are shown in Fig.~\ref{fig:gluon_overlap}. We will use $N_\text{tr}=30720$, $N_\text{BC}=30720$ in the following $p$/$z$-prediction.

\begin{figure}[htbp]
	\centering
	\includegraphics[width=0.49\textwidth]{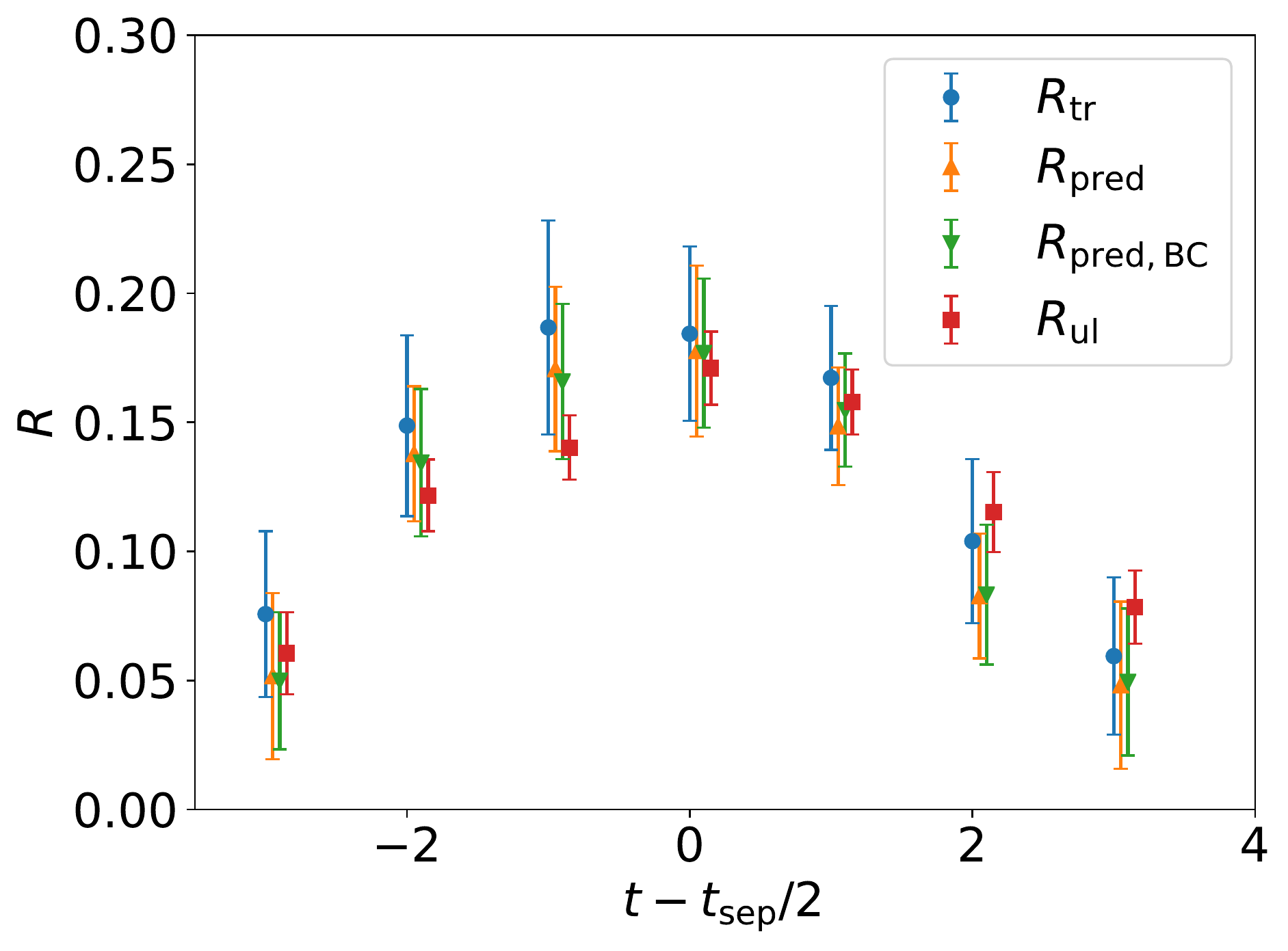}
	\includegraphics[width=0.49\textwidth]{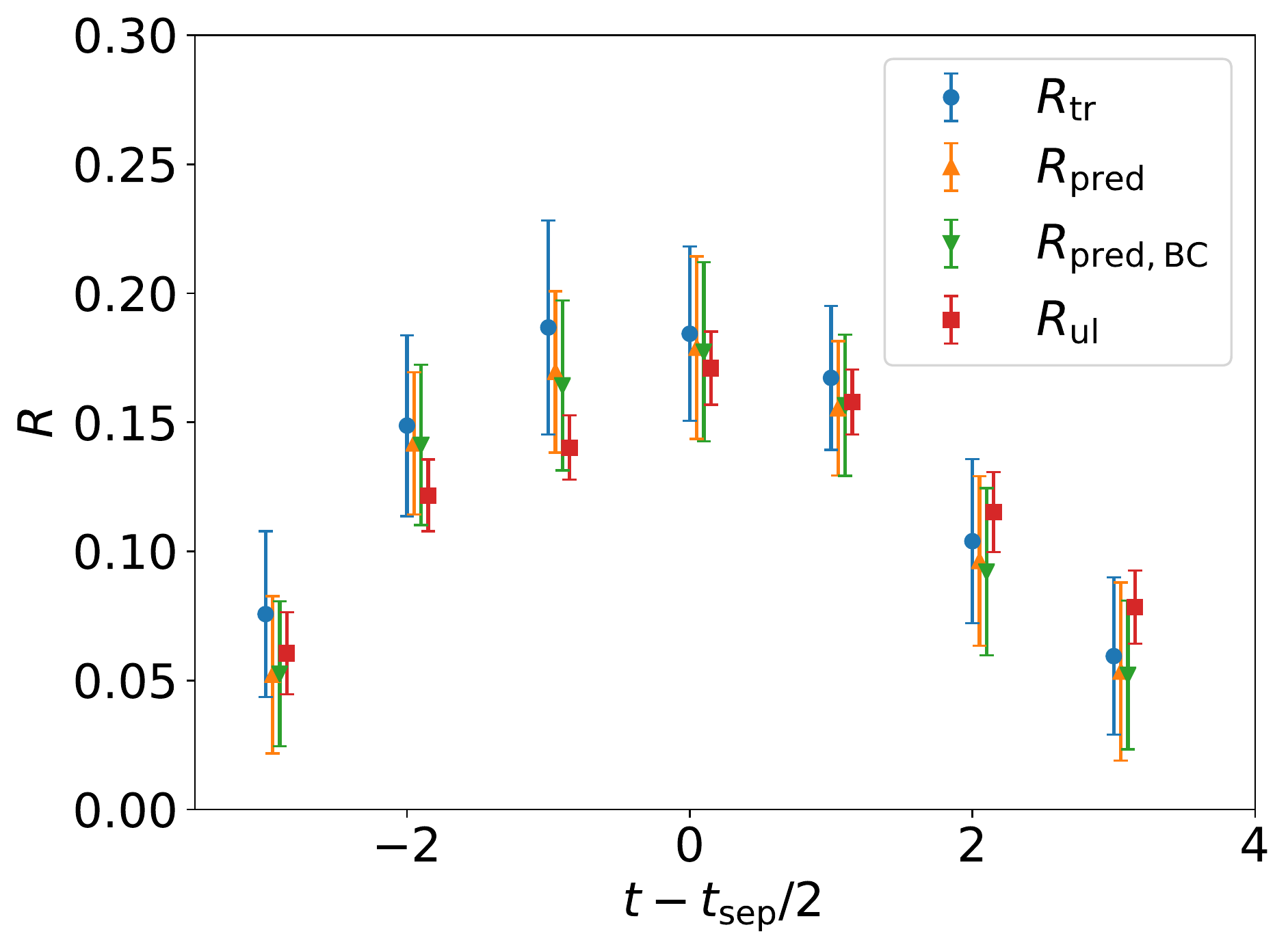}
	\includegraphics[width=0.49\textwidth]{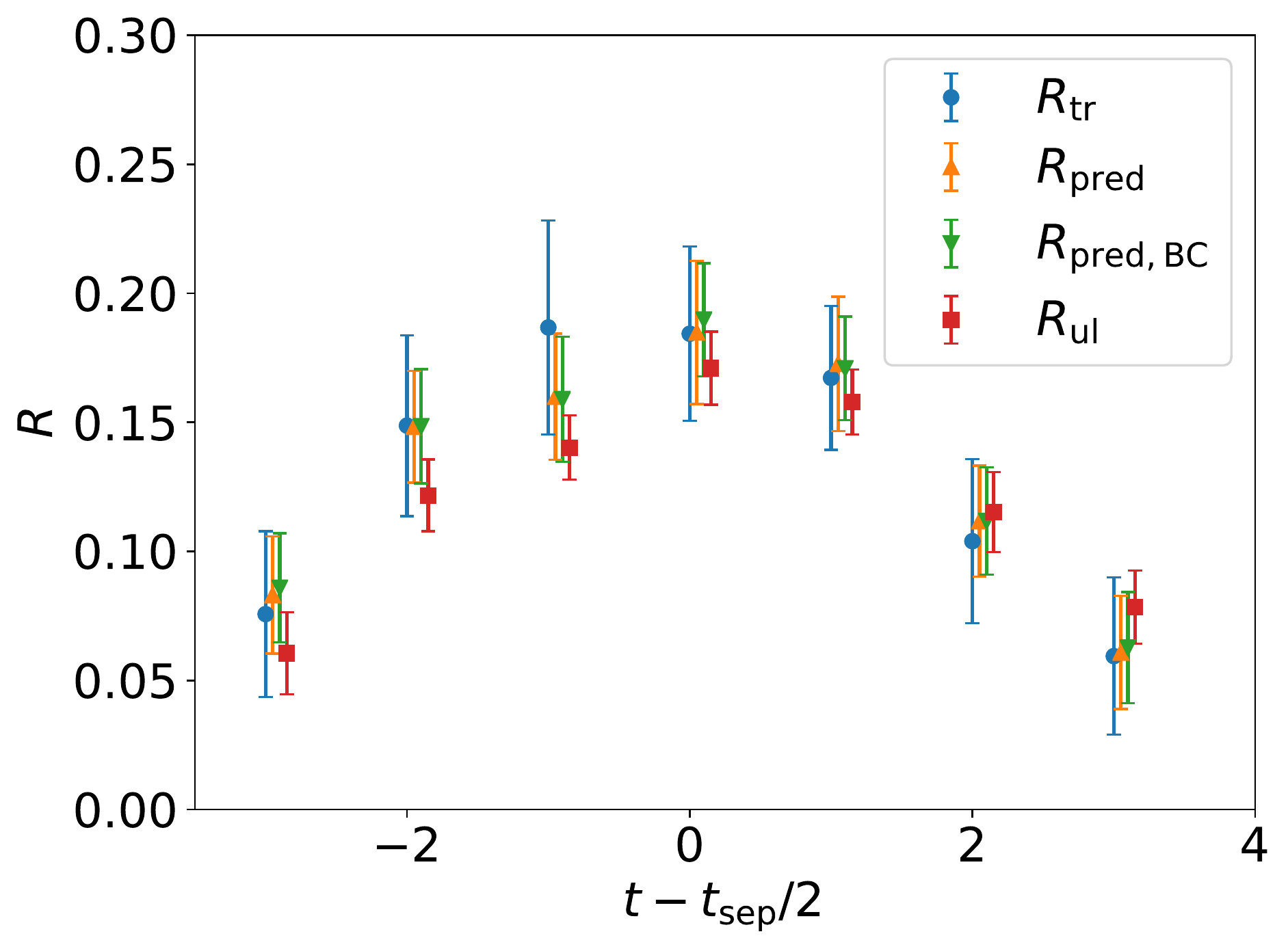}
	\includegraphics[width=0.49\textwidth]{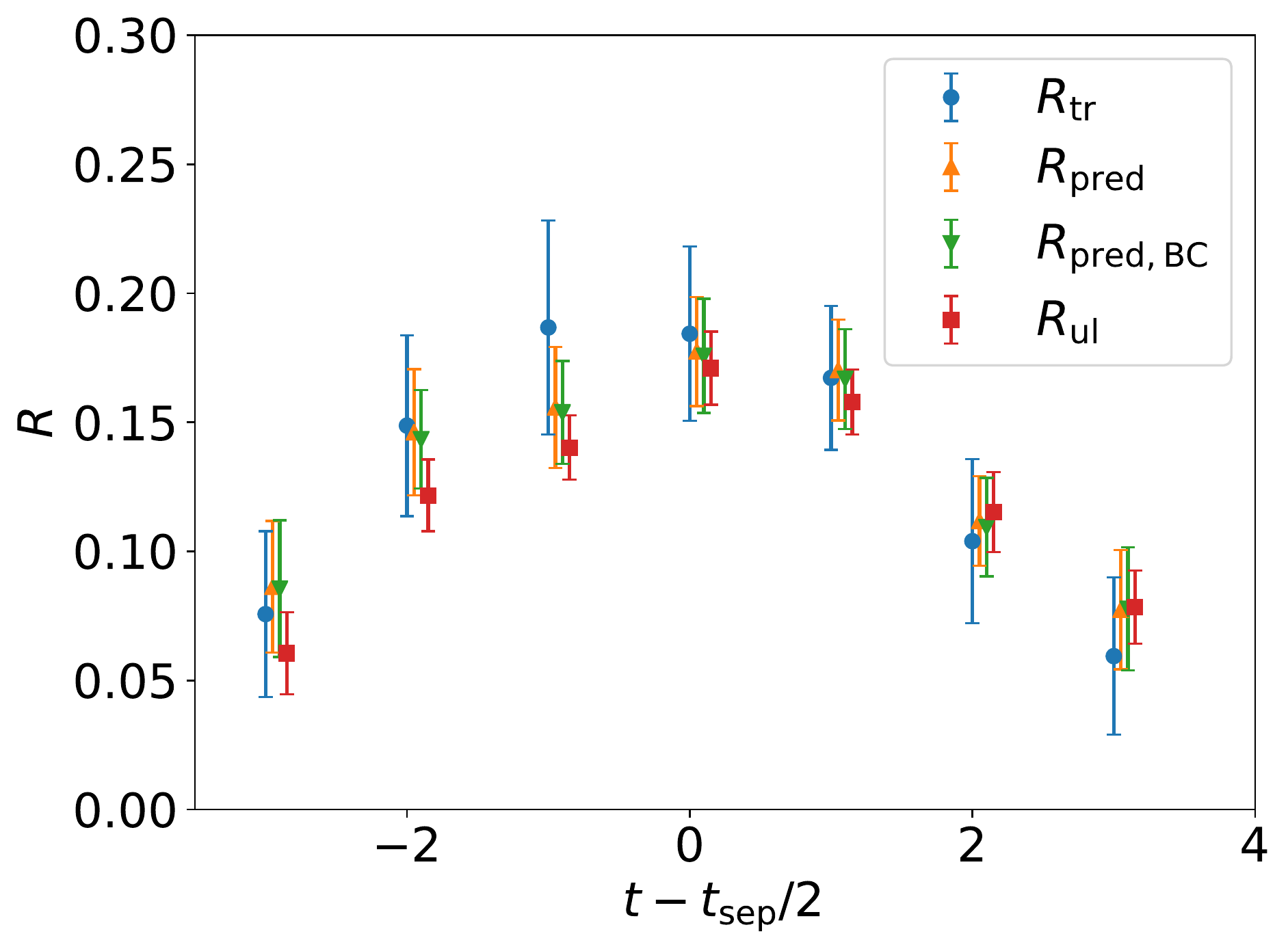}
	\caption{The observed/predicted gluon correlator $C_\text{3pt}$ and $C_\text{2pt}$ ratio of the overlap valence fermions lattice ensemble at $p_\text{pred}=2$, $z_\text{pred}=3$ from $p_\text{in}=1$, $z_\text{in}=3$ (upper) and $p_\text{pred}=2$, $z_\text{pred}=3$ from $p_\text{in}=2$, $z_\text{in}=2$ (lower) by using $N_\text{tr}=30720$, $N_\text{BC}=30720$, $N_\text{ul}=1433600$, $r=0.02$, and $N_\text{est}=150$. The GBT and linear regressor results are shown on the left and right, respectively. The predictions with bias correction do not improve much over the raw predictions.}
	\label{fig:gluon_pred_overlap}
\end{figure}

With the ML model parameters and the dataset we obtained from the overlap-fermion ensembles, we show the result of our prediction along with the observed datasets for both $p_\text{pred}$ and $z_\text{pred}$ predictions in Fig.~\ref{fig:gluon_pred_overlap}. In the prediction, we can use any $p_\text{in} < p_\text{pred}$ or $z_\text{in} < z_\text{pred}$ for prediction. In Table~\ref{table-overlap}, two-point and three-point correlator data at $p_\text{in}=1$, $z_\text{in}=3$, $t_\text{sep}=8$ or $p_\text{in}=2$, $z_\text{in}=2$, $t_\text{sep}=8$ are used for predicting the $p_\text{pred}= 2$, $z_\text{pred}=3$, $t_\text{sep}=8$ ratio. The data for insertion time $t=4$ are shown. From the table we can see that the $p$-predictions are bad for both models, because the correlations are weak, as shown in Fig.~\ref{fig:gluon_pred_overlap}. The $z$-predictions are better than $p$-predictions, and the linear model performs better than GBT.

\begin{table}[h]
	\centering
	\begin{tabular}{@{} |c|c|c|c|c|c|c|c| @{}}
		\hline
		Type& Input &Method   & $R_\text{tr}$ & $R_\text{pred}$ & $R_{\text{pred},\text{BC}}$ & $R_\text{ul}$  &$F_v$ \\
		\hline
		\multirow{2}{*}{$p$-pred}& \multirow{2}{*}{$p_\text{in}=1,z_\text{in}=3$}
		&GBT & 0.184(34) & 0.178(33) & 0.177(29) & 0.171(14) & 0.07(18)\\
		\cline{3-8}
		& &linear & 0.184(34) & 0.179(35) & 0.177(35) & 0.171(14) & -0.05(38)\\
		\hline
		\multirow{2}{*}{$z$-pred}& \multirow{2}{*}{$p_\text{in}=2,z_\text{in}=2$}
		&GBT & 0.184(34) & 0.185(28) & 0.189(22) & 0.171(14) & 0.53(12)\\
		\cline{3-8}
		& &linear & 0.184(34) & 0.177(21) & 0.176(22) & 0.171(14) & 0.665(79)\\
		\hline
	\end{tabular}
	\caption{Observations and predictions of gluon-correlator ratios for the overlap valence fermions observations and predictions at $p_\text{pred}=2$, $z_\text{pred}=3$, $t_\text{sep}=8$, $t=4$ by using $N_\text{tr}=30720$, $N_\text{BC}=30720$, $N_\text{ul}=1433600$, $r=0.02$, and $N_\text{est}=150$. For the $z$-predictions, the linear model shows a better fit variance than GBT. The $p$-predictions are bad for both models, because the correlations are poor, as shown in Fig.~\ref{fig:gluon_pred_overlap}.}
	\label{table-overlap}
\end{table}

\subsubsection{Predictions of the gluon correlators for clover valence fermions}

\begin{figure}[htb]
	\centering
	\includegraphics[width=0.49\textwidth]{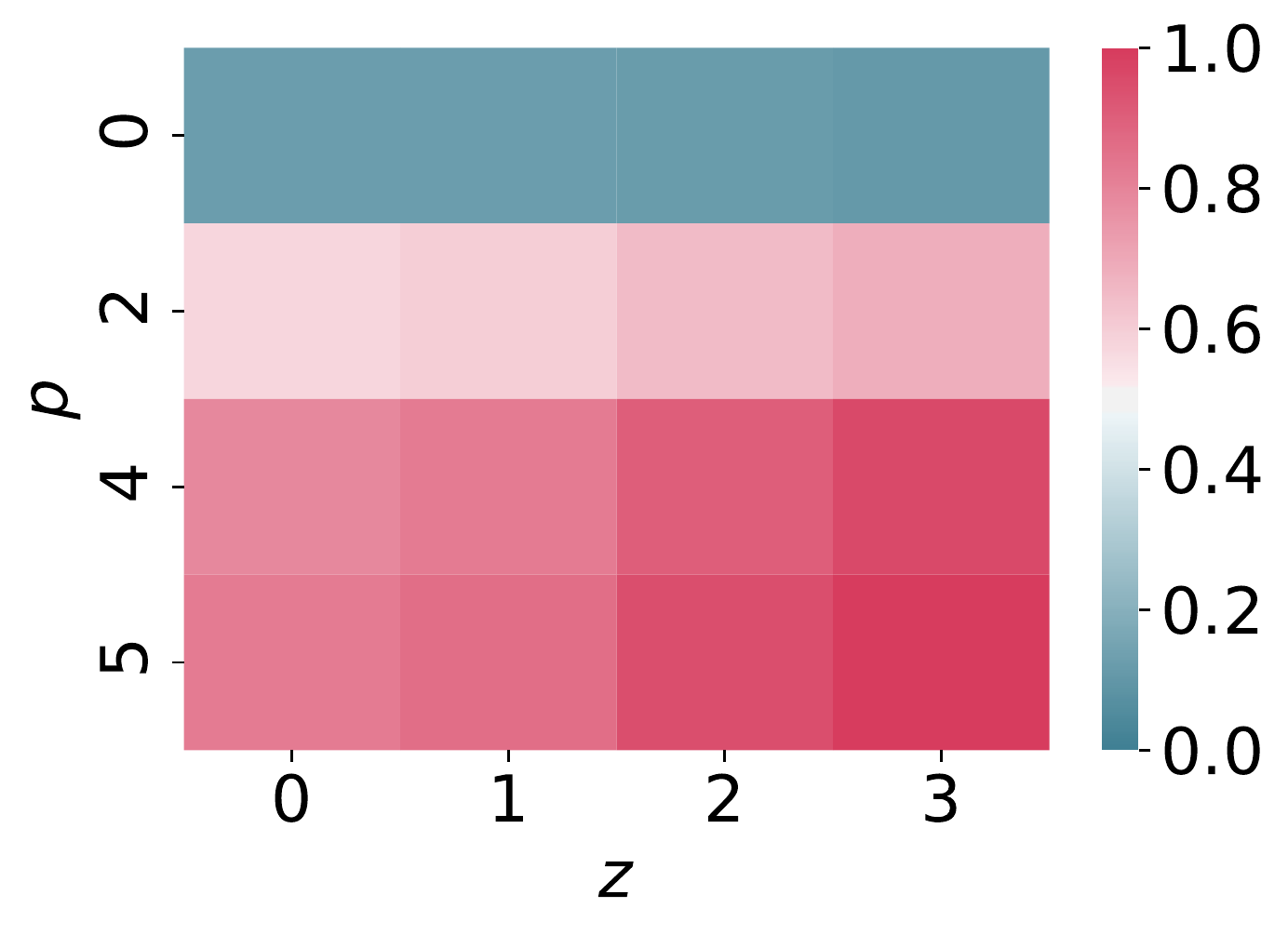}
	\caption{Correlation coefficients between the three-point correlation functions at $p_\text{pred}=5$, $z_\text{pred}=3$ at various values of $p_\text{in}=\{0,2,4\}$, $z_\text{in}=\{0,1,2,3\}$ calculated using the clover valence fermions. Different $z$ at the same $p$ cases show higher correlation than different $p$ at the same $z$ cases.}
	\label{fig:p3corr_clover}
\end{figure}

\begin{figure}[htbp]
	\centering
	\includegraphics[width=0.49\textwidth]{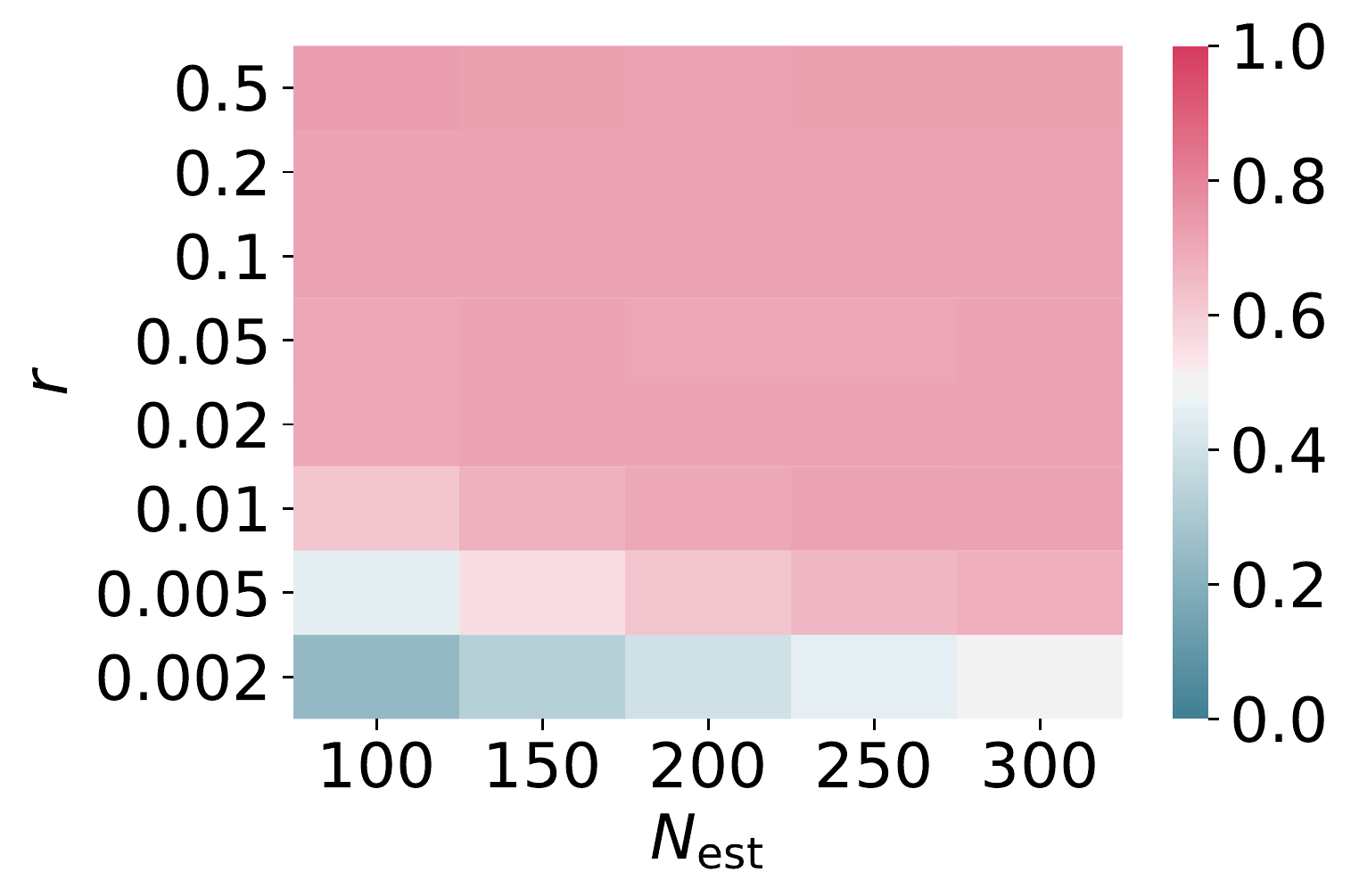}
	\includegraphics[width=0.49\textwidth]{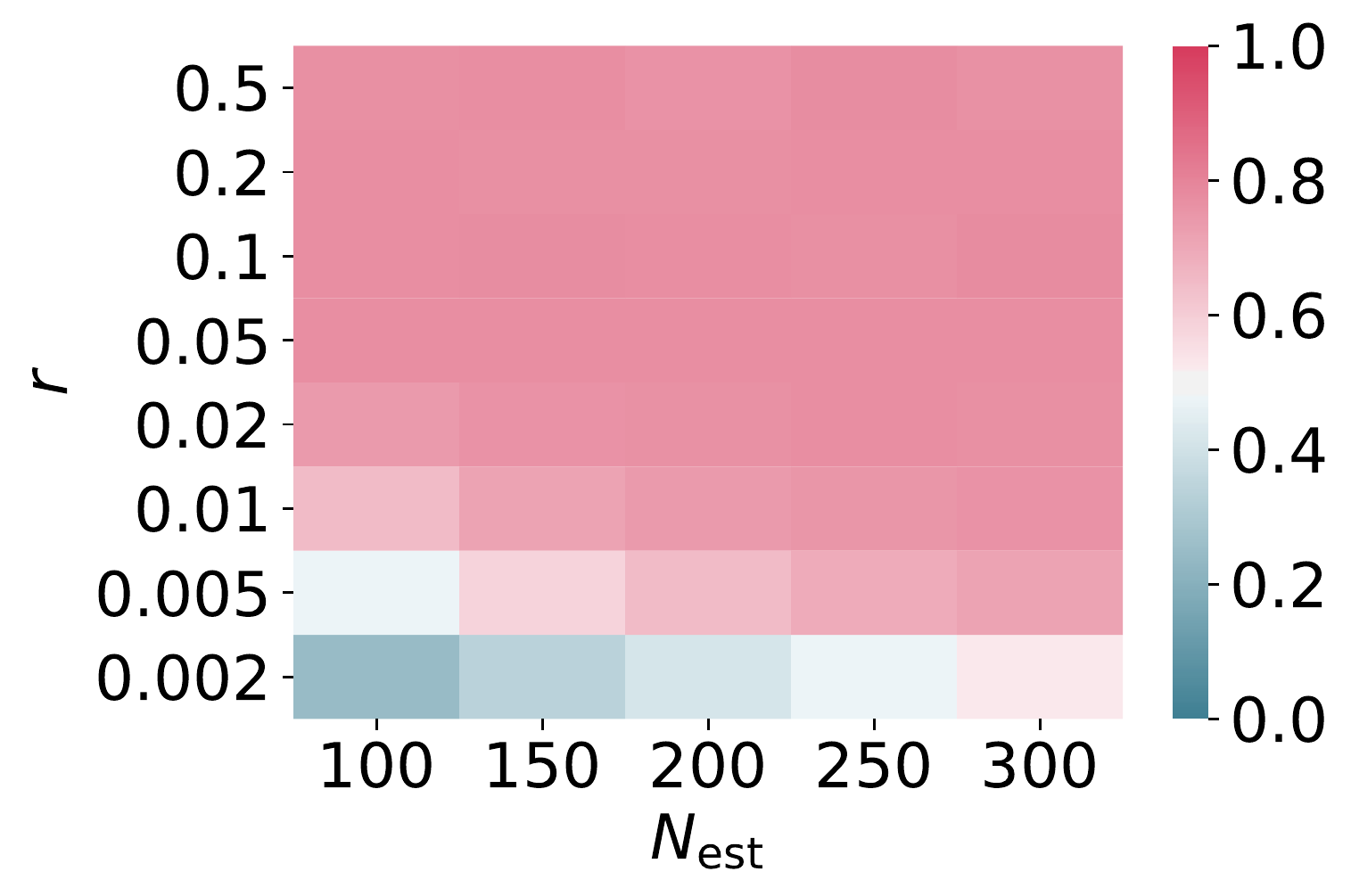}
	\caption{Gluon-correlator ratio fit variance for the $z$-prediction (left) and $p$-prediction (right) for the clover valence fermions at $p_\text{pred}=5$, $z_\text{pred}=3$, $t_\text{sep}=8$ at $t=4$ from $p_\text{in}=5$, $z_\text{in}=2$, $t_\text{sep}=8$ and $p_\text{in}=4$, $z_\text{in}=3$, $t_\text{sep}=8$ using $N_\text{tr}=2880$, $N_\text{BC}=2880$, and $N_\text{ul}=23040$. With a stronger correlation between input and target data, smaller learning rate and number of estimators are needed to have good fit variance score.}
	\label{fig:gluon-fitvar-clover}
\end{figure}

We repeat the procedure we established from the overlap valence fermions for the clover fermions, checking the correlations among correlators with different momenta and link lengths. In Fig.~\ref{fig:p3corr_clover}, we show the correlations between the three-point correlation functions at $p_\text{pred}=5$, $z_\text{pred}=3$ at various values of $p_\text{in}=\{0,2,4\}$, $z_\text{in}=\{0,1,2,3\}$. The source-sink time separation is fixed $t_\text{sep}=8$. The correlations between different momenta are much stronger than in the overlap case, which leads to a much higher $p$-prediction fit variance, as shown in Fig.~\ref{fig:gluon-fitvar-clover}. The reason that the correlations of clover fermion case are stronger than overlap fermion case is the construction of the sources of proton correlator are different in two cases. In overlap fermion, we use grid spatial source which needs gauge-averaging to get consistent correlators that dues to weak correlation properties. While the clover fermion does't have this kind of problem because of using one spatial location per time source.

We use the same fit-variance $F_v$ estimation as in the overlap case. The target measurement is $p_\text{in}= 4$, $p_\text{pred} = 5$, $z_\text{in}= 2$, $z_\text{pred} = 3$, $t_\text{sep} = 8$, and $t = 4$. We obtain $r=0.2$, $N_\text{est}=200$ as the parameters we will use in the following process from Fig.~\ref{fig:gluon-fitvar-clover}. These two figures indicate stronger correlations between input and target data are needed to obtain good results for the fit variance.

\begin{figure}[htbp]
	\centering
	\includegraphics[width=0.49\textwidth]{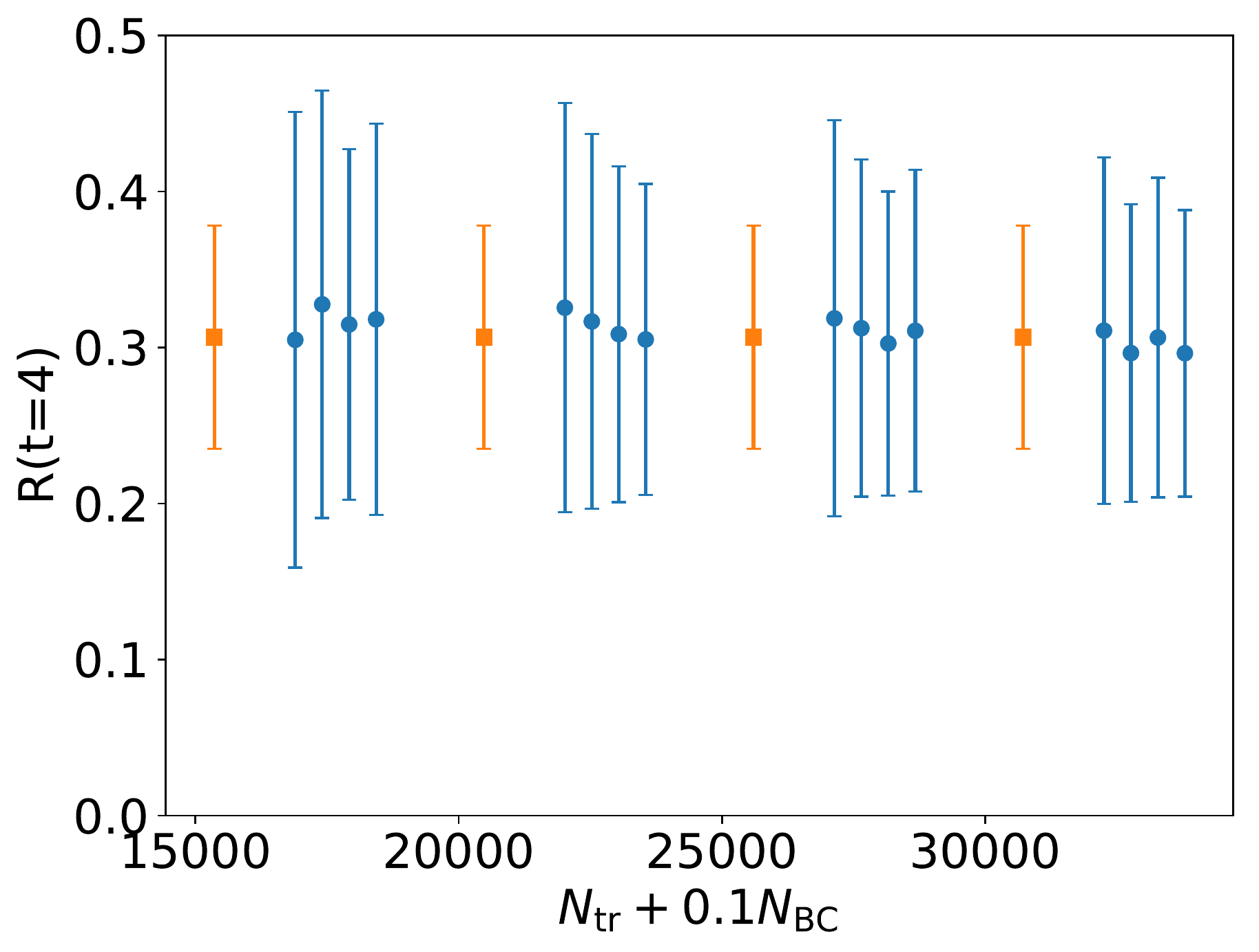}
	\includegraphics[width=0.49\textwidth]{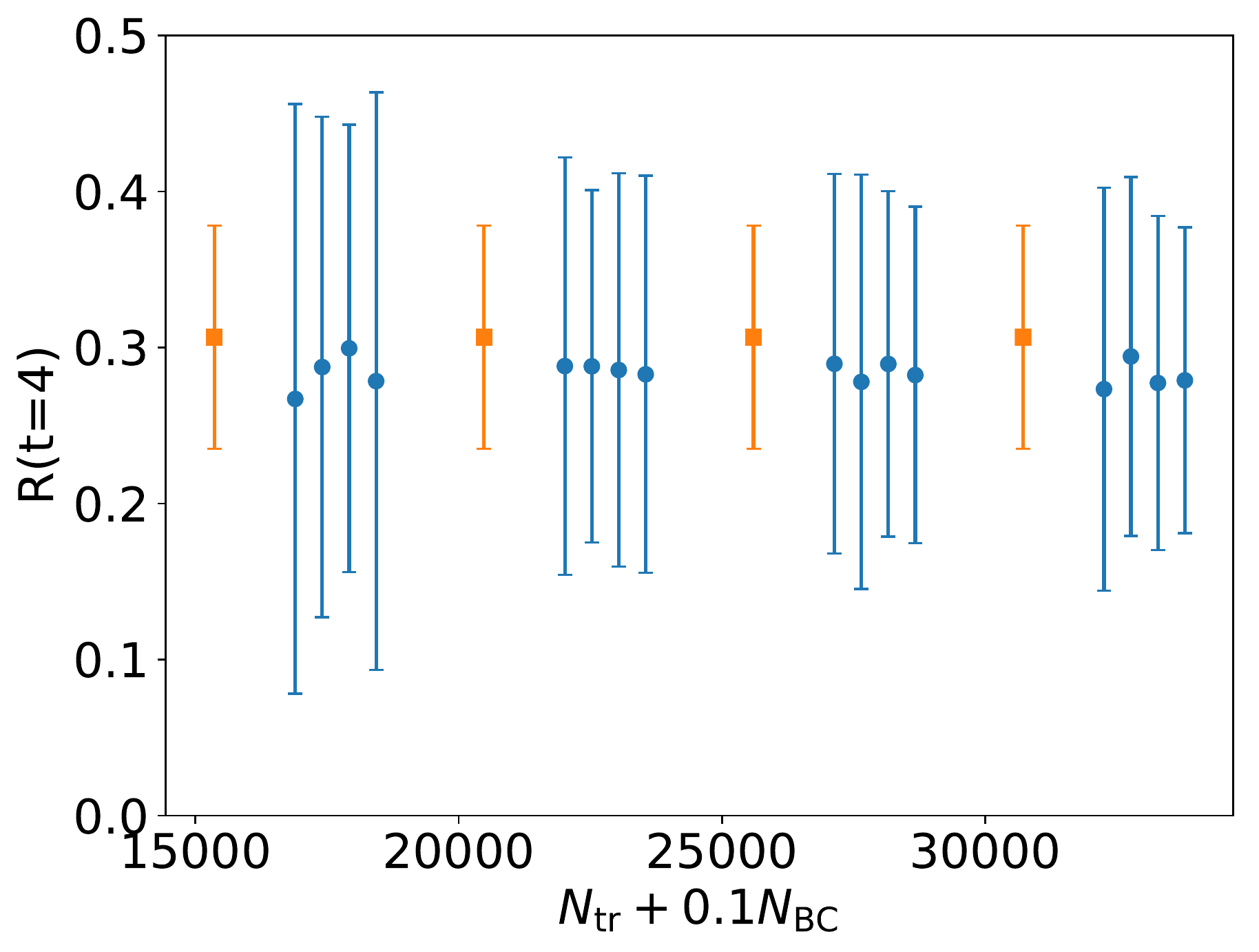}
	\includegraphics[width=0.49\textwidth]{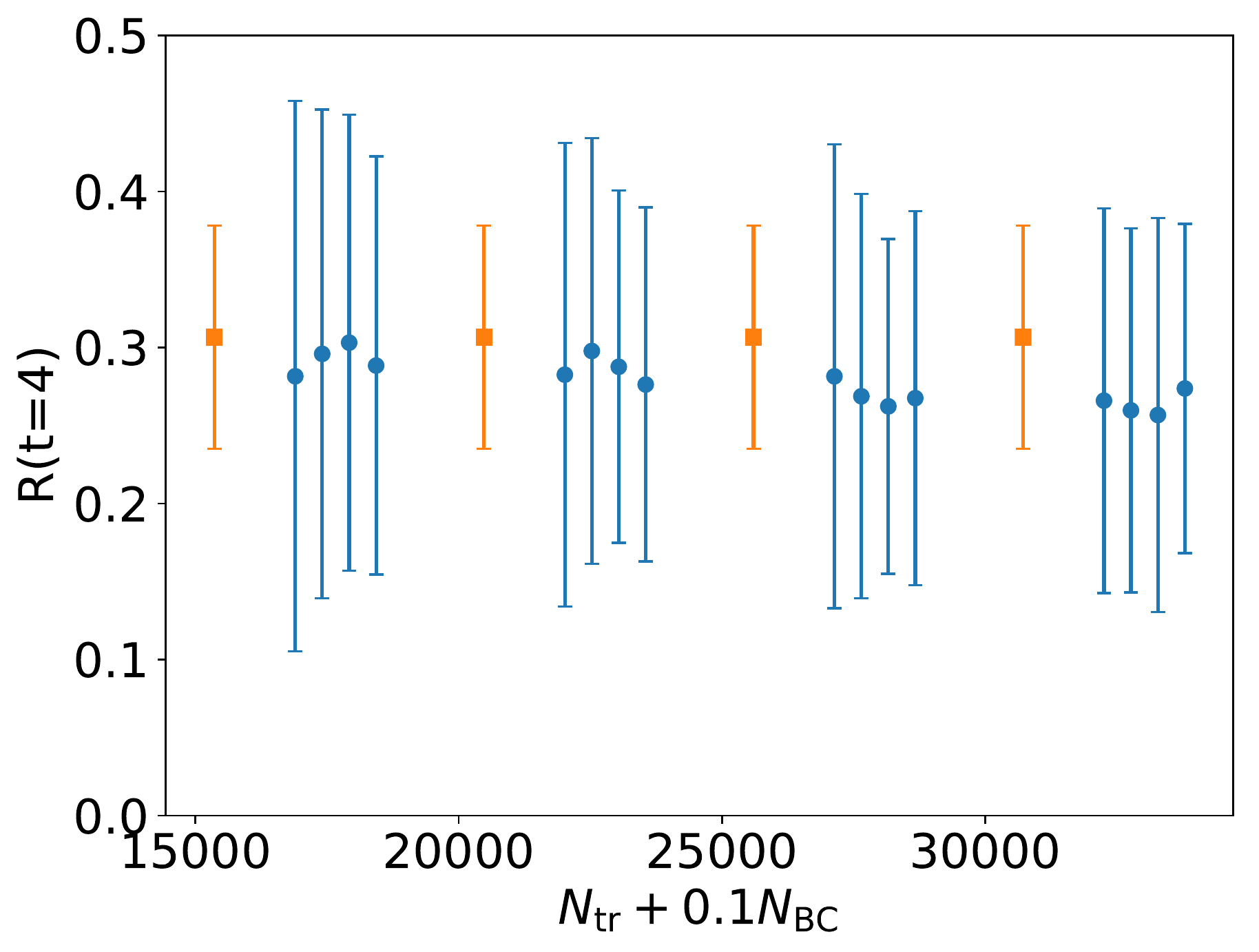}
	\includegraphics[width=0.49\textwidth]{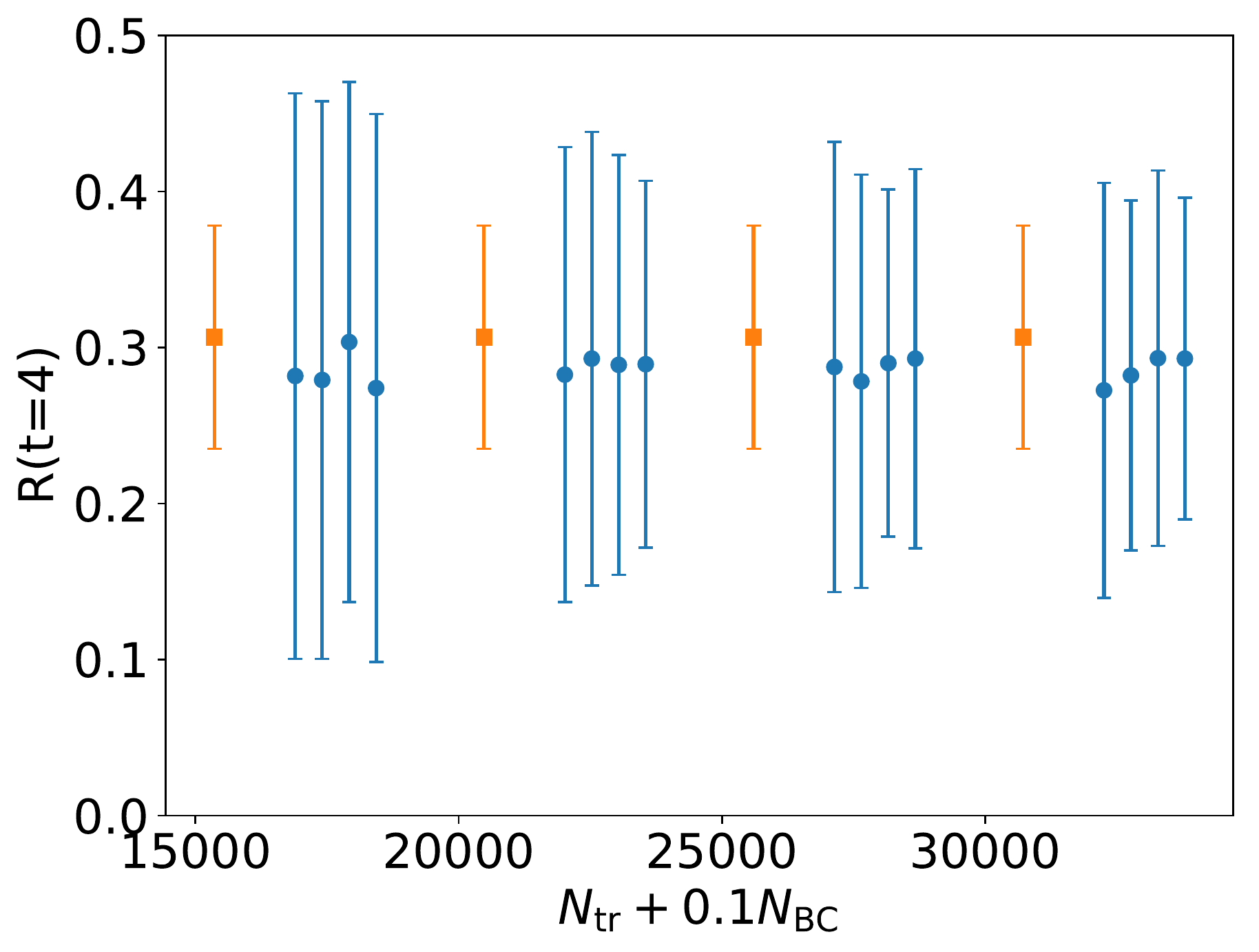}
	\caption{The observed/predicted gluon correlator $C_\text{3pt}$ and $C_\text{2pt}$ ratio of the clover valence fermions $p_\text{pred}=5$, $z_\text{pred}=3$ at $t_\text{sep}=8$ by using $r=0.2$ and $N_\text{est}=200$ for different counts of training data and bias-correction data. The horizontal axis is $N_\text{tr}+0.1N_\text{BC}$, with $N_\text{ul}=23040$ fixed. The GBT and linear-regressor results are shown on the left and right, respectively. The blue points are predictions with bias correction for the unlabeled test data, and the orange points are observations for unlabeled test data.}
	\label{fig:gluon_clover}
\end{figure}

Again, to compare their performance we varied the number of training data and bias correction data from 1440 to 2880, while keeping the number of unlabeled test data $N_\text{ul}=23040$ fixed. The observed, $p$- and $z$-predicted gluon correlator $C_\text{3pt}$ and $C_\text{2pt}$ ratio of the clover valence fermions $p_\text{pred}=5, z_\text{pred}=3$ at $t_\text{sep}=8$ are shown in Fig.~\ref{fig:gluon_clover}. Comparing with these results, we will use $N_\text{tr}=2880$, $N_\text{BC}=2880$ in the following $p$- and $z$-predictions.

\begin{figure}[htbp]
	\centering
	\includegraphics[width=0.49\textwidth]{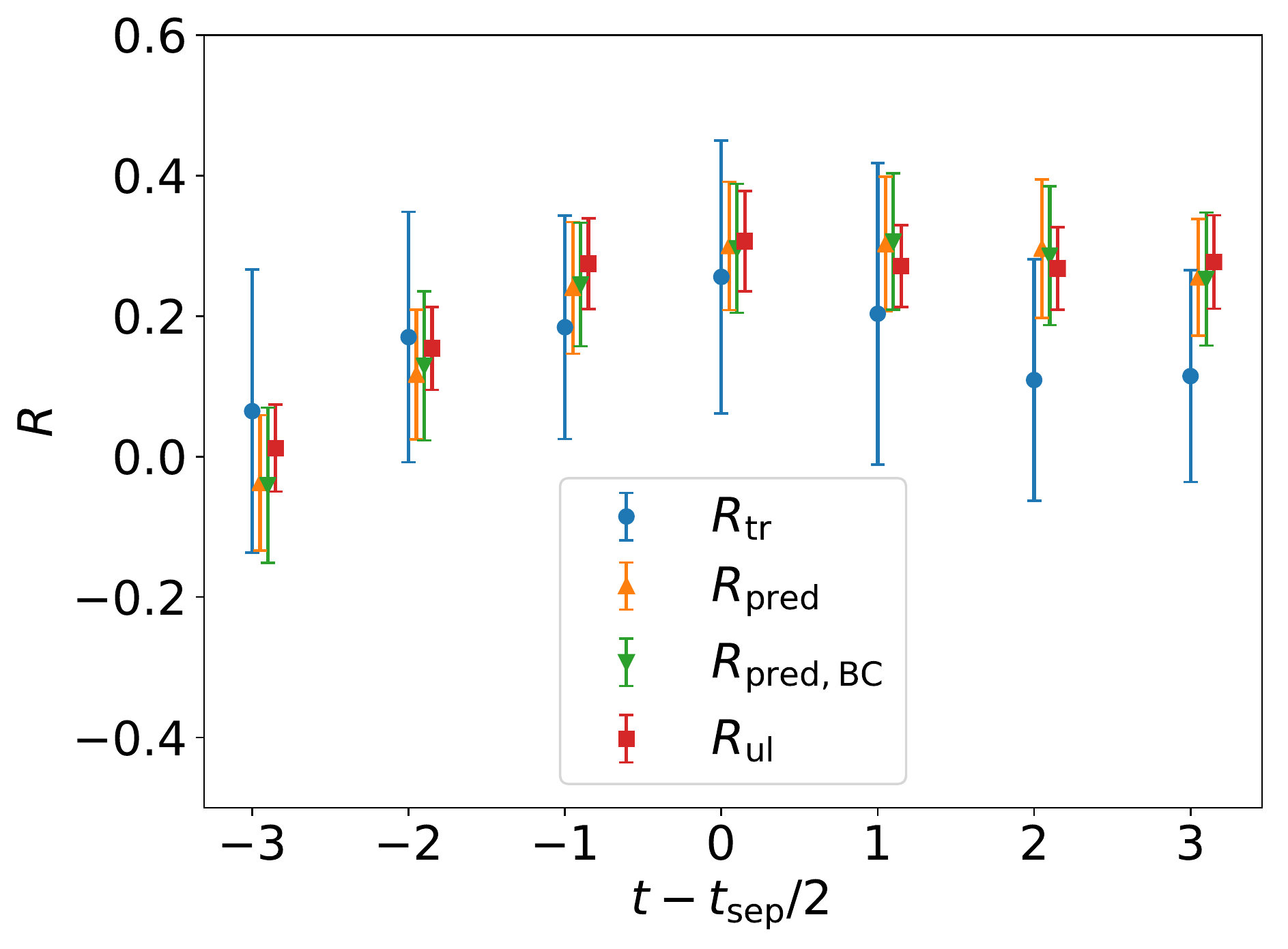}
	\includegraphics[width=0.49\textwidth]{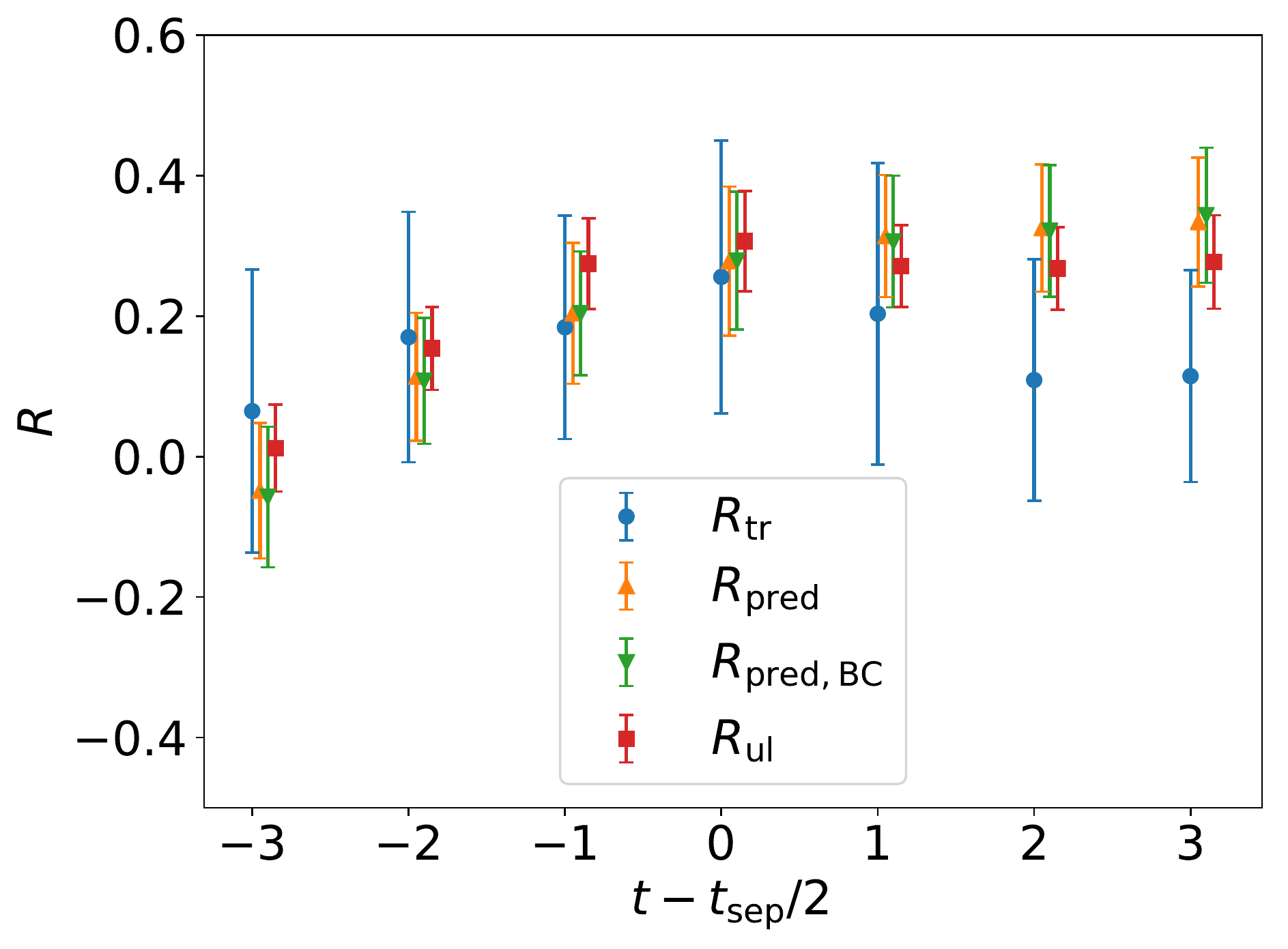}
	\includegraphics[width=0.49\textwidth]{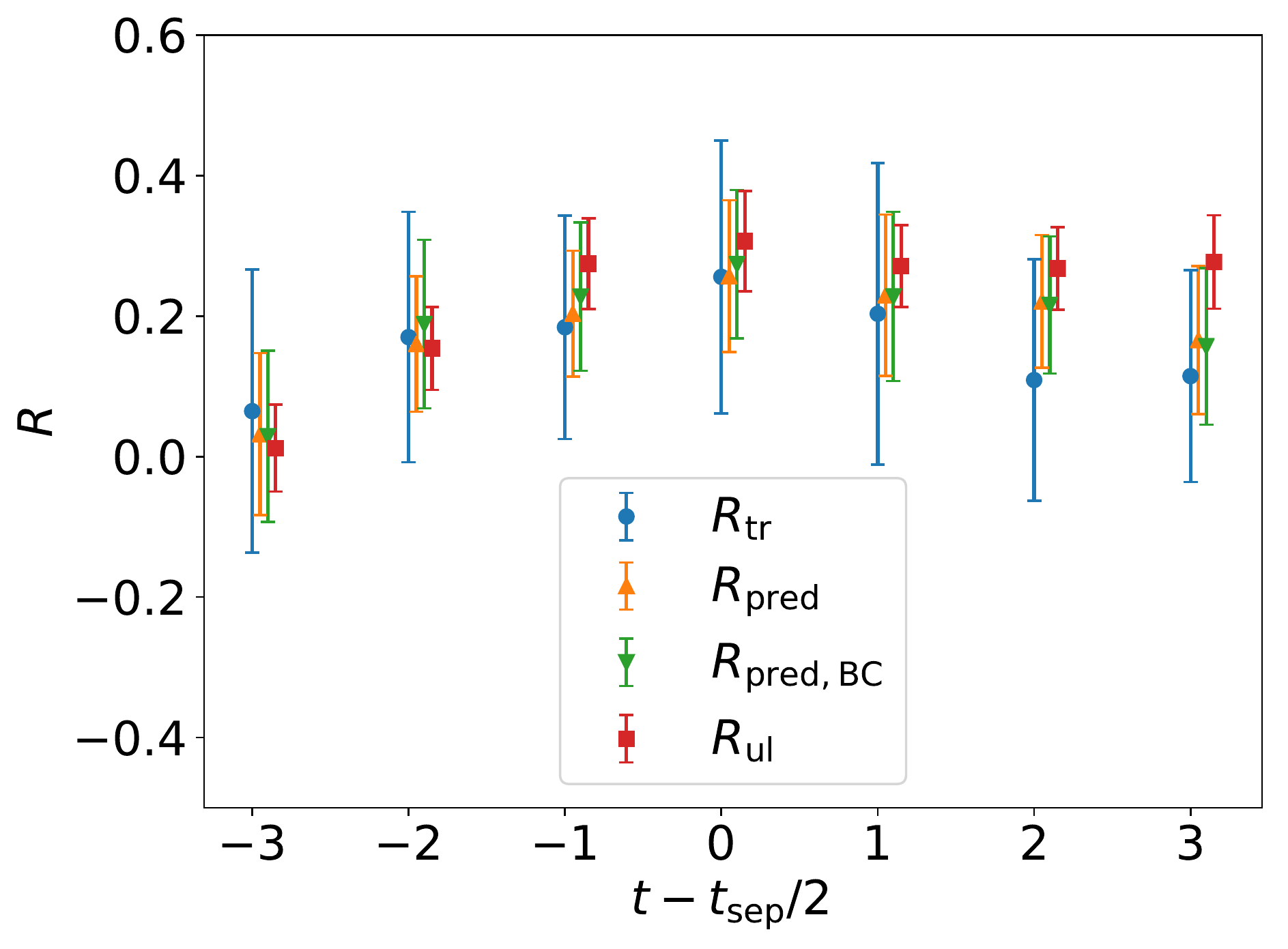}
	\includegraphics[width=0.49\textwidth]{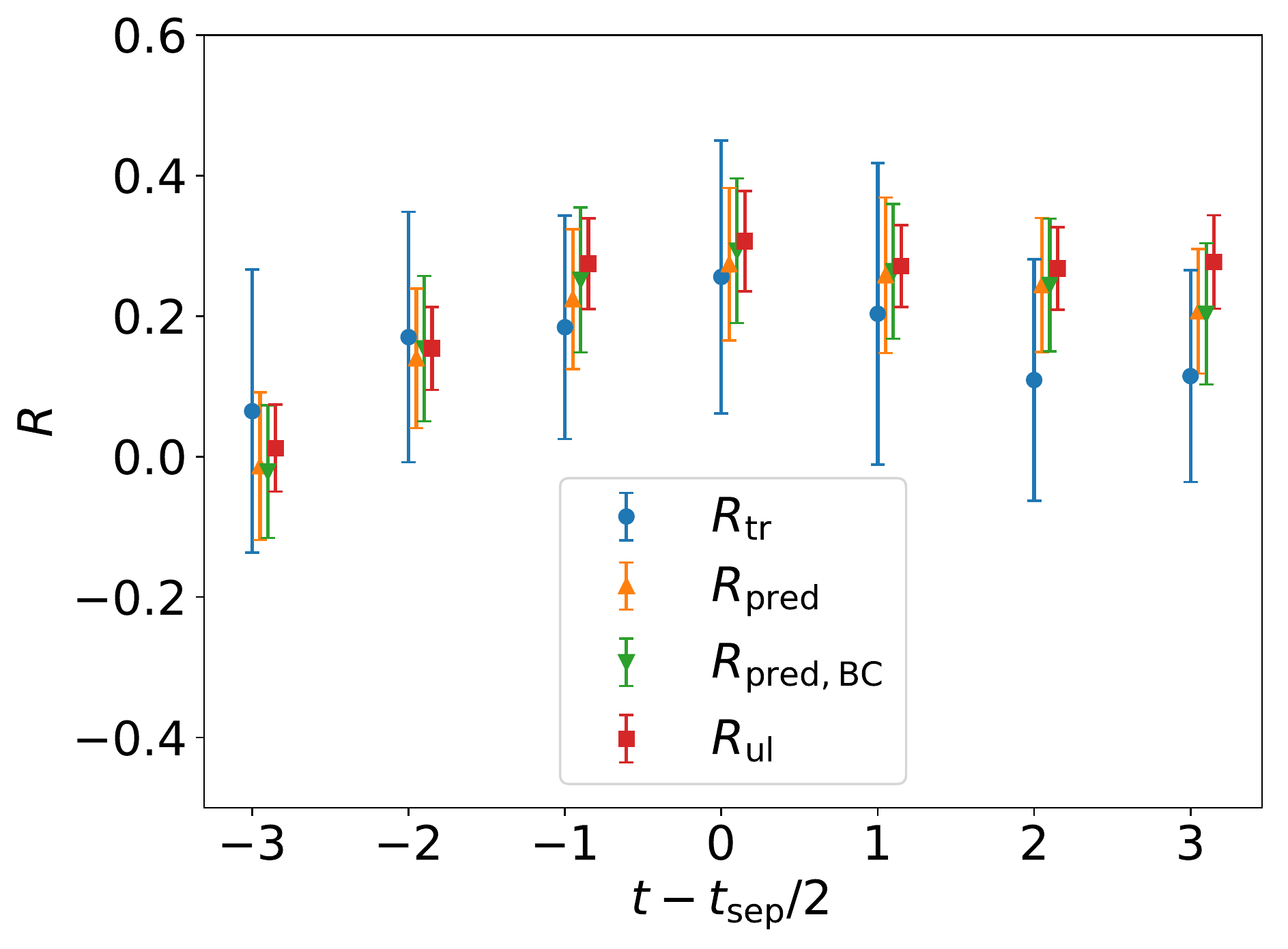}
	\caption{The observed/predicted gluon correlator ratios of the clover valence fermions at $p_\text{pred}=5$, $z_\text{pred}=3$ from $p_\text{in}=4$, $z_\text{in}=3$ (upper) and $p_\text{pred}=5$, $z_\text{pred}=3$ from $p_\text{in}=5$, $z_\text{in}=2$ (lower) by using $N_\text{tr}=2880$, $N_\text{BC}=2880$,  $N_\text{ul}=230400$, $r=0.2$, and $N_\text{est}=200$. The GBT and linear-regressor results are shown in the left and right columns, respectively. The predictions with bias correction do not much improve the raw predictions.}
	\label{fig:gluon_pred_clover}
\end{figure}

The observed/predicted gluon correlator ratios of the clover valence fermions of the GBT and linear regressor model at $p_\text{pred}=5$, $z_\text{pred}=3$ are shown in Fig.~\ref{fig:gluon_pred_clover}. The linear model gives a slightly better results. In Table~\ref{table-clover}, two-point and three point correlator data at $p_\text{in}= 4$, $z_\text{in}=2$, $t_\text{sep}=8$ are used for predicting $p_\text{pred}= 2$, $z_\text{pred}=3$, $t_\text{sep}=8$ correlator. The data at insertion time $t=4$ are shown. Compared with the overlap-fermion result in Table~\ref{table-overlap}, the fit variance is much higher, due to the input data having stronger correlations with the target data.

\begin{table}[h]
	\centering
	\begin{tabular}{@{} |c|c|c|c|c|c|c|c| @{}}
		\hline
		Type& Input&Method   & $R_\text{tr}$ & $R_\text{pred}$ & $R_{\text{pred},\text{BC}}$ & $R_\text{ul}$  &$F_v$ \\
		\hline
		\multirow{2}{*}{$p$-pred}& \multirow{2}{*}{$p_\text{in}=4,z_\text{in}=3$}
		&GBT & 0.26(19) & 0.300(91) & 0.296(92) & 0.307(72) & 0.733(62)\\
		\cline{3-8}
		& &linear & 0.26(19) & 0.28(11) & 0.279(98) & 0.307(72) & 0.845(60)\\
		\hline
		\multirow{2}{*}{$z$-pred}& \multirow{2}{*}{$p_\text{in}=5,z_\text{in}=2$}
		&GBT & 0.26(19) & 0.26(11) & 0.27(11) & 0.307(72) & 0.704(62)\\
		\cline{3-8}
		& &linear & 0.26(19) & 0.27(11) & 0.29(10) & 0.307(72) & 0.819(51)\\
		\hline
	\end{tabular}
	\caption{Observations and predictions of gluon correlator ratio for the clover valence fermions and predictions at $p_\text{pred}=5$, $z_\text{pred}=3$, $t_\text{sep}=8$, $t=4$ using $N_\text{tr}=2880$, $N_\text{BC}=2880$, $N_\text{ul}=23040$, $r=0.2$, and $N_\text{est}=200$. The linear model shows a better fit variance than GBT.}
	\label{table-clover}
\end{table}

\section{Summary}\label{sec:summary}
In this article, we applied the ML technique to quasi-DA and quasi-PDF correlators. Using both GBT model and linear model, we tried to predict the $C_\text{2pt}$ for meson quasi-DAs and the $C_\text{3pt}$ for meson and gluon quasi-PDFs at larger momenta and link lengths, which are noisier and need more computational resources. By predicting from the computationally less expensive data, we are able to reduce the computational cost. Systematic uncertainties from the ML prediction errors are converted to the statistical uncertainties by using the bias correction procedure. With the full bootstrap resampling, we effectively estimated and compared the errors of different model predictions.

Table~\ref{tab:summary_label} summarizes the best fit variances $F_v$ of all predictions we investigated. 
It is observed that for meson datasets, the data from different links are more correlated than those of different momenta. Consequently, the $z$-predictions for both models work much better than $p$-predictions. The ML approach on the $z$-prediction
of quasi-DAs and meson quasi-PDFs is very precise, while the $p$-predictions and the predictions for gluon quasi-PDFs show relatively worse precision. By comparing two ML regression models, we find that the linear model is preferred on cleaner datasets when the correlations between input data and target data are good enough, such as the $z$-prediction of meson-DAs and meson PDFs. On the other hand, the GBT model is more
robust to noisy and less-obviously correlated inputs. For the $p$-prediction of meson quasi-PDFs, both models are able to give a computational cost reduction of $16\%$.

\begin{table}[ht]
    \centering
    \begin{tabular}{|c|c|c|c|c|c|} 
    \hline
         Type & Method & $\eta_s$DA & kaon PDF & overlap gluon PDF& clover gluon PDF \\
            \hline
        \multirow{2}{*}{$z$-pred}& GBT & 0.62(14)& 0.890(26)& 0.53(12)& 0.704(62)\\
            \cline{2-6}
            &linear & 0.99935(40)& 0.998(1)& 0.665(79)& 0.819(51)\\
            \hline
        \multirow{2}{*}{$p$-pred}& GBT& 0.50(13)& 0.692(14)& 0.07(18)& 0.733(62)\\
            \cline{2-6}
            &linear & 0.911(43)& 0.772(29)& $-0.05(38)$& 0.845(60)\\
         \hline
    \end{tabular}
    \caption{Summary of fit variance $F_v$ for all the cases we investigate. The larger value of $F_v$ indicates the better predictions, and a perfect prediction yields $F_v=1.0$. In general, $z-$predictions work better than $p-$predictions, and linear model shows better performance than GBT on our dataset.}
    \label{tab:summary_label}
\end{table}
\section*{Acknowledgments}
We thank the MILC Collaboration and RBC Collaboration for sharing the lattices used to perform this study. The LQCD calculations were performed using the Chroma software
suite~\cite{Edwards:2004sx}.
This research used resources of the National Energy Research Scientific Computing Center, a DOE Office of Science User Facility supported by the Office of Science of the U.S. Department of Energy under Contract No. DE-AC02-05CH11231 through ERCAP; 
facilities of the USQCD Collaboration, which are funded by the Office of Science of the U.S. Department of Energy, 
and supported in part by Michigan State University through computational resources provided by the Institute for Cyber-Enabled Research. 
RL, ZF, HL and RZ are supported by the US National Science Foundation under grant PHY 1653405 ``CAREER: Constraining Parton Distribution Functions for New-Physics Searches''.
BY is supported by the U.S. Department of Energy, Office of Science, Office of High Energy Physics under Contract No. 89233218CNA000001 and by the Los Alamos National Laboratory (LANL) LDRD program. BY also acknowledges support from the U.S. Department of Energy, Office of Science, Office of Advanced Scientific Computing Research and Office of Nuclear Physics, Scientific Discovery through Advanced Computing (SciDAC) program.

\bibliographystyle{apsrev4-1}

\end{document}